\documentclass[preprint]{aastex}

\begin{document}

\shortauthors{Welty, Xue, \& Wong}
\shorttitle{Magellanic Clouds \ion{H}{1} and H$_2$}


\title{Interstellar \ion{H}{1} and H$_2$ in the Magellanic Clouds:  An Expanded Sample Based on UV Absorption-Line Data\footnotemark}
\footnotetext{Based in part on observations made with the NASA/ESA {\it Hubble Space Telescope}, obtained from the data archive at the Space Telescope Science Institute. STScI is operated by the Association of Universities for Research in Astronomy, Inc. under NASA contract NAS 5-26555.
This paper also includes archived data obtained through the Australia Telescope Online Archive (http://atoa.atnf.csiro.au), the MAST archive at STScI ({\it FUSE} data), and the University of Bonn (LAB and GASS 21~cm surveys).}

\author{Daniel E. Welty\altaffilmark{2,3}, Rui Xue\altaffilmark{2}, and Tony Wong\altaffilmark{2}}

\altaffiltext{2}{University of Illinois at Urbana/Champaign, Astronomy Department, 1002 W. Green St., Urbana, IL 61801}
\altaffiltext{3}{current address:  University of Chicago, Astronomy \& Astrophysics Center, 5640 S. Ellis Ave., Chicago, IL 60637; dwelty@oddjob.uchicago.edu}
\begin{abstract} 

We have determined column densities of \ion{H}{1} and/or H$_2$ for sight lines in the Magellanic Clouds from archival {\it HST} and {\it FUSE} spectra of \ion{H}{1} Lyman-$\alpha$ and H$_2$ Lyman-band absorption.
Together with some similar data from the literature, we now have absorption-based $N$(\ion{H}{1}) and/or $N$(H$_2$) for 285 LMC and SMC sight lines (114 with a detection or limit for both species) -- enabling more extensive, direct, and accurate determinations of molecular fractions, gas-to-dust ratios, and elemental depletions in these two nearby, low-metallicity galaxies.
For sight lines where the $N$(\ion{H}{1}) estimated from 21~cm emission is significantly higher than the value derived from Lyman-$\alpha$ absorption (presumably due to emission from gas beyond the target stars), integration of the 21~cm profile only over the velocity range seen in \ion{Na}{1} or H$_2$ absorption generally yields much better agreement. 
Conversely, $N$(21~cm) can be lower than $N$(Ly-$\alpha$) by factors of 2--3 in some LMC sight lines -- suggestive of small-scale structure within the 21~cm beam(s) and/or some saturation in the emission.
The mean gas-to-dust ratios obtained from $N$(H$_{\rm tot}$)/$E(B-V)$ are larger than in our Galaxy, by factors of 2.8--2.9 in the LMC and 4.1--5.2 in the SMC -- i.e., factors similar to the differences in metallicity.
The $N$(H$_2$)/$E(B-V)$ ratios are more similar in the three galaxies, but with considerable scatter within each galaxy.
These data may be used to test models of the atomic-to-molecular transition at low metallicities and predictions of $N$(H$_2$) based on comparisons of 21~cm emission and the IR emission from dust.

\end{abstract}


\section{INTRODUCTION}
\label{sec-intro}

Atomic and molecular hydrogen are the primary constituents of the interstellar medium (ISM) in galaxies.
Knowledge of the abundance and distribution of \ion{H}{1} and H$_2$ is thus important for understanding all aspects of the ISM -- e.g., kinematics and dynamics (on both large and small scales), metallicities, the depletions of heavier elements, and chemistry (in both diffuse and denser gas).
Moreover, understanding the relationship between \ion{H}{1} and H$_2$ is crucial for understanding the formation of molecular clouds -- with consequent implications for star formation and galactic evolution.
The abundance of H$_2$ is typically set by a balance between formation on grains and photodissociation by ambient UV radiation in the gas phase. 
Because the dissociation happens via line transitions, self-shielding is possible once sufficient H$_2$ column density is achieved -- leading to rapid, dramatic increases in the molecular fraction.
Theoretical models thus suggest a fairly sharp transition between locally \ion{H}{1}-dominated and H$_2$-dominated regions.

In extragalactic systems, the distribution of atomic gas is usually determined from radio observations of \ion{H}{1} 21~cm emission.
Because H$_2$ is difficult to measure directly in emission, its abundance is usually inferred from tracer species such as CO, often via the so-called ``$X$-factor'' derived from observations of dense molecular clouds (e.g., Dame, Hartmann, \& Thaddeus 2001).
Such estimates depend on several assumptions -- e.g., that the clouds are virialized and that the H$_2$ and CO are more or less coextensive -- which may not always be satisfied.
Moreover, the $X$-factor may vary with metallicity (e.g., Israel 1997; Leroy et al. 2007, 2011).

Both theoretical models of interstellar chemistry and observations of the Galactic ISM suggest that the relationship between CO and H$_2$ can be rather complex.
While CO also can self-shield, C and O are much less abundant than H -- so the H$_2$ will self-shield first (at lower overall column densities), and the regions where CO is the dominant carbon-containing species will be fewer and smaller than those where hydrogen is mostly molecular.
There thus may be substantial amounts of ``dark'' H$_2$ not traced by CO emission even in the nearby Galactic ISM (e.g., Goldsmith et al. 2008; Wolfire, Hollenbach, \& McKee 2010); such regions are expected to be even more extensive in lower-metallicity systems, due to both reduced dust shielding and stronger UV radiation fields (Maloney \& Black 1988; Pak et al. 1998).
Conversely, measurable CO emission can arise even from relatively diffuse molecular gas characterized by modest molecular fractions $f$(H$_2$) = 2$N$(H$_2$)/[$N$(\ion{H}{1})+2$N$(H$_2$)] (Liszt, Pety, \& Lucas 2010) -- and thus does not necessarily signify the presence of dense, predominantly molecular gas.

The distribution of H$_2$ has also been inferred by comparing observations of the IR emission from dust (which is thought to trace both atomic and molecular gas) and the 21~cm emission from the atomic component.
While this method also involves several assumptions -- e.g., equal dust-to-gas ratios for both the \ion{H}{1}- and H$_2$-dominated gas and optically thin 21~cm emission -- such analyses have provided additional evidence for significant amounts of H$_2$ not traced by CO emission (e.g., Leroy et al. 2007, 2009; Bernard et al. 2008; Roman-Duval et al. 2010).

The abundances of both \ion{H}{1} and H$_2$ may also be measured and compared directly, via their UV absorption features in the spectra of suitable background targets.
For respective column densities greater than about 10$^{18}$ cm$^{-2}$, both the \ion{H}{1} Lyman-$\alpha$ line at 1216 \AA\ and the strongest rotational transitions (from $J$=0,1) in the far-UV Lyman and Werner bands of H$_2$ exhibit damped profiles, from which accurate total $N$(\ion{H}{1}) and $N$(H$_2$) may be obtained.
Such measurements can be more sensitive than the emission-line observations and are not complicated by differences in beam size -- but they are limited to individual sight lines and generally cannot distinguish individual velocity components.
General surveys of Galactic absorption, based on UV spectra obtained with the {\it Copernicus}, {\it International Ultraviolet Explorer (IUE)}, and/or {\it Far-Ultraviolet Spectroscopic Explorer (FUSE)} satellites, have been undertaken for both \ion{H}{1} (Bohlin, Savage, \& Drake 1978; Shull \& van Steenberg 1985; Diplas \& Savage 1994a) and H$_2$ (Savage et al. 1977; Rachford et al. 2002, 2009; Gillmon et al. 2006) -- though such surveys have generally been limited to within several kpc in the Galactic plane due to extinction by dust.

The relatively nearby Magellanic Clouds, characterized by metallicities roughly 0.5$\times$solar (LMC) and 0.2$\times$solar (SMC) (e.g., Smith 1999; see also references in appendices to Welty et al. 1997, 1999), provide convenient venues for exploring both local and global relationships between atomic and molecular gas in lower-metallicity systems.
High-resolution surveys of the emission from \ion{H}{1} (Stanimirovi\'{c} et al. 1999; Kim et al. 2003), CO (Fukui et al. 2008; Wong et al. 2011), and dust (Leroy et al. 2007; Bernard et al. 2008) have been performed for both the LMC and SMC -- enabling detailed inter-comparisons of the distribution and abundances of those tracers for diverse locations and environments in the Magellanic Clouds. 
Comparisons of the emission maps with absorption-line data (from UV/optical/radio spectra of targets both within and beyond the Magellanic Clouds) can provide both information on the 3-dimensional structure of the ISM and an environmental context for the elemental abundances and physical conditions derived from the spectroscopic data.
Such multi-faceted studies of the ISM in the Magellanic Clouds should aid in understanding more distant low-metallicity systems, where only lower-resolution emission maps and/or few (if any) absorption-line probes will be available.

Unfortunately, the existing UV absorption-line measurements of \ion{H}{1} and H$_2$ in the Magellanic Clouds are somewhat limited (in number and/or accuracy).
Most previous determinations of $N$(\ion{H}{1}) from Lyman-$\alpha$ absorption, for example, were based on relatively low-resolution, low-S/N spectra obtained with {\it IUE} (e.g., Fitzpatrick 1985a, 1985b, 1986) -- allowing only fairly rough overall estimates for dust-to-gas ratios, with no information on H$_2$.
Accurate values for $N$(H$_2$), derived from far-UV spectra obtained with {\it FUSE}, have since been reported for about 80 MC sight lines (Tumlinson et al. 2002; Cartledge et al 2005).
Tumlinson et al. (2002) found both generally lower molecular fractions $f$(H$_2$) and generally higher H$_2$ rotational excitation in the Magellanic Clouds, compared to typical values in the local Galactic ISM.
Those characteristics seem to require both lower H$_2$ formation rates (due largely to the lower dust-to-gas ratios) and enhanced destruction of H$_2$ (due to stronger UV radiation fields). 
Because of the paucity of reliable Lyman-$\alpha$ data then available, however, Tumlinson et al. (2002) used estimates for $N$(\ion{H}{1}) derived from 21~cm emission, with an uncertain average adjustment for foreground-background and small-scale structure effects, in order to obtain molecular fractions.
Moreover, most of the sight lines in that study are fairly lightly reddened and exhibit fairly small molecular fractions, as they were chosen for studies of the properties of the target stars themselves.
In the course of examining the far-UV extinction behavior in the Magellanic Clouds, Cartledge et al. (2005) subsequently estimated H$_2$ column densities for a small number of more heavily reddened sight lines, including two sight lines with $f$(H$_2$) $>$ 0.5.

Since those studies, suitable UV spectra of many more LMC and SMC sight lines have been obtained with {\it Hubble Space Telescope (HST)} (now totalling more than 200 sight lines) and {\it FUSE} (nearly 300 sight lines) -- with significant overlap between the two data sets.
In this paper, we present an expanded sample of \ion{H}{1} and H$_2$ column densities (particularly for the higher column density regime in H$_2$) for the Magellanic Clouds, combining values from our analyses of the \ion{H}{1} Lyman-$\alpha$ and H$_2$ Lyman-band absorption in those archived UV spectra with previously published values.
For each sight line in our sample, we also list new determinations of $E(B-V)$ (derived from spectral types and photometry from the literature) and estimates of the total sight line \ion{H}{1} column density derived from the ATCA+Parkes 21~cm emission spectra.
Section 2 describes the UV spectral data and methods used to determine the \ion{H}{1} and H$_2$ column densities.
Section 3 compares the \ion{H}{1} column densities obtained from Lyman-$\alpha$ absorption and 21~cm emission, examines the gas-to-dust ratios [both average values and variations; as estimated from $N$(H$_{\rm tot}$) and $E(B-V)$], and briefly explores some implications of these data for understanding the atomic-to-molecular transition in low-metallicity gas and for evaluating predictions of H$_2$ abundances based on comparisons of IR emission from dust and \ion{H}{1} 21~cm emission.
Section 4 summarizes our results.
 
\section{DATA}
\label{sec-data}

\subsection{Stellar Sample}
\label{sec-stel}

\begin{figure}
\epsscale{0.9}
\plotone{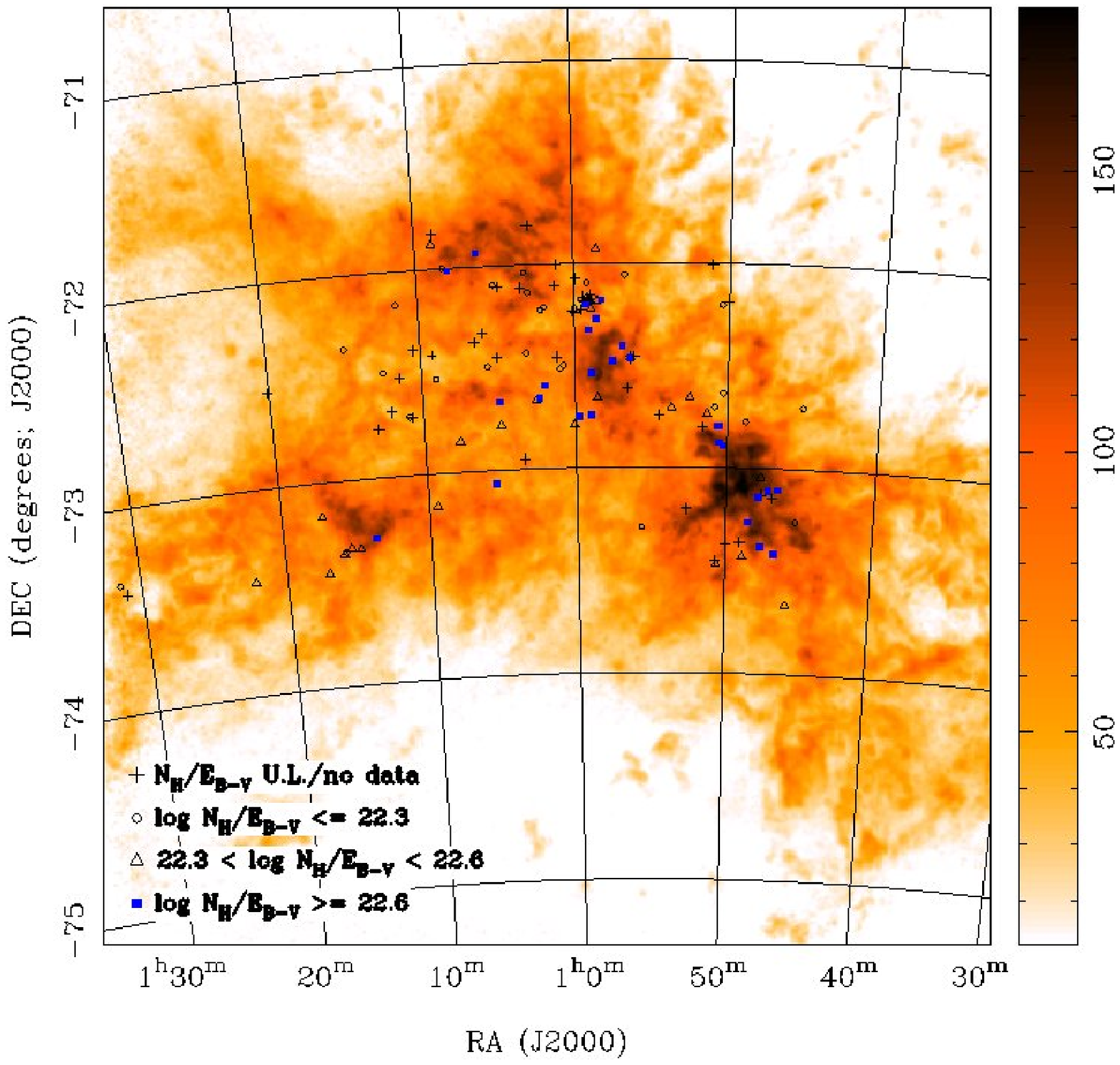}
\caption{SMC sight lines included in this survey, plotted on a map of the peak brigthness temperature of the \ion{H}{1} 21~cm emission (Stanimirovi\'{c} et al. 1999).
The sight lines are coded according to the gas-to-dust ratio $N$(H$_{\rm tot}$)/$E(B-V)$ (in three ranges); sight lines with no data for $N$(\ion{H}{1}) and/or $E(B-V)$ or with $E(B-V)$ $\le$ 0.03 are denoted by crosses.
A wide variety of regions/environments are sampled; there are no obvious large-scale regional variations in the gas-to-dust ratio.}
\label{fig:smclos}
\end{figure}

\begin{figure} 
\epsscale{0.9}
\plotone{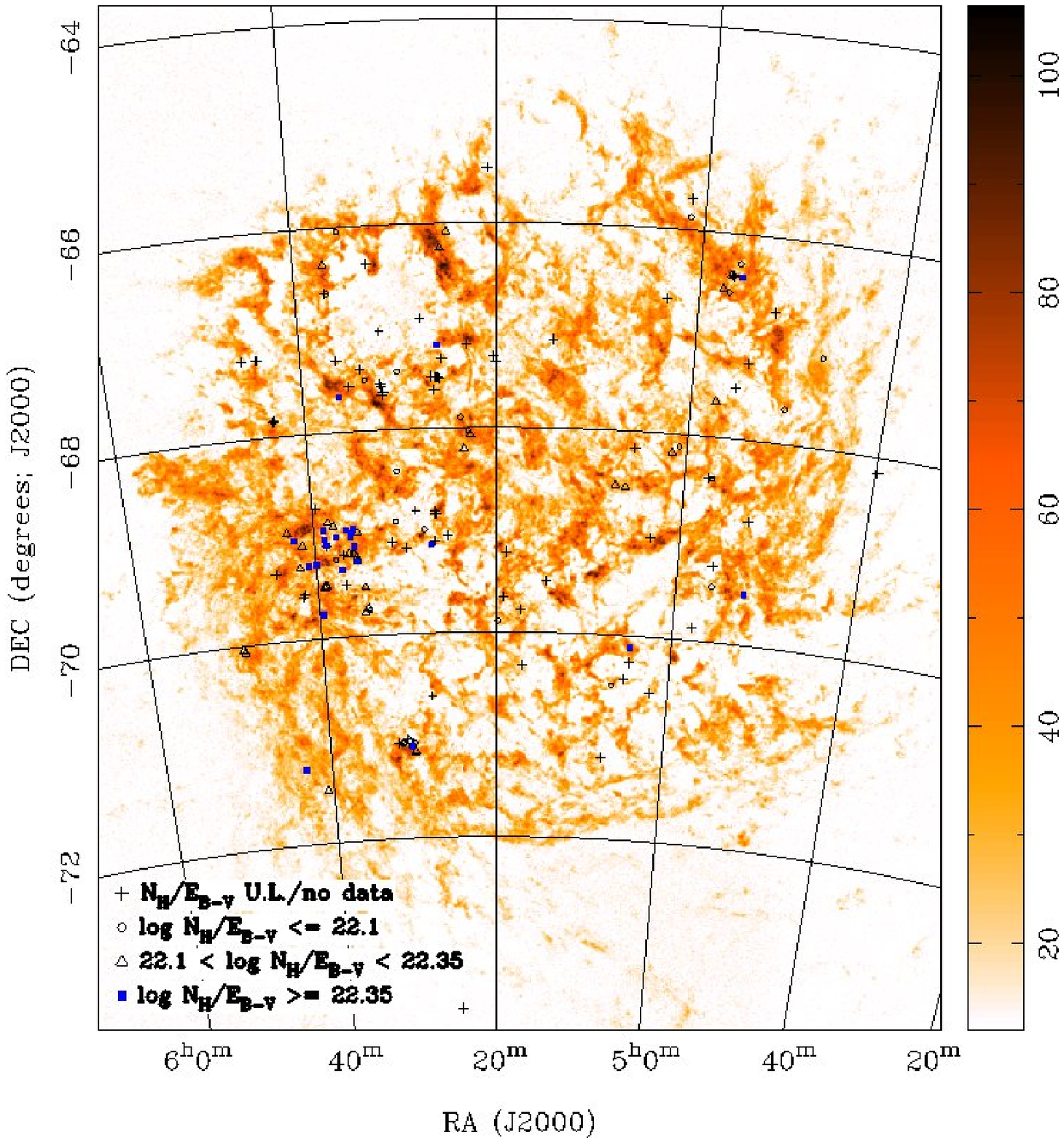}
\caption{LMC sight lines included in this survey, plotted on a map of the peak brightness temperature of the \ion{H}{1} 21~cm emission (Kim et al. 2003).
The sight lines are coded according to the gas-to-dust ratio $N$(H$_{\rm tot}$)/$E(B-V)$ (in three ranges); sight lines with no data for $N$(\ion{H}{1}) and/or $E(B-V)$ or with $E(B-V)$ $\le$ 0.03 are denoted by crosses.
A wide variety of regions/environments are sampled; there are no obvious large-scale regional variations in the gas-to-dust ratio.}
\label{fig:lmclos}
\end{figure}

Tables~\ref{tab:smclos}~and~\ref{tab:lmclos} list the 126 SMC and 159 LMC sight lines included in this survey.
For each sight line, the tables give equatorial coordinates (J2000)\footnotemark, spectral types, $V$, $B-V$, $E(B-V)$ (both total and MC), the column densities of \ion{H}{1} (from Lyman-$\alpha$ or Lyman-$\beta$ absorption), H$_2$, and \ion{H}{1} (from 21~cm emission), the H$_2$ rotational temperature $T_{01}$, and the peak 21~cm brightness temperature $T_{\rm pk}$.
\footnotetext{The coordinates are from the catalogs compiled by B. Skiff, available at ftp.lowell.edu/pub/bas/starcats.}
Figures~\ref{fig:smclos} and \ref{fig:lmclos} show the locations of the sight lines on maps of $T_{\rm pk}$(21~cm) (Stanimirovi\'{c} et al. 1999; Kim et al. 2003), with the symbols coded by $N$(H$_{\rm tot}$)/$E(B-V)$ (see Sec.~\ref{sec-gdr}).
A wide variety of regions in the two galaxies has been sampled, with broad coverage of both the main stellar ``bar'' and the ``wing'' region in the SMC and many locations across the LMC; a number of sight lines in the NGC~346 region of the SMC and in the LH10, LMC~4, and 30~Dor regions of the LMC are included.
The sight lines sample a variety of environmental conditions --- as indicated by overall column densities, molecular fractions, proximity to \ion{H}{2} regions and molecular clouds, UV extinction characteristics (e.g., Gordon et al. 2003), and inferred local physical conditions (e.g., Welty et al. 2006).
It is not an unbiased sample, however, as many of the \ion{H}{1} targets were originally chosen for studies of stellar properties (and are typically lightly reddened), while others were chosen for studies of interstellar extinction (and thus are more heavily reddened).
The sight lines with new measurements of molecular hydrogen were selected, via inspection of the spectra in the {\it FUSE} Magellanic Clouds Legacy archive (Blair et al. 2009), specifically to have relatively high $N_{\rm MC}$(H$_2$).
For the SMC, there are now 119 sight lines with a detection or upper limit for \ion{H}{1} (92 new) and 65 sight lines with a detection or limit for H$_2$ (39 new); 57 sight lines have a detection or limit for both species.
For the LMC, there are now 136 sight lines with a detection or limit for \ion{H}{1} (89 new) and 80 sight lines with a detection or limit for H$_2$ (30 new); 57 sight lines have a detection or limit for both species.

Spectral types, $V$ magnitudes, and $B-V$ colors for the target stars were taken from the references given in footnotes to the tables.
Nearly all of the types are based on slit spectra.
Most of the stars are O or early B supergiants, with 11.0 $\la$ $V$ $\la$ 15.0; some appear to be unresolved binary or multiple systems.
In general, we preferred photometry derived from ccd observations (e.g., Massey 2002), which should allow more reliable corrections for near neighbors and local background fluctuations than is generally possible for aperture photometry.
For $V$ $\la$ 15.0 mag, the uncertainties on $B$ and $V$ (and $B-V$) are typically $\le$ 0.03 mag (Massey 2002).

In most cases, the total $E(B-V)$ color excesses (including both Galactic and Magellanic Clouds contributions) were obtained using the intrinsic colors adopted by FitzGerald (1970) or Walborn (2002).  
Uncertainties in $(B-V)_0$ corresponding to those in assigned spectral type are typically $\la$ 0.01--0.02 mag for the early-type stars in this sample.
While intrinsic colors could be estimated from theoretical spectral energy distributions for some of the Wolf-Rayet stars (e.g., Crowther 2007), the contributions of the emission lines to the broadband fluxes render such estimates rather uncertain for some of the other W-R subtypes.
Given the variations both in the Galactic foreground contributions and in the strength of the 2175 \AA\ extinction bump in the SMC and LMC (e.g., Gordon et al. 2003), estimates of W-R reddening based on ``removing the 2175 \AA\ bump'' from the observed stellar fluxes also are likely to be rather uncertain.
Color excesses thus are not given for some of the W-R stars in our sample.

For the SMC sight lines, values of 0.03 to 0.04 mag were adopted for the Galactic portion of the total $E(B-V)$ (Schlegel, Finkbeiner, \& Davis 1998; adjusted based on the strength of the Galactic 21~cm emission in each case).
For the LMC sight lines, Galactic values ranging from 0.02 to 0.09 mag were estimated from Figure 13 of Staveley-Smith et al. (2003) [which is based on \ion{H}{1} 21~cm emission data and the relationship between $N$(H) and $E(B-V)$ for Galactic halo sight lines found by Burstein \& Heiles 1978].  
The second $E(B-V)$ entry listed for each sight line represents the portion due to dust in the SMC or LMC (total minus Galactic); negative values of $E(B-V)_{\rm MC}$ obtained for several sight lines have been retained in the tables.
The Magellanic Clouds color excesses range from $-$0.03 to 0.33 for the SMC sight lines and from 0.00 to 0.74 for the LMC sight lines;
only 12 SMC and 32 LMC sight lines have $E(B-V)_{\rm MC}$ greater than 0.2 mag.
The uncertainties in $E(B-V)_{\rm MC}$ -- including contributions from the photometry, spectral typing, and Galactic foreground -- are typically 0.03--0.04 mag.

\subsection{Observations and Data Processing} 
\label{sec-obs}

The {\it HST} data for the \ion{H}{1} Lyman $\alpha$ absorption lines analyzed in this survey were obtained under various observing programs, many of which were aimed at understanding the properties of the target stars themselves (Appendix Table~\ref{tab:prog}).
Table~\ref{tab:prog} lists the instrumental configurations employed in those programs, with effective spectral resolutions at Lyman $\alpha$ ranging from 0.011 \AA\ (2.75 km~s$^{-1}$) for STIS/E140H to $\sim$1.2 \AA\ ($\sim$300 km~s$^{-1}$) for STIS/G140L.
For comparison, the resolutions characterizing the high- and low-resolution spectra of Lyman-$\alpha$ obtained with {\it IUE} are of order 0.1 and 6.0 \AA, respectively.
The pipeline-processed spectra were retrieved from the MAST archive\footnotemark.
\footnotetext{http://archive.stsci.edu/hst}
For the highest resolution STIS echelle spectra, the multiple spectral orders within the range 1170--1260 \AA\ were smoothed and combined, using a linear ``tilt'' within each order (when necessary) to match the flux levels in adjacent orders.
A constant background, gauged from any non-zero flux in the saturated core of the Lyman $\alpha$ line (and not due to geocoronal emission), was subtracted from some of the lower resolution spectra (e.g., FOS/G130H).

Far-UV spectra of nearly 300 LMC and SMC targets are available through the {\it FUSE} Magellanic Clouds Legacy Project (Blair et al. 2009).
The {\it FUSE} data for \ion{H}{1} Lyman-$\beta$ and H$_2$ absorption, for sight lines exhibiting strong Magellanic Clouds H$_2$ absorption in the $J$ = 0 and 1 lines of the strongest Lyman bands, were retrieved from the Legacy Program website\footnotemark.
\footnotetext{http://archive.stsci.edu/prepds/fuse\_mc}
The LiF1a spectra were smoothed by a factor of 3 to achieve approximately optimal sampling, assuming a resolution of $\sim$ 20 km~s$^{-1}$.
Low order polynomial fits to line-free continuum regions were used to normalize the spectra near the strong R0, R1, and P1 lines of the H$_2$ Lyman bands chosen for fitting.

Several surveys of \ion{H}{1} 21~cm emission from the Magellanic Clouds are now available:  the high-resolution surveys of the SMC and LMC based on data from the Australia Telescope Compact Array (ATCA) and the 64-m Parkes telescope (FWHM $\sim$ 1.0--1.5 arcmin; Stanimirovi\'{c} et al. 1998, 1999; Kim et al. 2003; Staveley-Smith et al. 2003) and the lower-resolution but more extensive Leiden-Argentina-Bonn (LAB; FWHM = 30--36 arcmin; Kalberla et al. 2005) and Galactic All-Sky Surveys (GASS; FWHM = 14.4 arcmin; Kalberla et al. 2010).
The emission profiles for all the sight lines in our sample were retrieved from the respective survey web sites\footnotemark.
\footnotetext{http://www.atnf.csiro.au/research/smc\_h1/get\_spectrum.html; http://www.astro.uni-bonn.de/hisurvey/profile/}
For the ATCA+Parkes data, the SMC spectra cover the velocity range from about 88 to 215 km~s$^{-1}$ (sampled every 1.65 km~s$^{-1}$) on 60\arcsec pixels; the LMC spectra cover the velocity range from about 190 to 386 km~s$^{-1}$ on 80\arcsec pixels.
The Magellanic Clouds \ion{H}{1} column density and peak brightness temperature derived from the ATCA+Parkes 21~cm profile for the nearest neighbor grid point are listed for each of the SMC and LMC sight lines in Tables~\ref{tab:smclos} and \ref{tab:lmclos}.
The LAB and GASS spectra cover the velocity range from $-$400 to 400 km~s$^{-1}$ (sampled every 1.03 km~s$^{-1}$) -- thus including the Galactic emission; interpolated profiles are produced for each queried sight line.

\subsection{Determination of Column Densities}
\label{sec-cd}

\subsubsection{\ion{H}{1} Lyman-$\alpha$ Absorption}
\label{sec-h1abs}

Most of the new \ion{H}{1} column densities given in Tables~\ref{tab:smclos} and \ref{tab:lmclos} were derived by applying the standard continuum reconstruction method (Bohlin et al. 1978; Diplas \& Savage 1994a) to the observed Lyman-$\alpha$ profiles, using atomic data from Morton (2003).
In that method, the observed profile is divided by a theoretical profile composed of one or more Voigt profiles, varying the parameters of the theoretical profile until the division yields a ''reasonable'' restored continuum.
In general, the best-fit $N_{\rm MC}$(\ion{H}{1}) is determined by the width of the saturated core of the Lyman-$\alpha$ line and by the quality of the reconstructed continuum shortward of that core, as the opposite (red-ward) wing of the line is often blended with the \ion{N}{5} stellar wind P Cygni profile in these early-type stars.
In all cases, the absorption was assumed to be due to two components: a Galactic foreground component and a component due to gas in the SMC or LMC.
For each sight line, the column density of the Galactic component was obtained from the GASS 21~cm emission profile, assuming optically thin emission.
The foreground $N$(\ion{H}{1}) thus range from 2.6--3.7 $\times$ 10$^{20}$ cm$^{-2}$ toward the SMC and from 3.3--7.4 $\times$ 10$^{20}$ cm$^{-2}$ toward the LMC.
These values are in good agreement with the $N_{\rm MW}$(\ion{H}{1}) = 3--4 $\times$ 10$^{20}$ cm$^{-2}$ adopted by Fitzpatrick (1986) for SMC sight lines, and with the values from 4--8 $\times$ 10$^{20}$ cm$^{-2}$ estimated from fig.~13 of Staveley-Smith et al. (2003) toward the LMC.
The velocity separation between the Galactic and Magellanic Clouds components was estimated from observations of other species which typically are concentrated in the main \ion{H}{1} component(s) [e.g., \ion{Na}{1} (Welty \& Crowther, in prep.) and/or H$_2$], for the same or other nearby sight lines.
Use of more complex component structures -- e.g., as determined from high-resolution UV spectra of \ion{Zn}{2} absorption -- yields total $N$(\ion{H}{1}) essentially indistinguishable from the values derived in the two-component fits (for the few cases where such UV spectra are available).

\begin{figure}
\epsscale{0.8}
\plotone{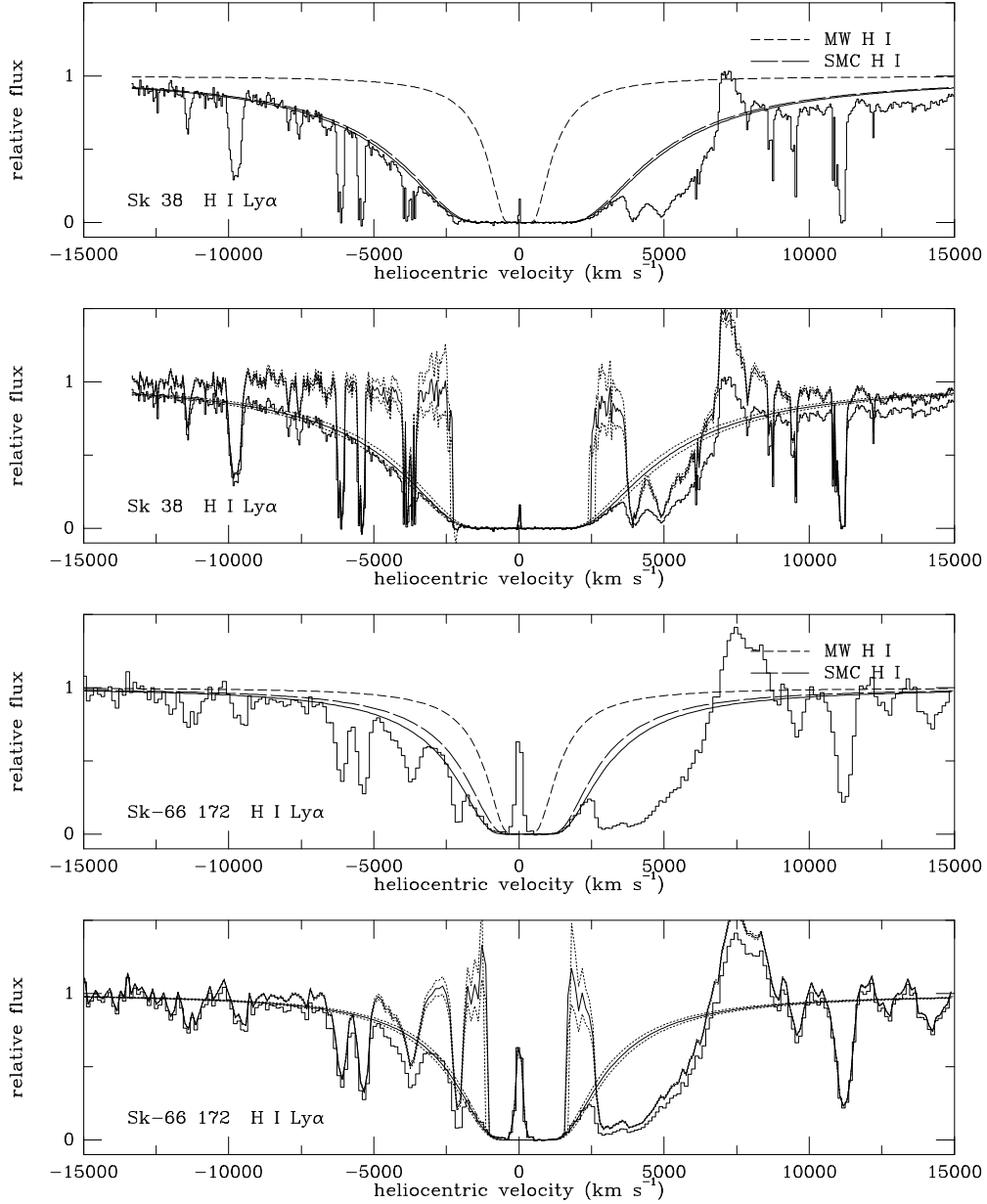}
\caption{Fits to {\it HST} spectra of the region around the \ion{H}{1} Lyman-$\alpha$ line near 1216 \AA\ (solid histograms), for the sight lines toward Sk~38 [SMC; $N_{\rm SMC}$(\ion{H}{1}) = 6.2$\pm$0.6 $\times$ 10$^{21}$ cm$^{-2}$; STIS E140M; upper two panels] and Sk$-$66~172 [LMC; $N_{\rm LMC}$(\ion{H}{1}) = 1.7$\pm$0.2 $\times$ 10$^{21}$ cm$^{-2}$; FOS G130H; lower two panels].
In each pair of panels, the upper panel shows the best fit to the observed profile (solid line) and the individual contributions from our Galaxy (short-dashed line) and from the SMC or LMC (long-dashed line).
In the lower panel of each pair, the smooth solid lines show the best fit to the profile and the corresponding reconstructed continuum; the dotted lines show the profiles and reconstructed continua for the estimated upper and lower limits.
Note the geocoronal Lyman-$\alpha$ emission near 0 km~s$^{-1}$ and the \ion{N}{5} stellar wind P Cygni profile superposed on the red-ward wing of the interstellar \ion{H}{1} absorption.}
\label{fig:lyafit}
\end{figure}

Figure~\ref{fig:lyafit} gives two examples of this procedure -- for the sight lines toward Sk~38, with $N_{\rm SMC}$(\ion{H}{1}) = 6.2 $\times$ 10$^{21}$ cm$^{-2}$, and Sk$-$66~172, with $N_{\rm LMC}$(\ion{H}{1}) = 1.7 $\times$ 10$^{21}$ cm$^{-2}$.
Sk~38 was observed with STIS/E140M, at a resolution of about 6.6 km~s$^{-1}$; Sk$-$66~172 was observed with FOS/G130H, at a resolution of about 225 km~s$^{-1}$.
The upper panel for each sight line shows the adopted best fit to the observed profile (solid lines), with the Galactic and Magellanic Clouds contributions given by the short-dashed and long-dashed lines, respectively.
In both cases, the absorption is dominated by the Magellanic Clouds component; note also the \ion{N}{5} stellar wind P Cygni profile blended with the red-ward wing of the interstellar Lyman-$\alpha$ line.
The narrow geocoronal Lyman-$\alpha$ emission near 0 km~s$^{-1}$ seen in these spectra can be much broader in spectra obtained through wider slits -- making it difficult to determine accurate column densities from the lower-resolution spectra of some of the lower-$N$(\ion{H}{1}) sight lines, as the width of the geocoronal emission can be comparable to that of the saturated core of the interstellar absorption feature in such cases.
The lower panel for each sight line shows the adopted best fit and corresponding reconstructed continuum (solid lines), as well as the fits and continua for Magellanic Clouds column densities higher and lower by about 10\% (dotted lines) -- which is typical of the uncertainties adopted for $N_{\rm MC}$(\ion{H}{1}) $\ga$ 10$^{21}$ cm$^{-2}$.

The \ion{H}{1} column densities derived from fitting the Lyman-$\alpha$ profiles range from about 6 to 110 $\times$ 10$^{20}$ cm$^{-2}$ in the SMC and from about 3.5 to 150 $\times$ 10$^{20}$ cm$^{-2}$ in the LMC.
For individual sight lines observed with more than one {\it HST} instrumental configuration, the differences in the $N_{\rm MC}$(\ion{H}{1}) derived from those multiple observations are generally less than about 5\% (0.02 dex).
Where the possible Magellanic Clouds contribution is less than about 3--4 $\times$ 10$^{20}$ cm$^{-2}$ (i.e., comparable to the Galactic contribution toward the SMC and the minimum Galactic contribution toward the LMC), a limit 3 times that possible $N_{\rm MC}$ is adopted.
Limits are also given for target star spectral types later than B3 (for supergiants) or B1 (for main-sequence stars), where there can be significant stellar contributions to the Lyman-$\alpha$ absorption (e.g., Diplas \& Savage 1994a).

\begin{figure}
\epsscale{0.9}
\plotone{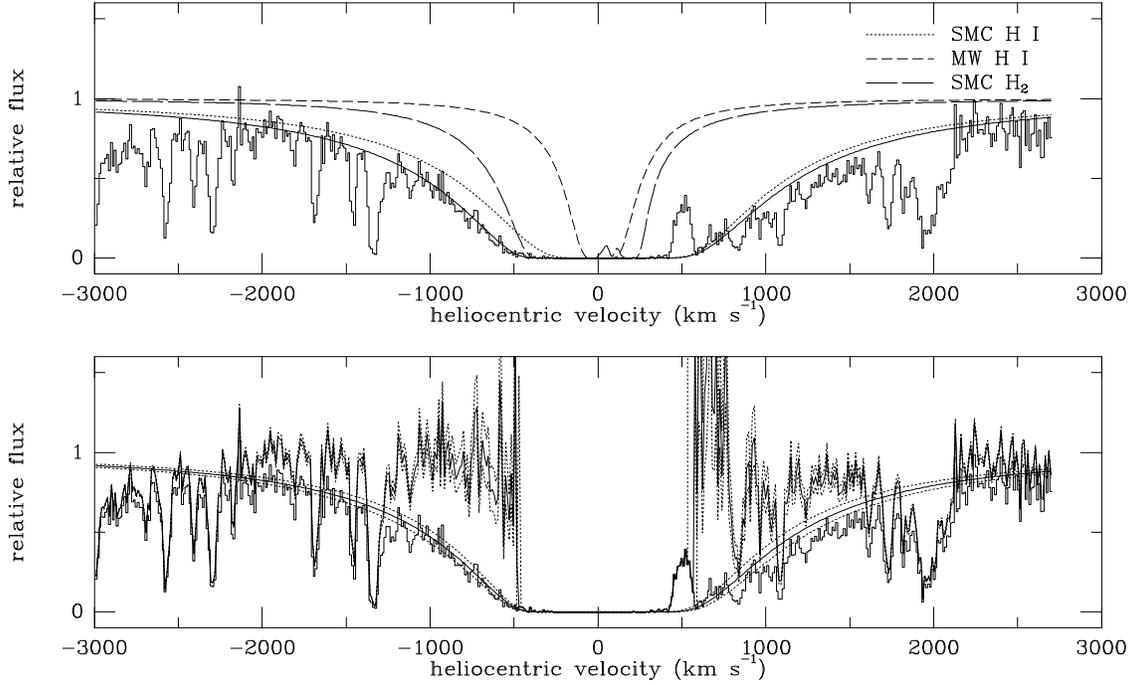}
\caption{Fit to {\it FUSE} spectrum of the region around the \ion{H}{1} Lyman-$\beta$ line near 1025 \AA\ (solid histogram), for the sight line toward the SMC star Sk~18.
The top panel shows the adopted best fit to the profile (solid line), plus the individual contributions from the SMC \ion{H}{1} (dotted line), the Galactic \ion{H}{1} (short-dashed line), and the SMC (and Galactic) H$_2$ [R0, R1, and P1 for the Lyman (6-0) band; $N_{\rm SMC}$(H$_2$) = 4.3 $\times$ 10$^{20}$ cm$^{-2}$; long-dashed line].
The bottom panel shows the continuum reconstruction for the adopted SMC $N$(\ion{H}{1}) = 6.5 $\times$ 10$^{21}$ cm$^{-2}$ (middle dotted line) and for 7.5 and 5.5 $\times$ 10$^{21}$ cm$^{-2}$ (upper and lower dotted lines); the smooth solid lines show the corresponding profiles for those three values of $N$(\ion{H}{1}).
Telluric \ion{O}{1} emission lines are present near 500 and 700 km~s$^{-1}$; the geocoronal Lyman-$\beta$ emission observed in the core of the interstellar absorption has been excised.}
\label{fig:lybfit}
\end{figure}

Where {\it HST} spectra of Lyman $\alpha$ were not available, an attempt was made to determine \ion{H}{1} column densities from the Lyman-$\beta$ line in the {\it FUSE} LiF1a spectra near 1025 \AA\ -- again using the continuum reconstruction method. 
In general, column densities derived from fits to the Lyman-$\beta$ line have somewhat larger uncertainties -- as the damping wings are weaker than for Lyman $\alpha$ and as there can be significant blending with both stellar features (absorption for very early O stars, emission for B supergiants), interstellar H$_2$ absorption [the strong R0, R1, and P1 lines of the Lyman (6-0) band], and telluric emission from several transitions of \ion{O}{1} (in addition to the geocoronal Lyman-$\beta$ emission).
Figure~\ref{fig:lybfit} gives an example of the fit to the Lyman-$\beta$ line toward the SMC star Sk~18.
The upper panel shows the overall best fit to the profile (smooth solid line), as well as the individual contributions from the SMC \ion{H}{1} (dotted line), the Galactic \ion{H}{1} (short-dashed line), and the SMC (and Galactic) H$_2$ (long-dashed line); note that the geocoronal Lyman-$\beta$ line has been removed.
The best fit to the profile yielded $N_{\rm SMC}$(\ion{H}{1}) = 6.5 $\times$ 10$^{21}$ cm$^{-2}$ -- somewhat higher than the value derived from fits to the corresponding Lyman-$\alpha$ profile (5.0 $\times$ 10$^{21}$ cm$^{-2}$).
The lower panel shows the fits and the corresponding reconstructed continua for the adopted $N_{\rm SMC}$(\ion{H}{1}) and for values roughly 15\% higher and lower.

\subsubsection{H$_2$ Absorption}
\label{sec-h2abs}

For sight lines exhibiting fairly strong Magellanic Clouds H$_2$ absorption, column densities for the lowest two rotational levels ($J$ = 0 and 1) were determined in two ways.
The first method employed a slightly modified version of the IDL-based program {\bf h2gui}, which was constructed and used by Tumlinson et al. (2002). 
For strong lines with column densities $\ga$ 10$^{18}$ cm$^{-2}$, the program determines total H$_2$ column densities and rotational excitation temperatures $T_{01}$ via iterative Voigt-profile fits to the lines from $J$ = 0 and 1 in the normalized spectra.
(Other, weaker H$_2$ lines from higher rotational levels are also fitted, but the derived column densities of those lines depend rather sensitively on the generally poorly determined $b$-value -- and are thus unreliable.) 
As the program fits absorption from various transitions at a single velocity, any Galactic absorption features or weaker Magellanic Clouds components must be masked in the fits.
The Lyman (4-0) band of H$_2$ was chosen as the primary band for fitting, as it is a strong band [$f$(R0) = 0.023; Abgrall \& Roueff 1989] in a fairly clean region of the spectrum -- exhibiting little blending with stellar or other interstellar features.
Similar fits to the weaker Lyman (1-0) band [$f$(R0) = 0.0058] generally yielded consistent results, with little dependence on $b$, for sight lines with $N$(H$_2$) $\ga$ 10$^{19}$ cm$^{-2}$; the values derived for lower column density sight lines were more sensitive to the derived $b$, however.

\begin{figure}
\epsscale{0.9}
\plotone{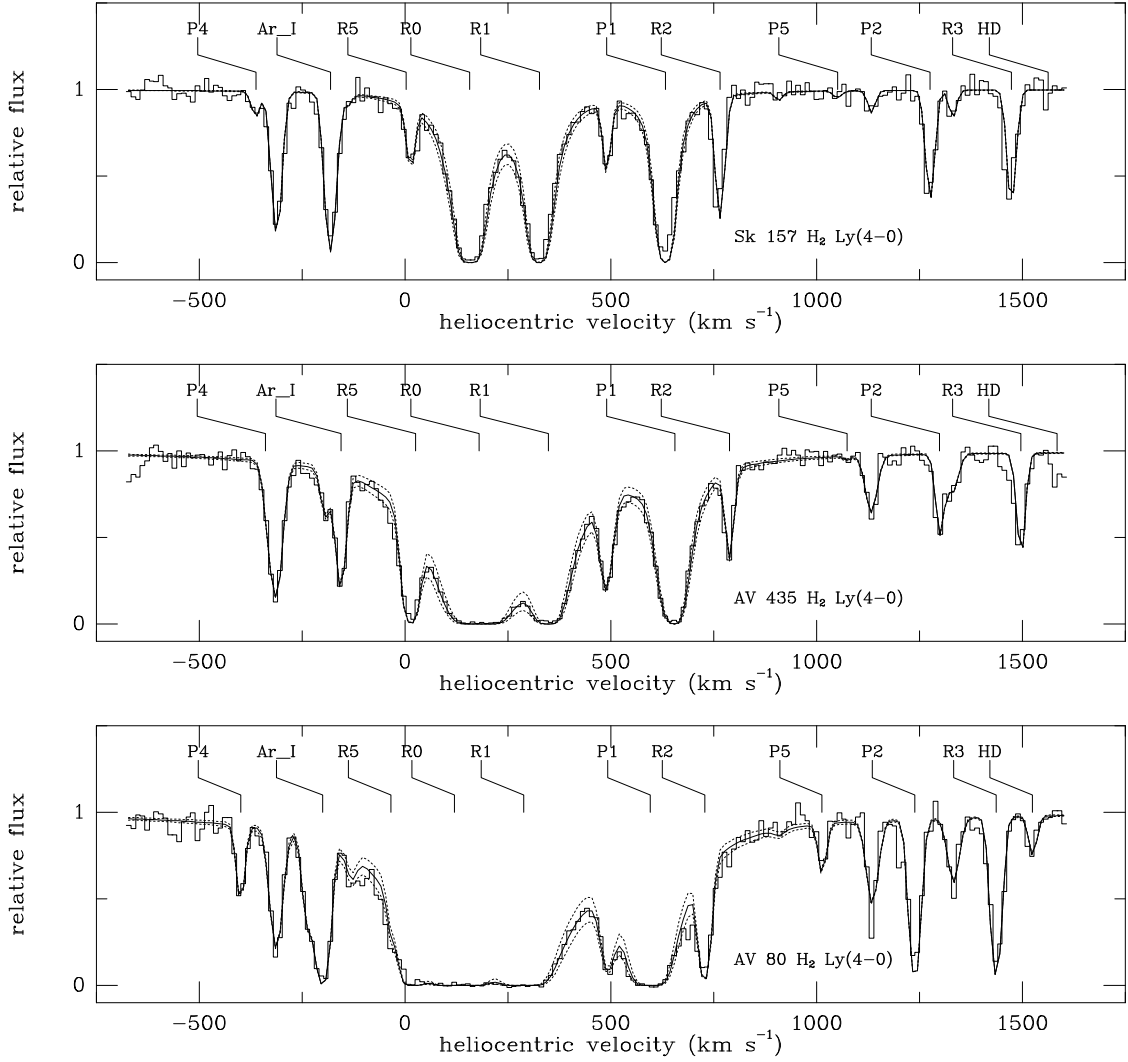}
\caption{Fits to {\it FUSE} spectra of the region around the R0, R1, and P1 lines of the H$_2$ Lyman (4-0) band near 1050 \AA\ (solid histograms), for the sight lines toward the SMC stars Sk~157 [log$N_{\rm SMC}$(H$_2$) $\sim$ 19.2], AV~435 [log$N_{\rm SMC}$(H$_2$) $\sim$ 19.9], and AV~80 [log$N_{\rm SMC}$(H$_2$) $\sim$ 20.1].
In each case, the solid line shows the fit for the adopted SMC $N$($J$=0) and $N$($J$=1); the dotted lines show the profiles for changes in the adopted $N$($J$=0) and $N$($J$=1) by $\pm$ 20\% (0.08 dex).
The upper and lower tick marks denote Galactic and SMC absorption, respectively; all lines (except \ion{Ar}{1} and HD) are H$_2$ rotational transitions from the Lyman (5-0) and (4-0) bands.
The velocity scale is relative to the rest wavelength of the (4-0) R0 line (1049.3674 \AA); the resolution of the {\it FUSE} spectra is $\sim$ 20 km~s$^{-1}$.}
\label{fig:h2fit}
\end{figure}

For $\sim$40 of the sight lines, more detailed fits were performed with the program {\bf fits6p} (e.g., Welty et al. 2003) to provide a further check on the results from the {\bf h2gui} fits.
The {\bf fits6p} fits include the absorption from Galactic H$_2$ (and any other species) and also allow for multiple Magellanic Clouds components.
Figure~\ref{fig:h2fit} shows the detailed fits to a portion of the Lyman (4-0) H$_2$ band toward three SMC stars, with $N_{\rm SMC}$(H$_2$) ranging from about 1.5 to 12 $\times$ 10$^{19}$ cm$^{-2}$. 
In the figure, the smooth solid lines show the adopted best fits to the profiles; the dotted lines show the profiles for changes in $N$($J$=0) and $N$($J$=1) by $\pm$20\% (0.08 dex).
The fits are particularly sensitive in the wings of the profiles of the strong R0, R1, and P1 lines -- especially in the regions where those lines overlap.
Note also the differences in velocity separation between the SMC and Galactic absorption for the three sight lines -- which also can affect those overlap regions.
For sight lines with $N$(H$_2$) between about 10$^{18}$ and 10$^{19}$ cm$^{-2}$, both the Lyman (4-0) and (1-0) bands were fitted, varying the $b$-value (usually between 2 and 10 km~s$^{-1}$) until consistent H$_2$ column densities were obtained from the two bands.
In most cases, the total H$_2$ column densities derived using {\bf h2gui} are within $\pm$0.1 dex of the values obtained via these more detailed fits.

For our full sample, the derived $N$(H$_2$) range from 0.3--90 $\times$ 10$^{19}$ cm$^{-2}$ for the SMC and from 0.2--25 $\times$ 10$^{19}$ cm$^{-2}$ for the LMC.
Uncertainties are typically of order 10--20 per cent for $N$(H$_2$) $\ga$ 10$^{19}$ cm$^{-2}$, but are somewhat larger for lower $N$(H$_2$), where the damping wings for the lines from $J$=0,1 are weak (even for the strongest bands) and the derived column densities are more sensitive to $b$.
Where possible, we adopt the values derived using {\bf fits6p}, which account explicitly for blending of Galactic and Magellanic Clouds absorption and possible multiple velocity components in the Magellanic Clouds.

Approximately 130 {\it FUSE} sight lines with weaker Magellanic Clouds H$_2$ absorption are not included in this study. 
For about half of those sight lines, only upper limits (typically below 10$^{15}$ cm$^{-2}$) will be obtainable.
Determination of accurate $N$(H$_2$) for the rest will require more detailed profile fitting, using component structures derived from higher resolution spectra and/or detailed curve of growth analyses, based on measurements of as many H$_2$ lines as possible, from all rotational levels.
We expect, however, that the total SMC or LMC $N$(H$_2$) generally will be less than 10$^{18}$ cm$^{-2}$, and that H$_2$ will be at most a minor contributor to the total hydrogen column densities in those sight lines.
For any such low-$N$(H$_2$) sight lines included in Tables~\ref{tab:smclos} and \ref{tab:lmclos}, the H$_2$ column densities are indicated approximately as moderate (``mod''), weak (``wk''), or upper limit (``ul'').

\subsubsection{\ion{H}{1} 21~cm Emission}
\label{sec-h1em}

For any given sight line, the 21~cm emission seen in the ATCA+Parkes, LAB, and GASS surveys covers very similar velocity ranges, with broadly similar overall shape.
There can be significant differences in detailed profile structure, however, with the lower resolution GASS and LAB spectra generally exhibiting smoother profiles.
Total Magellanic Clouds \ion{H}{1} column densities were derived for each sight line from each of the three data sets by simple integration over the full range of SMC or LMC velocities included in the line profiles, assuming the emission to be optically thin.
The $N_{\rm MC}$(\ion{H}{1}) thus obtained from the nearest-neighbor ATCA+Parkes (AP) spectra range from about 17 to 130 $\times$ 10$^{20}$ cm$^{-2}$ in the SMC and from about 2 to 60 $\times$ 10$^{20}$ cm$^{-2}$ in the LMC.
[The column densities obtained from interpolated AP spectra (using the four nearest neighbors on 1 arcmin grids) are very similar, with average differences less than 0.002 dex and rms deviations less than 0.02 dex for both galaxies.]
Integrations over the corresponding lower resolution GASS and LAB 21~cm profiles generally yielded slightly lower $N$(\ion{H}{1}) for the higher column density sight lines and somewhat higher $N$(\ion{H}{1}) for the lower column density sight lines, with larger differences for the LAB data (Fig.~\ref{fig:21comp}; see Sec.~\ref{sec-emcomp} below).
The Galactic column densities derived from the GASS and LAB profiles (for $v$ $\le$ 60 km~s$^{-1}$ toward the SMC and $v$ $\le$ 100 km~s$^{-1}$ toward the LMC) are generally in much better agreement, however, and the GASS and LAB values toward the LMC both agree well with those obtained from fig.~13 of Staveley-Smith et al. (2003).
Approximate 1-$\sigma$ uncertainties for $N$(21~cm) were estimated by integrating the typical rms noise values cited for each survey over the full LMC or SMC velocity ranges.
The resulting uncertainties (in cm$^{-2}$) for the AP spectra are 3.3$\times$10$^{19}$ (SMC) and 7.5$\times$10$^{19}$ (LMC); for the GASS spectra are 1.5$\times$10$^{18}$ (SMC) and 1.7$\times$10$^{18}$ (LMC); and for the LAB spectra are 2.2$\times$10$^{18}$ (SMC) and 2.4$\times$10$^{18}$ (LMC).
Corresponding 3-$\sigma$ upper limits for several low-$N$(\ion{H}{1}) sight lines in the LMC are given as 2.3$\times$10$^{20}$ cm$^{-2}$ (20.35 dex).

\section{RESULTS / DISCUSSION}
\label{sec-res}

\subsection{Comparisons with Previous Studies}
\label{sec-prev}

Most of the previously published \ion{H}{1} column densities for the Magellanic Clouds were derived from high- or low-dispersion spectra of Lyman-$\alpha$ absorption obtained with {\it IUE} (Bouchet et al. 1985; Fitzpatrick 1985a, 1985b, 1986; Gordon et al. 2003); a smaller number were derived from analyses of {\it FUSE} spectra of Lyman $\beta$ and/or {\it HST} spectra of Lyman $\alpha$ (Evans et al. 2004a, 2004b, 2004c; Crowther et al. 2002).
Unfortunately, the high-dispersion {\it IUE} spectra generally have very low S/N ratios near Lyman $\alpha$, and the combined effects of Galactic absorption, geocoronal emission, and the \ion{N}{5} stellar wind lines make it difficult to derive precise and accurate $N_{\rm MC}$(\ion{H}{1}) from the low-dispersion (6 \AA) spectra; Fitzpatrick (1985a) estimated uncertainties of $\pm$50\%, for example.
Comparison of Fitzpatrick's results with the more precise values derived from the {\it HST} spectra, however, indicates generally good agreement -- with slope $\sim$ 1.0 and rms deviation $\sim$ 0.1 dex for the twelve sight lines in common.
Our new \ion{H}{1} column densities also exhibit reasonably good agreement with the values from Bouchet et al. (1985), Evans et al. (2004a, 2004b, 2004c), and Crowther et al. (2002), if those previous values are adjusted for contributions from Galactic \ion{H}{1} absorption. 
Our $N$(\ion{H}{1}) also agree in most cases with those found by Gordon et al. (2003), after adjusting their differential values for the Magellanic Clouds \ion{H}{1} observed toward their comparison stars (which can be as high as 4 $\times$ 10$^{21}$ cm$^{-2}$).
We find a much lower $N$(\ion{H}{1}) toward the SMC star Sk~143, however (see also Howk et al. 2010). 
For some sight lines (particularly in the SMC), the $N$(\ion{H}{1}) derived in this paper differ slightly from those listed in Welty \& Crowther (2010), due to small differences in the adopted Galactic foreground contributions (Sec.~\ref{sec-h1abs}).

For sight lines with appreciable amounts of molecular material [$N_{\rm MC}$(H$_2$) $\ga$ 3 $\times$ 10$^{18}$ cm$^{-2}$] previously analyzed by Tumlinson et al. (2002), both the total $N_{\rm MC}$(H$_2$) and the $T_{01}$ obtained in our fits are generally in good agreement with the published values (rms deviation $\sim$ 0.1 dex).
Because Tumlinson et al. fitted a number of the H$_2$ bands [instead of just the Lyman (4-0) and (1-0) bands analyzed in this paper], we adopt their values for all but three of the sight lines in common.
For those three sight lines (Sk~10, AV~47, and Sk~116), the H$_2$ column densities derived from our fits to the (1-0) and (4-0) bands are consistent for smaller $b$-values [and thus significantly higher $N$(H$_2$)] than those found by Tumlinson et al. via curves of growth.
Even the (4-0) band $J$=0,1 lines are at most weakly damped in those three sight lines, however, and the derived column densities (2-3 $\times$ 10$^{18}$ cm$^{-2}$) are rather uncertain.
The agreement is slightly poorer (rms deviation $\sim$ 0.15 dex) with the values for the sight lines reported by Cartledge et al. (2005); for example, we find a significantly higher $N$(H$_2$) toward Sk$-$68~140.
 
\subsection{Comparison of Three 21~cm Emission Surveys}
\label{sec-emcomp}

As noted above (Sec.~\ref{sec-h1em}), there are systematic differences in the Magellanic Clouds \ion{H}{1} column densities derived from the ATCA+Parkes, LAB, and GASS 21~cm emission surveys.
The values determined from the lower resolution LAB and GASS surveys are generally lower than those obtained from the AP spectra for the highest $N$(\ion{H}{1}) sight lines, and generally higher for the lowest $N$(\ion{H}{1}) sight lines (Fig.~\ref{fig:21comp}).
In the figure, the slopes for the LMC ratios are similar to those for the SMC, if the LMC sample is restricted to $N$(AP) $\ge$ 10$^{21}$ cm$^{-2}$ (more comparable to the range in the SMC sample):  for both SMC and LMC sight lines, the slopes are of order $-$0.2 for GASS/AP vs. AP; of order $-$0.4 for LAB/AP vs. AP; and of order $-$0.2 for LAB/GASS vs. GASS.

These differences presumably reflect the differences in spatial resolution -- and consequent sampling of the complex structure of the ISM in the SMC and LMC -- in the three surveys.
At 50 kpc (the approximate distance to the LMC), the resolutions of the three 21~cm data sets correspond to 14.5--21.8 pc (ATCA+Parkes), 210 pc (GASS), and 438 pc (LAB) -- scales over which significant variations would be expected.
For the sight lines with highest $N$(\ion{H}{1}) seen with ATCA+Parkes, the larger GASS and LAB beams thus must typically sample somewhat lower values (on average), and vice versa.
Although beyond the scope of this paper, more detailed analyses of such comparisons might yield information on the scales, amplitudes, and properties of structures characterizing the predominantly neutral gas in the SMC and LMC (e.g., Wakker et al 2011).
The inferred structural characteristics could then be compared with the structure seen in even higher resolution IR images of the Magellanic Clouds.

\subsection{$N$(Lyman-$\alpha$ Absorption) vs. $N$(21~cm Emission)}
\label{sec-la21}

Differences in the \ion{H}{1} column densities derived from absorption and emission can reflect the line-of-sight distribution of the atomic gas (background vs. foreground), small-scale spatial structure in the ISM (in the plane of the sky), and possible saturation in the 21~cm emission.
Figure~\ref{fig:lavs21} plots the Magellanic Clouds $N$(\ion{H}{1}) derived from the ATCA+Parkes 21~cm emission profiles versus the column density obtained from Lyman-$\alpha$ absorption, for sight lines in the SMC ({\it left}) and LMC ({\it right}).
For the SMC, $N$(21~cm) is generally comparable to $N$(Ly-$\alpha$) [= $N$(\ion{H}{1})] for the highest column density sight lines, but becomes progressively higher than $N$(Ly-$\alpha$) for the lower column density sight lines; the slope of the relationship is of order 0.4 (not considering upper limits).
For the LMC, $N$(21~cm) is again larger than $N$(Ly-$\alpha$) at low $N$(\ion{H}{1}), but there are a number of (primarily) higher column density sight lines where $N$(Ly-$\alpha$) exceeds $N$(21~cm) -- by factors up to $\sim$ 3.

Comparison of the emission- and absorption-line profiles can provide some insight into the reasons for the observed differences in derived column density.
In the SMC, the 21~cm emission profiles typically exhibit multiple peaks, in many cases with a local minimum at $v$ $\sim$ 150-160 km~s$^{-1}$. 
Some investigations of the structure of the SMC have interpreted that multi-peaked emission as arising from several distinct ``sheets'' of \ion{H}{1} (e.g., Songaila et al. 1986; Wayte 1990; and references therein).
More recent studies based on the higher resolution ATCA+Parkes data have suggested, however, that numerous expanding shells of gas may be responsible for the observed complex profile structure (e.g., Staveley-Smith et al. 1997; Stanimirovi\'{c}, Staveley-Smith, \& Jones 2004).
While the exact locations of the various neutral gas components and the physical depth of the SMC are somewhat uncertain, the structural studies have generally concluded that the gas at $v$ $\sim$ 160--200 km~s$^{-1}$ is more distant than the gas at 100--140 km~s$^{-1}$.
Although it is difficult to assign precise velocities to the broad damped Lyman-$\alpha$ absorption lines (especially in the lower resolution spectra), examination of the absorption from other species that should trace the strongest \ion{H}{1} components [e.g., \ion{Na}{1}, \ion{K}{1}, CH, H$_2$ (Tumlinson et al. 2002; Welty et al. 2006; this paper; Welty \& Crowther, in prep.)] indicates that the strongest absorption generally corresponds to the lower velocity emission peaks (e.g., for Sk~143 in Fig.~\ref{fig:emabs}) -- consistent with the results of the studies of SMC structure.
So while the higher 21~cm column densities at lower $N$(Ly-$\alpha$) could reflect the broader spatial sampling of the 21~cm beam (as discussed above), they more likely are due primarily to the inclusion of higher velocity emission beyond the target stars that is not sampled in absorption.
For sight lines in which the main SMC \ion{Na}{1} and/or H$_2$ absorption components lie within the range of the lower-velocity 21~cm component(s), the $N$(21~cm) derived from just those lower-velocity components are generally in much better agreement with $N$(Ly-$\alpha$) (see Fig.~\ref{fig:h1pred} and discussions in the next section).

While the 21~cm emission profiles in the LMC also are clearly complex, most do not exhibit the relatively well-separated peaks seen in many of the SMC sight lines.
Using 21~cm data from the Parkes single dish (resolution $\sim$ 15 arcmin), Luks \& Rohlfs (1992) identified two large-scale components in the LMC.
The main ``D'' component, seen throughout the LMC, can be modeled as a differentially rotating, nearly face-on disk; the lower velocity ``L'' component, seen primarily in the eastern part of the LMC, appears to be more distant.
As for the SMC, however, the higher resolution ATCA data for the LMC have revealed many localized shells and filaments contributing to the complex emission profiles (Kim et al. 1998, 2003), and it is generally difficult to locate individual velocity components along a given line of sight.
Toward SN 1987A, for example, where light echo data have provided additional distance constraints, there does not seem to be a monotonic relationship between the distance and velocity of the various interstellar components (Xu \& Crotts 1999; Welty et al. 1999a).
As in the SMC, however, the sight lines with $N$(21~cm) greater than $N$(Ly-$\alpha$) -- especially the lower-$N$(\ion{H}{1}) sight lines -- probably have a significant amount of neutral gas located beyond that seen in absorption.
For example, the relatively weak \ion{Na}{1} absorption observed toward BI~229 [Fig.~\ref{fig:emabs}; with $N$(21~cm)/$N$(Ly-$\alpha$) $\sim$ 1.5] is at a velocity well removed from the main \ion{H}{1} emission -- which thus must originate beyond the star.

Foreground-background effects do not account for the (mostly) higher-$N$(\ion{H}{1}) LMC sight lines where $N$(Ly-$\alpha$) is greater than $N$(21~cm), however.
In such cases, the lower column densities derived from the 21~cm emission must be due to the broader 21~cm beam sampling (on average) lower-$N$(\ion{H}{1}) material and/or to underestimation of the true column density from the emission-line observations (due to unrecognized saturation and/or self-absorption).
While the comparisons among the three 21~cm surveys (Sec.~\ref{sec-emcomp}) suggest that beam-size effects do play a role, there are indications that saturation effects may also be significant for some of the sight lines in our sample.
For many of the 16 LMC sight lines with $N$(Ly-$\alpha$) $\ga$ 1.5 $\times$ $N$(21~cm) [including the seven with highest $N$(Ly-$\alpha$)], most of the emission appears to be concentrated in 1--3 relatively strong, narrow components -- for which the optically thin assumption may underestimate the true column density.
Where absorption-line data are available for those sight lines, the main components seen in \ion{Na}{1} and/or $H_2$ absorption generally are found at very similar velocities to the main components in \ion{H}{1} emission (e.g., for BI~237 in Fig.~\ref{fig:emabs}) -- suggesting that the bulk of the neutral gas lies between us and the target star.
In addition to two obvious cases of \ion{H}{1} self-absorption -- in the ATCA+Parkes 21~cm profiles toward Mk~42 and Sk$-$69~243 (in the core of 30 Dor) -- there are seven other LMC sight lines where both the dominant absorption from \ion{Na}{1} and/or H$_2$ and the peak in the GASS 21~cm emission coincide with a possible self-absorption feature in the AP profile (e.g., for BI~184 in Fig.~\ref{fig:emabs}).
These sight lines generally have modest \ion{H}{1} column densities, $N$(GASS)/$N$(AP) ranging from about 1.5 to 11, and represent many of the low-$N$(AP) LMC sight lines with high $N$(GASS)/$N$(AP) and $N$(LAB)/$N$(AP) in Fig.~\ref{fig:21comp}.

While sight lines with $N$(Ly-$\alpha$) $>$ $N$(21~cm) may be found throughout the LMC, a number of them are located in or near regions of recent star-formation, nebulosity, and/or CO emission -- e.g., 30~Dor, LH~10, LH~90, LH~101, He~206.
In figure~\ref{fig:n206}, for example, sight lines with data for \ion{H}{1} and/or H$_2$ absorption near the star-forming region He~206 (in the southern LMC, near $\alpha$, $\delta$ = $5^{h}30^{m}$, $-$71$\arcdeg$) are noted on maps comparing the AP 21~cm emission ({\it right}) and the 8$\mu$m and CO emission ({\it left}; Meixner et al. 2006; Wong et al. 2011).
The sight lines to BI~184 and Sk$-$71~38 (northern-most adjacent triangles), located at the edge of a seeming ``hole'' in the AP \ion{H}{1} and 8$\mu$m distributions, both exhibit AP 21~cm profiles suggestive of self-absorption (Fig.~\ref{fig:emabs}).
Toward BI~184, there is fairly strong absorption from both \ion{H}{1} and H$_2$, with $N$(Ly-$\beta$) = 2.5 $\times$ 10$^{21}$ cm$^{-2}$ and $N$(H$_2$) = 4.5 $\times$ 10$^{19}$ cm$^{-2}$.
The GASS 21~cm profile, which samples both the ``hole'' and the surrounding stronger emission, has $N$(GASS) = 1.8 $\times$ 10$^{21}$ cm$^{-2}$ -- slightly smaller than $N$(Ly-$\beta$), but significantly larger than the $N$(AP) = 0.7 $\times$ 10$^{21}$ cm$^{-2}$. 
The AP 21~cm profile (Fig.~\ref{fig:emabs}) exhibits a clear minimum, with $T_{\rm br}$ $<$ 0, at about 240 km~s$^{-1}$ -- the same velocity as the GASS 21~cm emission, the main H$_2$ absorption, and the nearby CO emission (Fukui et al. 2008; Wong et al. 2011).
The significant molecular fraction [$f$(H$_2$) = 0.036], the relatively low $T_{01}$(H$_2$) = 55 K, and the possible self-absorption in \ion{H}{1} [though at modest $N$(\ion{H}{1})] suggest that this sight line intercepts a fairly small, cold, neutral cloud.
The sight lines to LMC1-233 and LMC1-246 (southwestern-most squares in the figure) have $N$(Ly-$\alpha$) = 3.7--4.5 $\times$ 10$^{21}$ cm$^{-2}$ -- roughly twice the value obtained from the corresponding AP and GASS 21~cm profiles; most of the \ion{H}{1} emission there is in a single relatively narrow component, also at about 240 km~s$^{-1}$.

Studies of \ion{H}{1} 21~cm absorption toward continuum sources behind the LMC and SMC have yielded similar indications of saturation in the 21~cm emission profiles.
Roughly half of the LMC (36 of 62) and SMC (13 of 28) sight lines examined so far exhibit 21~cm absorption (Dickey et al. 1994, 2000; Marx-Zimmer et al. 2000; Wong et al., in prep.).
For the LMC, the general conclusion is that the gas seen in absorption is either cooler (30--40 K) or more prevalent than in the Milky Way (Mebold et al. 1997; Marx-Zimmer et al. 2000).
In the SMC, the gas is again cooler ($\la$ 40 K), but seems less abundant ($<$ 15\% of the total \ion{H}{1}) (Dickey et al. 2000).
Both the low temperatures and the strength of the 21~cm absorption suggest that the corresponding 21~cm emission will be somewhat saturated, particularly where the observed 21~cm brightness temperature is significantly greater than the temperature of the cold gas.
Based on the results for the SMC, Stanimirovi\'{c} et al. (1999; see also Leroy et al. 2009) adopted an empirical column density ``correction factor'' $f_{c}$ which increases linearly from 1.0 to 1.4 as the optically thin estimate for log[$N$(21~cm)] increases from 21.4 to 22.0.
Using a somewhat different approach, Bernard et al. (2008) assumed a constant spin temperature of 60 K to estimate the saturation in the LMC emission profiles as a function of velocity, and obtained median column density correction factors of 1.3 and 3.9 for log[$N$(21~cm)] = 21.0 and 21.3, respectively. 
Ratios $N$(Ly-$\alpha$)/$N$(21~cm) $\sim$ 2--3 seen for some LMC sight lines thus may be due entirely to uncorrected saturation in the 21~cm emission.

\subsection{Estimating $N$(\ion{H}{1}) without Lyman-$\alpha$ Absorption}
\label{sec-predh1}

While observations of damped Lyman-$\alpha$ absorption yield the most direct, sensitive, and accurate interstellar \ion{H}{1} column densities, the requisite UV spectra are not always available or usable (e.g., for main-sequence stars later than B1 or giants later than B3, where the stellar \ion{H}{1} absorption becomes significant; Diplas \& Savage 1994a).
In order to enable estimates of $N$(\ion{H}{1}) for a wider variety of sight lines (e.g., for comparison with absorption-line data for other species), it would thus be advantageous to identify suitable proxies for Lyman-$\alpha$ absorption -- i.e., other, more readily observable quantities that are well-correlated with $N$(Ly-$\alpha$).
Several such surrogates have been suggested and employed in various investigations of interstellar material:  the color excess $E(B-V)$; the column densities of dominant ions of little-depleted elements (e.g., \ion{O}{1}, \ion{S}{2}, \ion{Zn}{2}); the relative abundances of species characterized by different levels of depletion (Jenkins 2009); the column densities of trace neutral species (e.g., \ion{Na}{1}, \ion{K}{1}); and the equivalent width of the diffuse interstellar band at 5780.5 \AA\ (Herbig 1993; Friedman et al. 2011).
Unfortunately, most of these trace the total (atomic plus molecular) hydrogen column density better than that of \ion{H}{1} [so that some estimate for $N$(H$_2$) is also needed], their behavior in lower-metallicity systems may not be well established, and each has its own particular additional shortcomings.
Observations of 21~cm emission provide direct measures of \ion{H}{1}, but (as discussed above) those measures can include background gas not seen in absorption and can be affected both by saturation and by small-scale structure within the radio beam.

Because Lyman-$\alpha$ data were not then available for most of the 70 sight lines in their survey of H$_2$, Tumlinson et al. (2002) estimated \ion{H}{1} column densities from the ATCA+Parkes 21~cm profiles -- adopting 0.75$\pm$0.25 $\times$ $N$(21~cm) to account in an average sense for those effects.
Now that more Lyman-$\alpha$ spectra have been obtained, those estimates may be compared with values or limits derived from the absorption-line data for 55 of the sight lines (top panel of Fig.~\ref{fig:h1pred}).
While at least eighteen of the estimates are consistent with the measured values (within the fairly generous assumed uncertainties), at least fourteen of the sight lines exhibit differences greater than a factor of 3; on average, the differences are larger for the SMC.
The estimated values are greater than 1.5 $\times$ $N$(Ly-$\alpha$) in more than half of the sight lines (including thirteen of the fourteen most discrepant) -- presumably due to significant amounts of \ion{H}{1} lying beyond the target stars.
These comparisons indicate that the molecular fractions derived for the LMC and SMC were thus somewhat underestimated (on average and with increased scatter) -- particularly for the SMC.

In principle, it should be possible to significantly improve the $N$(\ion{H}{1}) estimated from 21~cm emission by eliminating (as much as possible) the contributions from neutral gas beyond the target stars.
For example, moderate- to high-resolution observations of absorption lines from species that are well-correlated with \ion{H}{1} (or that at least trace the main neutral components) can reveal the dominant foreground velocity components.
For now, we take that velocity information from existing observations of H$_2$ (Tumlinson et al. 2002; this paper) and \ion{Na}{1} (Vladilo et al. 1993; Cox et al. 2006, 2007; Welty et al. 2006; Welty \& Crowther, in prep.).\footnotemark
Integration of the 21~cm emission profiles only over the velocities seen for those absorption-line tracers then should provide better estimates for the intervening $N$(\ion{H}{1}).
\footnotetext{While \ion{Ca}{2} has been used in some of the structural studies of the SMC and LMC, it is not as reliable a tracer of the main neutral components, due to the large variations in calcium depletion.
The \ion{Ca}{2} absorption often covers a much wider velocity range than the \ion{Na}{1} absorption, but the larger \ion{Ca}{2}/\ion{Na}{1} ratios in those ``extra'' components suggest that calcium is just much less depleted there -- not that there is a significant amount of neutral gas.
In many cases, there is no discernible 21~cm emission corresponding to those ``extra'' \ion{Ca}{2} components (Welty \& Crowther, in prep.).}

As noted above (Sec.~\ref{sec-la21}), in many of the SMC sight lines, the absorption from \ion{Na}{1} and H$_2$ appears to be associated with the lower-velocity component(s) seen between about 100 and 150 km~s$^{-1}$ in the 21~cm emission profiles.
The middle panel of Fig.~\ref{fig:h1pred} compares the SMC \ion{H}{1} column densities estimated from just the lower-velocity 21~cm emission (for sight lines where the \ion{Na}{1} and H$_2$ absorption is seen at those lower velocities) with the values obtained from Lyman-$\alpha$ absorption (cf. the left-hand panel of Fig.~\ref{fig:lavs21}, where all the 21~cm emission is included).
For each sight line, the integration over the 21~cm profile is taken from about 90 km~s$^{-1}$ to the velocity corresponding to the local minimum in emission between the lower- and higher-velocity component(s) (typically near 150 km~s$^{-1}$).
Even for this relatively crude restriction in velocity, the correlation between the two estimates for $N$(\ion{H}{1}) is quite good (linear correlation coefficient $r$ = 0.855), the relationship is nearly linear (slopes $\sim$ 1.0--1.2 for weighted and unweighted fits), and the scatter about the best-fit lines is fairly small (rms deviation $\sim$ 0.11--0.12 dex).

The bottom panel of Fig.~\ref{fig:h1pred} shows a similar comparison for a slightly different way of restricting the 21~cm velocity interval, for sight lines in the SMC and LMC with existing high-resolution \ion{Na}{1} spectra (Vladilo et al. 1993; Cox et al. 2006, 2007; Welty et al. 2006; Welty \& Crowther, in prep.).
In this case, the integration limits are set to slightly beyond the actual velocity intervals exhibiting \ion{Na}{1} absorption (typically 5 km~s$^{-1}$ beyond the range in fitted \ion{Na}{1} components), to allow for the somewhat broader \ion{H}{1} profiles for each component (Table~\ref{tab:21vel}).
Again, the correlation is fairly good ($r$ = 0.792), the relationship is nearly linear (slopes $\sim$ 0.8--1.0), and the scatter is fairly small (rms $\sim$ 0.12--0.13 dex).
Only four of the 40 sight lines have differences between $N$(Ly-$\alpha$) and $N$(21~lim) greater than a factor of 2.
The one significant positive outlier, AV~321 in the SMC, has a weak \ion{Na}{1} component at the edge of the (very) dominant higher-velocity 21~cm emission peak; in this case, the integration over the 21~cm profile (though restricted) is likely to include a significant contribution from background gas.

These simple exercises suggest that reasonably accurate foreground $N$(\ion{H}{1}) may be derived from 21~cm emission profiles if the velocity range of the foreground gas can be determined from absorption-line observations of suitable tracers.
Of course, high spatial resolution should be employed at 21~cm to minimize the effects of small-scale structure within the radio beam.
This method may yield poorer estimates for $N$(\ion{H}{1}), however, if there is significant saturation in the 21~cm emission and/or if foreground and background components overlap significantly in velocity (as for the one discrepant sight line in the bottom panel of Fig.~\ref{fig:h1pred}).
Detailed multi-component fits to the 21~cm profiles, using component structures derived from high-resolution spectra of the absorption from tracers of the neutral gas (e.g., Fitzpatrick \& Spitzer 1997), can yield estimates for the \ion{H}{1} in individual components (that are unresolved in the broad, saturated Lyman-$\alpha$ profiles) -- and might enable overlapping foreground and background components to be identified and separated.
 
\subsection{$N$(\ion{H}{1}), $N$(H$_2$), $N$(H$_{\rm tot}$), and $E(B-V)$:  Gas-to-Dust Ratios}
\label{sec-gdr}

The interstellar gas-to-dust ratio (GDR) can be estimated using observations of either the absorption or the emission from the gas and dust.
In general, the GDR depends primarily on the metallicity of the ISM and the fraction of heavy elements in the dust (i.e., the overall level of depletion), but also on the composition and size distribution of the dust grains (which can affect the relationships between the dust mass and the absorption, scattering, and emission characteristics of the dust).
The use of absorption data [typically $E(B-V)$ and $N$(H$_{\rm tot}$)] has the advantages that the data sample the same pencil beam volume toward the background target, accurate $N$(\ion{H}{1}) and $N$(H$_2$) are directly obtained, and the often poorly known dust emission properties are not required.
On the other hand, only specific sight lines can be probed, UV spectra of both \ion{H}{1} and H$_2$ are needed, and the color excesses can be somewhat uncertain.
[And while it might be better to use the total visual extinction $A_{\rm v}$ (which should be more directly related to the amount of dust present) instead of $E(B-V)$, values for $A_{\rm v}$ often are not known.]
The use of emission data (typically 21~cm, CO, and IR emission) can provide wider spatial information, but requires fairly broad wavelength coverage in the IR (for adequate coverage of the spectral energy distribution), accurate knowledge of the dust emission properties, and reliable conversion of the 21~cm emission to $N$(\ion{H}{1}) and of the CO emission to $N$(H$_2$).

For the local Galactic ISM, the average GDR is well determined from observations of \ion{H}{1} and H$_2$ absorption for sight lines with known $E(B-V)$:  $N$(H$_{\rm tot}$)/$E(B-V)$ $\sim$ 5.6--5.9$\times$10$^{21}$ cm$^{-2}$ mag$^{-1}$ (Bohlin et al. 1978; Shull \& van Steenberg 1985; Diplas \& Savage 1994b; Rachford et al. 2009; see also Table~\ref{tab:ratgal}); the corresponding average for just the atomic gas is $N$(\ion{H}{1})/$E(B-V)$ $\sim$ 4.4--5.2$\times$10$^{21}$ cm$^{-2}$ mag$^{-1}$.
For Galactic sight lines in which the hydrogen is predominantly neutral, $N$(H$_{\rm tot}$)/$E(B-V)$ is (on average) essentially independent of $E(B-V)$ for $E(B-V)$ $\la$ 1.0 mag; $N$(\ion{H}{1})/$E(B-V)$ is roughly constant for $E(B-V)$ $\la$ 0.5 mag, but appears to decline somewhat at higher reddenings (e.g., Figures~\ref{fig:rvsebv} and \ref{fig:avgrat}).
The reddening is better correlated with the total (atomic plus molecular) hydrogen column density (with linear correlation coefficient $r$ = 0.89) than with $N$(\ion{H}{1}) alone ($r$ = 0.81). 
There are indications that the GDR may vary, both regionally and in some individual sight lines (e.g., Burstein \& Heiles 1978; Diplas \& Savage 1994b).  
While the low values for some low-$N$(\ion{H}{1}) sight lines may be due to unaccounted-for \ion{H}{2}, and some of the high values to larger-than-average $R_{\rm v}$ = $A_{\rm v}$/$E(B-V)$ (e.g., for some sight lines in Orion and Sco-Oph), variations in the gas-to-dust mass fraction also seem to be required (Barbaro et al. 2004).
Differences in the GDR along longer sight lines sampling the inner Galaxy may reflect a Galactic metallicity gradient (e.g., Watson 2011).
 
While it has been evident that the gas-to-dust ratios in the Magellanic Clouds generally are higher than the average Galactic values, different studies have yielded somewhat different results.
The average $N$(\ion{H}{1})/$E(B-V)$ ratios determined from the earliest observations of Lyman-$\alpha$ absorption in the LMC and SMC range from about 22--24$\times$10$^{21}$ and 37--87$\times$10$^{21}$ cm$^{-2}$ mag$^{-1}$ (4.6--5.0 and 7.7--18 times the average Galactic value), respectively (Koornneef 1982; Bouchet et al. 1985; Fitzpatrick 1985a, 1985b, 1986).
Those studies employed relatively small samples, however, with $N$(\ion{H}{1}) determined from {\it IUE} spectra and with no data for H$_2$.
A more recent study by Gordon et al. (2003), also based primarily on {\it IUE} spectra, found both somewhat smaller average $N$(\ion{H}{1})/$E(B-V)$ and possible regional variations:  $\sim$ 19$\times$10$^{21}$ cm$^{-2}$ mag$^{-1}$ for the LMC2 region, $\sim$ 11$\times$10$^{21}$ cm$^{-2}$ mag$^{-1}$ for all other LMC sight lines, $\sim$36$\times$10$^{21}$ cm$^{-2}$ mag$^{-1}$ for sight lines in the main SMC ``bar'', and $\sim$ 15$\times$10$^{21}$ cm$^{-2}$ mag$^{-1}$ in the near ``wing'' region toward Sk~143.\footnotemark
\footnotetext{The UV extinction observed toward Sk~143 is unique (in the very small current sample of SMC sight lines) in having a Milky Way-like 2175 \AA\ bump and a shallower far-UV rise than the other SMC curves.  
Such extinction may not be characteristic of the SMC wing region, however, as examination of the low-resolution {\it IUE} spectra of other (somewhat less reddened) wing sight lines reveals no obvious indications of the 2175 \AA\ bump and suggests that the far-UV extinction is fairly steep (as in the SMC bar).}

More recent studies based on 21~cm, CO, and IR emission from the Magellanic Clouds have yielded both average values for the gas-to-dust ratios and indications of possible regional variations within each galaxy.
In those studies, the GDRs are often given as mass ratios, with the dust mass typically estimated from the optical depth at some far-IR wavelength (for an assumed dust model) and the total gas-phase hydrogen mass multiplied by 1.36 to account for helium.
For the SMC, Leroy et al. (2007) combined {\it Spitzer} maps at 24, 70, and 160 $\mu$m with the ATCA+Parkes 21~cm and NANTEN CO data to obtain an overall average GDR $\sim$ 1000 (10 times the adopted Galactic value of 100), with the $N$(\ion{H}{1})/$\tau$(160 $\mu$m) ratios about 70\% higher in the wing region than in the main SMC bar.
Leroy et al. (2011) incorporated additional {\it Spitzer} data, and found a lower overall GDR for the SMC (by a factor $\sim$ 3), with the total GDR in the wing {\it lower} than in the bar.
Bolatto et al. (2011) adopted an iterative approach to determining the SMC GDR (on scales of $\sim$ 200 pc), and obtained a median GDR roughly 14 times that found for Galactic cirrus clouds; again, the GDR values typically are higher in the wing than in the bar.
For the LMC, Bernard et al. (2008) analyzed data from {\it Spitzer} and {\it IRAS} and obtained an overall average GDR roughly 3 times the typical Galactic value, with values in the outlying regions $\sim$ 50\% higher than that.
Dobashi et al. (2008), using visual/near-IR extinctions derived from 2MASS photometric data, found an average LMC $N$(H$_{\rm tot}$)/$A_{\rm v}$ 3.6 times the Galactic value, with lower values in the 30 Dor region and higher values in the outlying regions.
Roman-Duval et al. (2010) have examined {\it Herschel} data at 250 and 350 $\mu$m and CO data from the MAGMA survey (Wong et al. 2011) for two molecular cloud complexes in the LMC and determined average GDRs of about 350 and 235, with little apparent variation in GDR within each cloud.
Initial results from the {\it Planck} satellite suggest average GDRs higher than Galactic by factors of 2.4 in the LMC and 13 in the SMC (Planck collaboration 2011b) -- reasonably consistent with the earlier studies.
The emission-based studies thus generally suggest values for the GDR that are 2--4 times the local Galactic average for the LMC and 8--15 times the Galactic average for the SMC.

\subsubsection{Average Gas-to-Dust Ratios} 
\label{sec-gdravg}

Table~\ref{tab:ratgal} lists the average values for $N$(\ion{H}{1})/$E(B-V)$, $N$(H$_{\rm tot}$)/$E(B-V)$, and several other ratios for our larger sample of Magellanic Clouds \ion{H}{1} and H$_2$ data, with corresponding values for the local Galactic ISM given for comparison.
The LMC and SMC samples are each divided into two regional subsets thought to be characterized by differences in the UV extinction curves (e.g., Gordon et al. 2003):  the LMC2 region (extending to the southeast of 30 Dor) and the rest of the LMC (here denoted ``LMC main''), and the main ``bar'' and ``wing'' regions of the SMC.
For several of the ratios, two values are given:  the ``wlog'' values are weighted means of log(ratio), while the ``nlin'' values are log(unweighted mean of ratio).
Sight lines with $E(B-V)$ $<$ 0.05 (where the uncertainties on the reddening are relatively large) and significant outliers ($>$ 2.5$\sigma$) have been omitted in calculations of the mean ratios for all three galaxies.
The ``wlog'' values are somewhat less sensitive to any remaining large positive outliers, but somewhat more sensitive to large negative outliers.
The median values for the $N$(\ion{H}{1})/$E(B-V)$ and $N$(H$_{\rm tot}$)/$E(B-V)$ ratios lie between the ``wlog'' and ``nlin'' values.
Inclusion of upper limits (at the limiting values) generally decreases the overall average values by less than about 10\%.
The ``$\Delta$'' value (given below each average ratio) gives the difference with respect to the corresponding average Galactic value.
Figure~\ref{fig:rvsebv} shows the ratios of $N$(\ion{H}{1})/$E(B-V)$, $N$(H$_2$)/$E(B-V)$, and $N$(H$_{\rm tot}$)/$E(B-V)$, versus $E(B-V)$, for individual sight lines in our Galaxy, the LMC, and the SMC, with the mean values for the \ion{H}{1} and H$_{\rm tot}$ ratios for the three galaxies shown by dotted horizontal lines.\footnotemark
\footnotetext{Uncertainties for the individual points are not given in the figure, in order to more clearly show the relative distributions of the ratios for the three galaxies.
Typical uncertainties for $E(B-V)$ are 0.03--0.04 mag, for $N$(\ion{H}{1}) and $N$(H$_{\rm tot}$) are 0.04--0.08 dex, and for $N$(H$_2$) are 0.1--0.2 dex.}
Figure~\ref{fig:avgrat} shows those three ratios averaged over 0.1-mag bins in $E(B-V)$ [giving a clearer view of the trends with $E(B-V)$]; Figure~\ref{fig:hist} shows histograms of the distributions of the three ratios (giving a clearer view of the scatter in each ratio).

For some of the sight lines in the Galactic, LMC, and SMC samples represented in Table~\ref{tab:ratgal} and Figures~\ref{fig:rvsebv}, \ref{fig:avgrat}, and \ref{fig:hist}, there are no data for H$_2$, and $N$(H$_{\rm tot}$) has been set equal to $N$(\ion{H}{1}).
In order to gauge the effects of that choice, $N$(H$_2$) has been estimated for those sight lines via the observed average Galactic relationship between $N$(H$_2$)/$E(B-V)$ and $E(B-V)$ (Fig.~\ref{fig:avgrat}) (likely giving overestimates for the LMC and SMC sight lines; see below).
For 19 of the 61 Galactic sight lines with no H$_2$ data, independent estimates for $N$(H$_2$) may be obtained from the observed correlation between $N$(H$_2$) and $N$(CH) (Danks, Federman, \& Lambert 1984; Mattila 1986; Rachford et al. 2002; Welty et al. 2006).
While the two estimates generally agree within a factor of 2, the $N$(H$_2$) estimated from $E(B-V)$ are much too high for several sight lines in the Orion Trapezium region, where strong radiation fields have significantly reduced the molecular abundances -- even for $E(B-V)$ $\ga$ 0.5 mag.
In most cases, the estimated values for H$_2$ suggest that $N$(H$_2$)/$N$(\ion{H}{1}) is less than about 0.1 (Milky Way) or less than about 0.05 (LMC, SMC), and imply increases in $N$(H$_{\rm tot}$) of more than 0.1 dex for only 11 of the 61 Galactic sight lines, 2 of the 51 LMC sight lines (Sk$-$66~21, LMC2-702), and 3 of the 40 SMC sight lines (Sk~36, Sk~101, AV~326); none of the LMC or SMC values for $N$(H$_{\rm tot}$) is increased by more than 0.15 dex. 
The corresponding increases in the average $N$(H$_{\rm tot}$)/$E(B-V)$ are at most 0.01 dex for all three galaxies (``H2est'' entries in Table~\ref{tab:ratgal}).

For $N$(\ion{H}{1})/$E(B-V)$, the new absorption-line data imply somewhat smaller average ratios for the Magellanic Clouds than those found previously (more consistent with the results of Gordon et al. 2003):  15--16$\times$10$^{21}$ cm$^{-2}$ mag$^{-1}$ for the LMC and 21--30$\times$10$^{21}$ cm$^{-2}$ mag$^{-1}$ for the SMC.
For $N$(H$_{\rm tot}$)/$E(B-V)$, the average values are 15--17$\times$10$^{21}$ cm$^{-2}$ mag$^{-1}$ for the LMC and 23--30$\times$10$^{21}$ cm$^{-2}$ mag$^{-1}$ for the SMC. 
The average $N$(\ion{H}{1})/$E(B-V)$ ratios thus are higher in the LMC and SMC by factors of 3.5 and 4.8--6.3, respectively, and the average $N$(H$_{\rm tot}$)/$E(B-V)$ ratios are higher by factors of 2.8--2.9 and 4.1--5.2, respectively (relative to the Galactic values given in the first column of Table~\ref{tab:ratgal}).
Kolmogorov-Smirnov (K-S) tests indicate that the $N$(H$_{\rm tot}$)/$E(B-V)$ ratios for our samples of Milky Way, LMC, and SMC sight lines are unlikely (probability $<$ 0.001) to be drawn from a common distribution -- and that the differences in the mean values are significant.
For both the LMC and SMC, the new data indicate that the differences in the gas-to-dust ratios are more comparable to the differences in metallicity (relative to our Galaxy) than had been suggested by many of the previous estimates -- particularly for the SMC.

The $N$(H$_2$)/$E(B-V)$ ratio behaves somewhat differently than the corresponding ratios involving $N$(\ion{H}{1}) and $N$(H$_{\rm tot}$).
First, while the Galactic $N$(\ion{H}{1})/$E(B-V)$ and $N$(H$_{\rm tot}$)/$E(B-V)$ ratios depend weakly (at most) on $E(B-V)$, the average $N$(H$_2$)/$E(B-V)$ increases by a factor of 3--4 as $E(B-V)$ increases from about 0.1 to 0.9 mag (Fig.~\ref{fig:avgrat}).
Second, the $N$(H$_2$)/$E(B-V)$ ratios in the LMC and SMC are generally more comparable to those seen in our Galaxy (as suggested by Gunderson et al. 1998, Richter 2000, and Cartledge et al. 2005 for much smaller samples) (Figs.~\ref{fig:rvsebv} and \ref{fig:hist}).
However, the new LMC and SMC samples are still fairly small, the scatter for each is large, and there are as yet no measurements of H$_2$ for sight lines with $E(B-V)$ $>$ 0.4 mag in the Magellanic Clouds (Fig.~\ref{fig:rvsebv}). 
If the Galactic sample is further restricted to $E(B-V)$ $\le$ 0.35 (similar to the range covered in the LMC and SMC), then the average log[$N$(H$_2$)/$E(B-V)$] = 20.80 is slightly smaller -- closer to the corresponding averages for the LMC and SMC (``EBV35'' entries in Table~\ref{tab:ratgal}) -- and the K-S probabilities that the LMC and SMC ratios could be drawn from the Galactic distribution increase to several percent.
[Note that only sight lines with $N$(H$_2$) $>$ 3 $\times$ 10$^{18}$ cm$^{-2}$ (where the H$_2$ is self-shielded and generally correlated with CH, \ion{Na}{1}, and \ion{K}{1}) have been included in calculations of the mean ratios.]

\subsubsection{Variations in Gas-to-Dust Ratios?}
\label{sec-gdrvar}

Besides providing better definition of the mean relationships between $N$(\ion{H}{1}), $N$(H$_2$), $N$(H$_{\rm tot}$), and $E(B-V)$ in the Magellanic Clouds, the new data enable both exploration of possible regional variations in the gas-to-dust ratios (as suggested by some of the studies of IR emission and/or UV extinction) and identification of individual ``discrepant'' sight lines.
In Figures~\ref{fig:smclos} and \ref{fig:lmclos}, the SMC and LMC sight lines included in this survey are coded by log(GDR), in three ranges.
The plots of the ratios for individual sight lines (Fig.~\ref{fig:rvsebv}) suggest that the scatter in the values for each galaxy is greater than the typical uncertainties of the individual values, particularly in the LMC and SMC.
For the GDR $N$(H$_{\rm tot}$)/$E(B-V)$, for example, the typical uncertainties are $\la$ 0.15 dex for LMC and SMC sight lines with $E(B-V)$ $>$ 0.1 mag, but the overall rms deviations are 0.22 and 0.34 dex, respectively (Table~\ref{tab:ratgal}).
While some of the scatter could be due to differences in $R_{\rm v}$ (as noted above for some Galactic sight lines), most of the $R_{\rm v}$ derived so far for LMC and SMC sight lines are between 2.3 and 3.7 (Gordon et al. 2003).
In the LMC, the ratios for both \ion{H}{1} and H$_2$ (and thus H$_{\rm tot}$) may be slightly higher (and exhibit less scatter) in the LMC~2 supershell region, but the ratios there generally are still consistent with the overall average values, within the mutual 1-$\sigma$ uncertainties (Table~\ref{tab:ratgal}).
The six sight lines in the LH~10 association (in the N11 star-forming region) all have lower than average ratios for H$_2$, and five of the six have $N$(H$_2$)/$E(B-V)$ $<$ 10$^{20}$ cm$^{-2}$ mag$^{-1}$ -- factors $\ga$ 4 smaller than the overall LMC average -- even though $E(B-V)$ $\ge$ 0.14 (Table~\ref{tab:lmclos}; Fig.~\ref{fig:rvsebv}).
Stronger than average radiation fields in LH~10 may suppress the H$_2$ abundance there (as may also be the case for some Galactic sight lines in Orion and Sco-Oph); determination of the relative populations in the higher H$_2$ rotational levels ($J$ = 4,5) could provide corroboration.
In the SMC wing region, the average ratio for \ion{H}{1} is a factor $\sim$2 lower than in the main SMC bar, but the average ratio for H$_2$ is a factor $\sim$2 higher.
There is significant scatter in both ratios, however; both SMC wing and SMC bar samples, for example, include sight lines with very low and very high $N$(H$_2$)/$E(B-V)$ ratios.
K-S tests suggest that there may be differences in $N$(H$_{\rm tot}$)/$E(B-V)$ between the LMC2 and LMC main subsamples (presumably reflecting the small dispersion in the LMC2 values), but are inconclusive for the SMC bar and wing subsamples.
At least some of the (relatively small) differences in the regional average values thus may just be due to local environmental effects and small-number statistics; the absorption-line data reveal no obvious very large-scale regional variations in the GDRs in the Magellanic Clouds.

Whether there are regional variations or not, there are individual sight lines (in all three galaxies) whose ratios differ significantly from the corresponding mean values.
For several of the sight lines with $E(B-V)$ $<$ 0.03 mag (AV~104, NGC330-B37, NGC346-E46), the high ratios for \ion{H}{1} and H$_{\rm tot}$ might be (at least partly) due to underestimated $E(B-V)$, as such high values are not seen for more reddened sight lines.
The high $N$(H$_2$)/$E(B-V)$ ratios for several more reddened sight lines (Sk~18, NGC346-12, Sk~143, Sk~150, AV~476, Sk~191, Sk$-$67~2), however, appear to be accurate. 
Several of these ratios are higher than those found for nearly all Galactic sight lines -- reflecting a combination of strongly self-shielded H$_2$ and lower dust-to-gas ratios.
On the other hand, there are also a number of LMC and SMC sight lines (covering a range in reddening) with values of $N$(\ion{H}{1})/$E(B-V)$ and/or $N$(H$_{\rm tot}$)/$E(B-V)$ that are comparable to or lower than the average Galactic values.
For the LMC WC4 stars (especially Sk$-$66~21, Sk$-$68~15, and Sk$-$69~234), the low values may be due in part to overestimated $E(B-V)$; the adopted $(B-V)_0$ = $-$0.40 may be too small.
The sight lines toward Sk$-$67~2 (LMC) and Sk~143 (SMC) [the only Magellanic Clouds sight lines in our sample with $f$(H$_2$) $>$ 0.5] have low values for $N$(\ion{H}{1})/$E(B-V)$, but high values for $N$(H$_2$)/$E(B-V)$. 
The resulting ratio for H$_{\rm tot}$ is consistent with the LMC average for Sk$-$67~2 (Cartledge et al. 2005), but is still closer to the Galactic average for Sk~143 [8.2 $\times$ 10$^{21}$ cm$^{-2}$ mag$^{-1}$ -- significantly lower than the value found (for \ion{H}{1} alone) by Gordon et al. 2003].
Of the other sight lines with low ratios for \ion{H}{1} and $E(B-V)$ $>$ 0.15 mag (Sk~5, Sk~36, Sk~101, Sk~142, Sk$-$69~142a, LMC1-548, LMC2-702), only Sk$-$69~142a has data for H$_2$ (an upper limit), but we consider it unlikely that inclusion of any molecular gas in those sight lines [all with $E(B-V)$ $<$ 0.3 mag] would increase $N$(H$_{\rm tot}$) by more than about 50\% (see above).\footnotemark
\footnotetext{The $E(B-V)$ for Sk$-$69~142a may be significantly overestimated.} 
Unlike Sk~143, none of the four other SMC sight lines with low $N$(\ion{H}{1})/$E(B-V)$ ratios shows any obvious indications of the 2175 \AA\ bump in the existing low-dispersion {\it IUE} spectra.
Sk~143 thus remains unique in the SMC in having both UV extinction and gas-to-dust ratio more like those in the Galactic ISM; those properties do not appear to be characteristic of the SMC wing.

The assumption of similar gas-to-dust ratios in both atomic and molecular gas (commonly adopted in studies attempting to infer the amount of H$_2$ from comparisons of IR, CO, and \ion{H}{1} emission; see below) seems consistent with the observed good correlations between $E(B-V)$ and the total hydrogen column density in our Galaxy, the LMC, and the SMC. 
It would not seem unreasonable, however, for the dust to be more closely associated with the H$_2$ than with the \ion{H}{1} -- as H$_2$ is formed on grain surfaces and as the grains would be expected to survive and grow more readily in denser gas, where the molecular fraction is likely to be higher.
For the simplest case of constant (but different) GDRs associated with atomic and molecular gas, we would expect to observe the atomic GDR for sight lines with low molecular fraction [e.g., $f$(H$_2$) $\la$ 0.1], with a gradual transition to the molecular GDR as $f$(H$_2$) approaches 1.0.
Figure~\ref{fig:rvsf} shows the $N$(H$_{\rm tot}$)/$E(B-V)$ ratio as a function of $f$(H$_2$), for sight lines in our Galaxy, the LMC, and the SMC.
While the scatter is large and there are few sight lines with $f$(H$_2$) $>$ 0.1 for the LMC and SMC, there is an intriguing possible trend of declining GDR for $f$(H$_2$) $>$ 0.1 in the Galactic data (Figure~\ref{fig:rvsf2}).
Fits to the Galactic data for both log[$N$(H$_{\rm tot}$)/$E(B-V)$] and log[$N$(H$_{\rm tot}$)/$A_{\rm v}$] yield slopes $\sim$ $-$0.2 over the range $-$1.0 $<$ log[$f$(H$_2$)] $<$ 0.0 -- suggesting that the GDR in Galactic molecular gas could be lower by a factor of 1.5--2.0 than in predominantly atomic gas.
Such a decrease in the gas-to-dust ratio in molecular gas would be consistent with a recent analysis of the dust emission from 350 $\mu$m to 3 mm in the Taurus molecular complex, measured with the {\it Planck} satellite, which finds a factor-of-2 increase in the ratio of dust optical depth to total hydrogen column density for molecular gas with $A_{\rm v}$ $\ga$ 1 (Planck collaboration 2011a).
It also appears to be at least qualitatively consistent with the increased dust-phase abundances of significant dust constituents (C, O, Mg, Si) in sight lines exhibiting the most severe depletions (Jenkins 2009) and with models of grain growth in dense interstellar clouds (e.g., Dwek 1998).
On the other hand, Roman-Duval et al. (2010) found no significant differences in GDR within two LMC molecular cloud complexes, ascribing variations in the ratio of far-IR emission to gas surface density to the presence of a significant amount of H$_2$ not traced by the CO emission (see discussion below).

\subsection{Depletions and Diffuse Cloud Chemistry}
\label{sec-depl}

Larger samples of accurate $N$(\ion{H}{1}) and $N$(H$_2$) provide reference abundances for broader studies of depletions and diffuse cloud chemistry in the Magellanic Clouds.
Preliminary values of some of the color excesses and \ion{H}{1} and H$_2$ column densities listed in Tables~\ref{tab:smclos} and \ref{tab:lmclos} were used in studies of CH, CH$^+$, CN, and several of the DIBs (Welty et al. 2006) and of the depletion of titanium (Welty \& Crowther 2010).
As noted above, some of the SMC $N$(\ion{H}{1}) have been adjusted slightly for differences in the adopted Galactic contribution to the Lyman-$\alpha$ absorption, and new or revised values for \ion{H}{1} and/or H$_2$ are now available for a number of the sight lines in Welty et al. (2006).
The revised values do not significantly change any of the results of the \ion{Ti}{2} survey:  the depletion of Ti remains less severe in the LMC and (especially) the SMC than in the local Galactic ISM, for any given $N$(H), $E(B-V)$, or molecular fraction $f$(H$_2$) (Welty \& Crowther 2010).

In principle, comparisons of the gas-to-dust ratios and metallicities in the Milky Way and Magellanic Clouds can yield constraints on the relative depletions of the major dust constituents (generally thought to be C, O, Mg, Si, Fe) in the three galaxies.
For the LMC, the mean $N$(H$_{\rm tot}$)/$E(B-V)$ for our sample is nearly 3 times the average local Galactic value, slightly larger than the factor-of-2 difference in metallicities -- perhaps indicative of generally somewhat less severe depletions in the LMC.
While that would be at least qualitatively consistent with the observed titanium depletions (Welty \& Crowther 2010), Ti is not a major dust constituent.
For the SMC, the mean $N$(H$_{\rm tot}$)/$E(B-V)$ for our sample is more comparable to the difference in metallicities -- despite the even less severe depletions of Ti in the SMC (Welty \& Crowther 2010) and suggestions that Mg and Si might be significantly less depleted in at least some SMC sight lines (Welty et al. 2001 and in prep.; cf. Sofia et al. 2006).
Such comparisons, however, might also be affected by differences in the relationships between $E(B-V)$ and the total amount of dust, differences in dust composition, and/or differences in the relative total abundances of the individual major dust constituents.
Surveys of additional refractory species (particularly the significant dust constituents) and determinations of total visual extinctions for more sight lines are needed to better understand the apparent differences in depletion patterns and gas-to-dust ratios in the Magellanic Clouds.

The only significant change regarding the study of the diatomic molecules and DIBs is the more clearly defined relationship in the LMC between the equivalent width of the $\lambda$5780.5 DIB and $N$(\ion{H}{1}) -- which is now seen to run roughly parallel to the corresponding Galactic relation (Herbig 1993; Friedman et al. 2011), with $W$(5780.5) smaller by a factor of $\sim$10 (Fig.~\ref{fig:5780} and Table~\ref{tab:ratgal}).
As that difference is much larger than the factor-of-2 difference in metallicity, the metallicity dependence of the $W$(5780.5)/$N$(\ion{H}{1}) ratio is not yet understood.
The average $N$(CH)/$N$(H$_2$) ratio in the LMC (six sight lines) remains very similar to that seen in the local Galactic ISM, despite the lower metallicity in the LMC.
Additional optical spectra of sight lines with higher $E(B-V)$, $N$(\ion{H}{1}), and/or $N$(H$_2$) would provide a clearer view of the relationships between those three quantities, the column densities of the diatomic molecules, and the DIBs in the Magellanic Clouds -- and thus a better understanding of diffuse cloud chemistry in low-metallicity galaxies.
While the new values for $N$(\ion{H}{1}) imply somewhat higher molecular fractions for a number of the sight lines in the H$_2$ survey of Tumlinson et al. (2002) (particularly in the SMC), the $f$(H$_2$) still require both reduced H$_2$ formation and enhanced H$_2$ destruction in the LMC and SMC, relative to the local Galactic ISM.

\subsection{Understanding the Atomic-to-Molecular Transition in Galaxies}
\label{sec-trans}

The formation of molecular gas is generally considered to be a key factor for determining the star formation rate -- and thus for driving galactic evolution.
A number of recent theoretical studies, incorporating detailed treatments of the formation, shielding, and destruction of H$_2$, therefore have attempted to understand how the relationship between atomic and molecular gas depends on such properties as the metallicity, dust content, gas column or volume density, and radiation field (e.g., Gnedin, Tassis, \& Kravtsov 2009; Pelupessy \& Papadopoulos 2009; Krumholz, McKee, \& Tumlinson 2009; McKee \& Krumholz 2010; Gnedin \& Kravtsov 2011).
These studies generally conclude that the molecular fraction $f$(H$_2$) depends primarily on the total column density and metallicity (or gas-to-dust ratio), with a somewhat weaker dependence on the ambient radiation field.
The predicted relationships seem reasonably consistent with the distributions of atomic and molecular gas (as traced by \ion{H}{1} 21~cm and CO emission) in nearby galaxies [e.g., Krumholz et al. 2009 (KMT09)].

Another approach to this issue, starting with the observed distributions of atomic and molecular gas in nearby galaxies, has revealed a correlation between the H$_2$/\ion{H}{1} ratio ($R_{\rm mol}$) and the total interstellar pressure ($P_{\rm tot}$) at the galactic midplane (e.g., Wong \& Blitz 2002; Blitz \& Rosolowsky 2004, 2006).
That empirical correlation is then taken to reflect a dependence of the molecular fraction on the total pressure -- at least on scales greater than several hundred pc.
Both the H$_2$ and the pressure are estimated indirectly, however, and the physical process(es) underlying the correlation have not been identified.
Recent work by Ostriker, McKee, \& Leroy (2010), considering the relationships between the relative amounts of dense (molecular) and diffuse (atomic) gas, the rate of star formation, the photoelectric heating of the gas, and the thermal pressure -- as set by the coupled conditions of thermal and dynamical equilibrium -- may provide some theoretical support for the empirical $R_{\rm mol}$--$P_{\rm tot}$ relationship.
Fumagalli, Krumholz, \& Hunt (2010) suggest that observations of low-metallicity dwarf galaxies with high stellar densities should allow some discrimination between the two approaches. 
For a given $N$(H$_{\rm tot}$), the models based on H$_2$ microphysics would predict lower molecular fractions (due to the low metallicities), while the pressure-driven models would predict higher molecular fractions (due to the high stellar densities).
Unfortunately, the \ion{H}{1}, CO, and IR emission data used to test these ideas all come with significant caveats, so that additional diagnostics would be valuable.

While the existing H$_2$ microphysics-based models are designed to describe the average or bulk behavior of molecular clouds, direct absorption-line measures of \ion{H}{1} and H$_2$ -- which sample specific lines of sight through the clouds -- can (in principle) provide such additional diagnostics.
Figure~\ref{fig:mckee} compares the relationships between $N$(H$_2$) and $N$(H$_{\rm tot}$) predicted by the simple analytic approximation given in eqn.~89 of McKee \& Krumholz (2010), for metallicities corresponding to the Milky Way (1.0 $\times$ solar; top panel, black curve), the LMC (0.5 $\times$ solar; bottom panel, green curve), and the SMC (0.2 $\times$ solar; bottom panel, red curve), with the column densities of H$_2$ and total hydrogen observed for sight lines in those three galaxies (black crosses, green circles, and red triangles, respectively).\footnotemark
The three model curves all exhibit the expected rapid rise in $N$(H$_2$) due to self-shielding, once sufficient total hydrogen column density has been reached; the predicted threshold for self-shielding occurs at higher $N$(H$_{\rm tot}$) for lower metallicities, due (presumably) to the lower dust abundances.
\footnotetext{Note that the figure shows the relationships in terms of column densities ($N$, in cm$^{-2}$), rather than surface densities ($\Sigma$, in M$_{\odot}$~pc$^{-2}$).
For reference, log[$\Sigma$(H)] = log[$N$(H)] $-$ 19.96 (which includes a factor of 1.36 for helium).}

There is some qualitative agreement between the measured column densities and the trends with metallicity predicted by the KMT09 models, in that higher $N$(H$_{\rm tot}$) are generally required for high $N$(H$_2$) in the lower metallicity LMC and SMC.
As recognized by Krumholz et al. (2009) for Galactic sight lines, however, many of the observed points (for all three galaxies) lie above and/or to the left of the predicted curves -- indicating higher $N$(H$_2$) than those predicted by the models.
While the observed points with $N$(H$_2$) $\la$ 10$^{16}$ cm$^{-2}$ represent the small amounts of unshielded H$_2$ commonly present in low-$N$(H$_{\rm tot}$) sight lines (and ignored in the KMT09 models), the points with $N$(H$_2$) $\ga$ 10$^{18}$ represent self-shielded H$_2$ -- and their presence at lower than predicted $N$(H$_{\rm tot}$) is opposite to what would be expected for absorption-line observations randomly sampling individual molecular clouds.
Simple simulations of such sampling of spherical clouds (over ranges of total cloud column densities and impact parameters), for example, suggest that sight lines near the cloud centers would probe slightly higher molecular fractions (relative to the cloud average values), while the more numerous sight lines farther from the centers would see lower than average molecular fractions.
The ensemble of observed column densities would thus populate a region largely to the right of the predicted curve -- bounded on the left by a curve starting at somewhat higher $N$(H$_{\rm tot}$) for low $N$(H$_2$) and increasing to just slightly above the predicted curve when $f$(H$_2$) exceeds $\sim$0.5 (see the discussion of similar simulations and associated fig.~12 in Krumholz et al. 2009).
Moreover, removal of any predominantly atomic gas that is unrelated to the H$_2$ [which is currently included in the observed $N$(H$_{\rm tot}$)] along the sight lines would move the observed points even further to the left in Figure~\ref{fig:mckee}.

The KMT09 models thus appear to overestimate the $N$(H$_{\rm tot}$) at which self-shielding becomes effective -- and consequently at which significant amounts of H$_2$ can be present.
For the Milky Way, that transition occurs at $E(B-V)$ $\sim$ 0.08 (Savage et al. 1977), corresponding to log[$N$(H$_{\rm tot}$)] $\sim$ 20.65 (for the average gas-to-dust ratio listed in Table~\ref{tab:ratgal}) -- quite consistent with the more extensive Galactic data shown in the top panel of Figure~\ref{fig:mckee}, but a factor of $\sim$3 lower than the column density threshold predicted by the solar metallicity model.
While the corresponding transition column densities are not yet as well characterized for the LMC and SMC, the presence of self-shielded H$_2$ at significantly lower $N$(H$_{\rm tot}$) than the corresponding predicted lower-metallicity thresholds suggests that the threshold values may be overestimated there as well.
This discrepancy between observed and predicted thresholds is at least qualitatively consistent with Leroy et al.'s (2009) observation that the $\Sigma$(H$_2$)/$\Sigma$(\ion{H}{1}) ratios inferred for the N83 region (in the SMC wing) were most consistent with the KMT09 models for metallicities somewhat higher than that of the SMC.
The simulations of Gnedin \& Kravtsov (2011) yield similar trends for the molecular fraction with dust-to-gas ratio (a proxy for the metallicity), but with broader transition regions [in $N$(H$_{\rm tot}$)] -- more like those observed for both the Galactic and Magellanic Clouds sight line samples in Fig.~\ref{fig:mckee}.
 
The lower metallicity KMT09 models also predict generally higher molecular fractions, for self-shielded H$_2$, than those observed so far in the LMC and SMC.
While part of the difference may reflect the set of sight lines selected for observation with {\it FUSE}, nearly all the observed Magellanic Clouds sight lines with strong H$_2$ absorption have been included in this study.
As concluded by Tumlinson et al. (2002), enhanced photodissociation of H$_2$ by stronger radiation fields appears to be required, in addition to reduced formation, to account for the abundance of H$_2$ in the Magellanic Clouds.

\subsection{Predicting $N$(H$_2$) from \ion{H}{1} 21~cm and Far-IR Emission}
\label{sec-predh2}

Because cold H$_2$ is difficult to measure directly in emission, the amount of cold molecular gas is often inferred from tracer molecules -- e.g., CO, typically assuming a ``standard'' value for the CO-to-H$_2$ ``$X$-factor'' derived from observations of Galactic molecular clouds.
Strictly speaking, the use of the $X$-factor assumes that the molecular material is in virial equilibrium (which may not always be the case).
Moreover, at low metallicities, the $X$-factor may differ significantly from the Galactic value (e.g., Israel 1997; Leroy et al. 2007, 2011).
An alternative approach is to use the far-IR emission from dust (thought to trace both atomic and molecular gas) to infer the total gas mass, and then to subtract the contribution from \ion{H}{1} (from 21~cm observations) to obtain the amount of H$_2$.
This approach also depends on several assumptions:  that the gas-to-dust ratio is the same in both predominantly atomic and predominantly molecular gas, that the 21~cm emission is optically thin, and that reference regions free of H$_2$ can be identified (in order to calibrate the gas-to-dust ratio).
Studies aplying this approach to the Magellanic Clouds generally find that use of a Galactic $X$-factor underestimates the H$_2$ mass by factors $\ga$5 in the LMC and $\ga$10 in the SMC (Bot et al. 2007; Bernard et al. 2008; Leroy et al. 2009) -- suggesting that substantial ``dark'' H$_2$ envelopes may surround the CO-emitting portions of molecular clouds.
Those results are qualitatively consistent with theoretical expectations that the more abundant H$_2$ should self-shield more readily, and thus occupy a larger volume than CO -- particularly in lower metallicity systems, where C, O, and the dust (for shielding the CO in more diffuse molecular gas) are all less abundant (e.g., Maloney \& Black 1988; Wolfire et al. 2010).
On the other hand, as discussed above, the gas-to-dust ratio may be somewhat lower in predominantly molecular gas, and the 21~cm emission may be somewhat saturated in regions near/surrounding the molecular clouds.
Adjusting for those effects would reduce the amount of H$_2$ inferred from the ``excess'' IR emission.

As one example of this approach to estimating the molecular content, we consider a recent detailed study of the N83/N84 star-forming region, located in the SMC wing (Figure~\ref{fig:leroy}).
Leroy et al. (2009) combined IR data from {\it Spitzer} and {\it IRAS}, the ATCA+Parkes 21~cm data, and CO 1-0 and 2-1 maps from Bolatto et al. (2003) to investigate the detailed structure and molecular content of this region.
Using an estimated gas-to-dust ratio of 5 $\times$ 10$^{22}$ cm$^{-2}$ mag$^{-1}$ (8.6 times the Galactic value), they inferred a substantial amount of H$_2$ outside the CO contours, with a CO-to-H$_2$ conversion factor (averaged over the entire complex) 20--55 times the typical Galactic value (but locally much lower near the peaks of the CO emission).

Absorption-line probes near the molecular cores seen in CO emission can provide checks of the $N$(H$_2$) inferred from the IR and 21~cm emission -- and can also yield estimates for abundances and physical conditions to aid in characterizing the diffuse molecular gas.
Table~\ref{tab:smclos} includes five sight lines in the vicinity of N83/N84 with data for $E(B-V)$, $N$(\ion{H}{1}), and $N$(H$_2$).
Two of these (AV~476, Sk~155) lie just within the CO emission contours; the three others (Sk~156, Sk~157, Sk~159) lie slightly to the east of the CO emission; Leroy et al. used data for Sk~159 (the farthest of the five) in estimating the gas-to-dust ratio in the region.
The heliocentric velocities of the various CO peaks identified by Bolatto et al. (2003) range from about 162 to 179 km~s$^{-1}$, which fall within the strong higher-velocity peak in the 21~cm emission in this region.
The available high-resolution spectra of interstellar \ion{Na}{1} and \ion{Ca}{2} absorption toward Sk~155, Sk~156, and Sk~159, however, indicate that the strongest absorption is at somewhat lower velocities (Wayte 1990; Welty et al. 2001, 2006; Andr\'{e} et al. 2004; Welty \& Crowther, in prep.) -- suggesting that those three stars lie in front of the N83/N84 complex, even though they all exhibit significant absorption from H$_2$ [with $N$(H$_2$) $\sim$ 10$^{19}$ cm$^{-2}$].

The sight line to AV~476 (closest to the strong CO emission) does appear to probe material in the N83/N84 complex, however.
This sight line has the highest $N$(H$_2$) in our SMC sample (at 9 $\times$ 10$^{20}$ cm$^{-2}$), and the corresponding absorption from CH and \ion{Na}{1} is at the same velocity (within about 1 km~s$^{-1}$) as the two nearest CO peaks (Bolatto et al. 2003; Welty et al. 2006).
The gas-to-dust ratio toward AV~476 is about 4.5 $\times$ 10$^{22}$ cm$^{-2}$ mag$^{-1}$ -- slightly higher than the values toward the other four stars in the region and for the SMC wing region as a whole (Table~\ref{tab:ratgal}), but close to the value estimated by Leroy et al. (2009).
The mass surface density of molecular gas $\Sigma$(H$_2$) toward AV~476 predicted from the IR and 21~cm emission is between 150 and 200 M$_{\odot}$ pc$^{-2}$ (Fig.~\ref{fig:leroy}), which corresponds to $N$(H$_2$) $\sim$ 7--9 $\times$ 10$^{21}$ cm$^{-2}$ (eqn.~11 of Leroy et al. 2009) -- nearly an order of magnitude higher than the value determined from UV absorption.

The significant difference between measured and inferred $N$(H$_2$) for this one sight line might (in principle) be ascribed to small-scale structure effects (unresolved at the 23 arcsec resolution of the CO 2-1 data and the 36 arcsec resolution of the 160$\mu$m data) and/or to AV~476 perhaps not lying completely behind the molecular material.
Indeed, Bolatto et al. (2003) concluded that the high CO(2-1)/CO(1-0) ratios in the region around AV~476 might indicate that the CO emission comes from many small ($r$ $\sim$ 0.1 pc) clumps.
On the other hand, as discussed above, the H$_2$ would likely be distributed more broadly than the CO (i.e., less strongly clumped).
Moreover, the \ion{H}{1} column densities determined from both Lyman-$\alpha$ absorption and 21~cm emission (at the 1.5 arcmin resolution of the AP 21~cm data) agree to better than 10\%.
Unless the 21~cm emission significantly underestimates the true $N$(\ion{H}{1}) -- which would also have implications for the $N$(H$_2$) inferred from the emission data -- there is thus likely to be little atomic gas beyond AV~476.
Similar comparisons, for other regions exhibiting CO emission in the LMC and SMC, would be valuable for gauging the general reliability of the $N$(H$_2$) inferred from the IR, CO, and 21~cm emission data.

\section{SUMMARY / CONCLUSIONS}
\label{sec-summ}

In order to examine the relationships between atomic gas, molecular gas, and dust in the ISM of the Magellanic Clouds, we have collected the following data for a set of 126 sight lines in the SMC and 159 sight lines in the LMC:\footnotemark
\footnotetext{These data (and future updates) may be obtained at http://astro.uchicago.edu/$\sim$dwelty/mcoptuv.html and coldens\_mc.html.}

\begin{itemize}

\item{Column densities of \ion{H}{1} for 255 sight lines (181 newly determined, mostly from archival {\it HST} spectra of Lyman-$\alpha$ absorption)}

\item{Corresponding estimates for $N$(\ion{H}{1}) from three recent 21~cm emission surveys covering the Magellanic Clouds (for all 285 sight lines)}

\item{Column densities of H$_2$ for 145 sight lines (69 newly determined from archival {\it FUSE} spectra of Lyman-band absorption)}

\item{Spectral types and $B$, $V$ photometry for nearly all of the 285 sight lines (from the literature) --- enabling new estimates for $E(B-V)$ (both Galactic and Magellanic Clouds contributions)}

\end{itemize}

Some of these data have already been used in an exploratory study of diffuse cloud chemistry and diffuse interstellar bands in the Magellanic Clouds (Welty et al. 2006) and in a survey of titanium depletions in the Milky Way and Magellanic Clouds (Welty \& Crowther 2010).
Revised values for $E(B-V)$, $N$(\ion{H}{1}), $N$(H$_2$) presented here yield no significant changes to the conclusions of those earlier papers, other than a more well defined relationship between the equivalent width of the DIB at 5780.5 \AA\ and $N$(\ion{H}{1}) for sight lines in the LMC.
Comparisons among these data examined in this paper indicate or suggest the following:

\begin{itemize}

\item{Systematic differences in the Magellanic Clouds $N$(\ion{H}{1}) derived from surveys of 21~cm emission undertaken at different spatial resolutions most likely reflect structure in the ISM in the LMC and SMC on scales smaller than the radio beams.}

\item{Differences in the Magellanic Clouds $N$(\ion{H}{1}) determined from 21~cm emission and Lyman-$\alpha$ absorption can be due to foreground-background effects [typically with $N$(21~cm) $>$ $N$(Ly-$\alpha$)], small-scale structure in the ISM, and/or possible saturation or self-absorption in the 21~cm data [typically with $N$(Ly-$\alpha$) $>$ $N$(21~cm)].
Comparisons between the 21~cm emission profiles and moderate-to-high resolution UV/optical absorption-line spectra of species such as H$_2$ and \ion{Na}{1} can aid in understanding those differences in $N$(\ion{H}{1}).
For some sight lines in the LMC, the $N$(21~cm) determined even from the high-resolution ATCA+Parkes data can be lower than the $N$(\ion{H}{1}) obtained from Lyman-$\alpha$ absorption by as much as a factor of 3 -- likely due to saturation in the 21~cm emission.
Some of the 21~cm profiles seem to suggest self-absorption in the \ion{H}{1}.}

\item{Use of information from \ion{Na}{1} and/or H$_2$ absorption to restrict the velocity range over which the 21~cm profiles are integrated generally yields $N$(21~cm) in better agreement with $N$(Ly-$\alpha$) -- minimizing foreground-background effects and thus providing a way of estimating more accurate $N$(\ion{H}{1}) from 21~cm data if Lyman-$\alpha$ data are not available. 
In principle, estimates for $N$(\ion{H}{1}) may also be obtained from the strength of the DIB at 5780.5 \AA, if the typical $N$(\ion{H}{1})/$W$(5780.5) ratio (dependent on metallicity and other factors) can be determined for a given system.}

\item{Comparisons of the column densities of \ion{H}{1}, H$_2$, and H$_{\rm tot}$ with the color excess $E(B-V)$ indicate that the best correlation is between $N$(H$_{\rm tot}$) and $E(B-V)$.
In the local Galactic ISM, the average gas-to-dust ratio $N$(H$_{\rm tot}$)/$E(B-V)$ $\sim$ 5.8 $\times$ 10$^{21}$ cm$^{-2}$ mag$^{-1}$, apparently independent of reddening for $E(B-V)$ $\la$ 1 mag.
In the Magellanic Clouds, that average ratio is higher by factors of 2.8--2.9 in the LMC and 4.1--5.2 in the SMC -- more similar to the differences in metallicity than indicated by previous such estimates -- especially for the SMC. 
While there are real variations in the GDR for individual sight lines within each galaxy, any very large-scale regional variations in the GDR in the LMC or SMC are difficult to discern, given the scatter in individual sight line values. 
In the Galactic ISM, the average $N$(H$_2$)/$E(B-V)$ ratio increases by a factor of 3--4 as $E(B-V)$ increases from 0.1 to 0.9 mag; on average, that ratio is more similar in the three galaxies, at any given $E(B-V)$.
The Galactic data for sight lines with molecular fraction $f$(H$_2$) $\ga$ 0.1 suggest that the GDR may be lower in predominantly molecular gas by a factor of 1.5--2, compared to its value in more diffuse, predominantly atomic gas.}

\item{Comparison of the column densities of \ion{H}{1}, H$_2$, and H$_{\rm tot}$ for the Milky Way, LMC, and SMC with the predictions of the theoretical models of McKee \& Krumholz (2010) for the relationship between atomic and molecular gas indicates that while the models correctly predict that the atomic-to-molecular transition occurs at higher total hydrogen column densities for lower-metallicity systems, they appear to overestimate the specific threshold $N$(H$_{\rm tot}$) for the Milky Way (and probably for the LMC and SMC as well) by a factor of at least 3, and they also may overestimate the molecular fractions in the lower-metallicity LMC and SMC.}

\item{For one sight line within the CO contours of the N83/N84 star-forming region in the SMC wing region, the $N$(H$_2$) determined directly from Lyman-band absorption is nearly an order of magnitude smaller than the value estimated from the IR and 21~cm emission along that sight line.}

\end{itemize}

This expanded set of column densities for \ion{H}{1} and H$_2$ provides reference abundances for future studies of the metallicity, depletions, and chemistry in the diffuse atomic and molecular gas in the Magellanic Clouds.
Some of the sight lines with relatively high $E(B-V)$ and/or $N$(H$_2$) would be good candidates for future observations of various molecular species and DIBs -- to better characterize diffuse cloud chemistry and the atomic-to-molecular transition in these lower-metallicity systems.
Determination of $N$(H$_2$) for the rest of the sight lines in the {\it FUSE} Magellanic Clouds Legacy Archive -- and of the higher-$J$ populations for the entire sample -- should help to better define and characterize the atomic-to-molecular transition in the Magellanic Clouds. 
More extensive comparisons with the predictions for H$_2$ based on considerations of 21~cm, CO, and IR emission should help to refine the predictions -- and thus more accurately gauge the amount of ``dark'' H$_2$ not traced by CO.

\acknowledgements

We thank Jason Tumlinson both for supplying the {\bf h2gui} package and for constructive comments on the paper and Mark Krumholz for several stimulating discussions.
The {\it HST} and {\it FUSE} data analyzed in this paper were obtained from the Multimission Archive at the Space Telescope Science Institute (MAST). STScI is operated by the Association of Universities for Research in Astronomy, Inc., under NASA contract NAS5-26555. Support for MAST for non-{\it HST} data is provided by the NASA Office of Space Science via grant NAG5-7584 and by other grants and contracts.
The Australia Telescope Compact Array and Parkes radio telescope are part of the Australia Telescope National Facility which is funded by the Commonwealth of Australia for operation as a National Facility managed by CSIRO.
Work on this project was begun under NASA archival grant HST-AR-10692.01-A from STScI, and finished under NASA ADAP grant 10-ADAP10-0137; DEW also acknowledges support from the CARMA project at UIUC, which is funded by NSF URO grant AST-0838226.

{\it Facilities:} \facility{HST (COS, FOS, GHRS, STIS)}, \facility{FUSE}, \facility{IUE}

\appendix

\section{HST Programs}
\label{sec-hstprog}

Table~\ref{tab:prog} lists the {\it HST} programs which obtained the spectra used in this study of Lyman-$\alpha$ absorption in the Magellanic Clouds.
For each program, the table lists the instrumental configuration, the spectral resolution at Lyman $\alpha$ (corresponding to the FWHM of the instrumental line spread function), the number of SMC and LMC targets used, and the principal investigator.

\section{21~cm Velocity Ranges}
\label{sec-21range}

Table~\ref{tab:21vel} lists the velocity ranges, derived from high-resolution observations of \ion{Na}{1} absorption (Vladilo et al. 1993; Cox et al. 2006, 2007; Welty et al. 2006; Welty \& Crowther, in prep.), over which the ATCA+Parkes 21~cm emission was integrated for comparison with column densities derived from Lyman-$\alpha$ absorption (see middle panel of Fig.~\ref{fig:h1pred}).
The velocity ranges were extended by 5 km~s$^{-1}$ at each end to account for the broader \ion{H}{1} profiles.



\clearpage

\begin{figure}
\epsscale{0.85}
\plotone{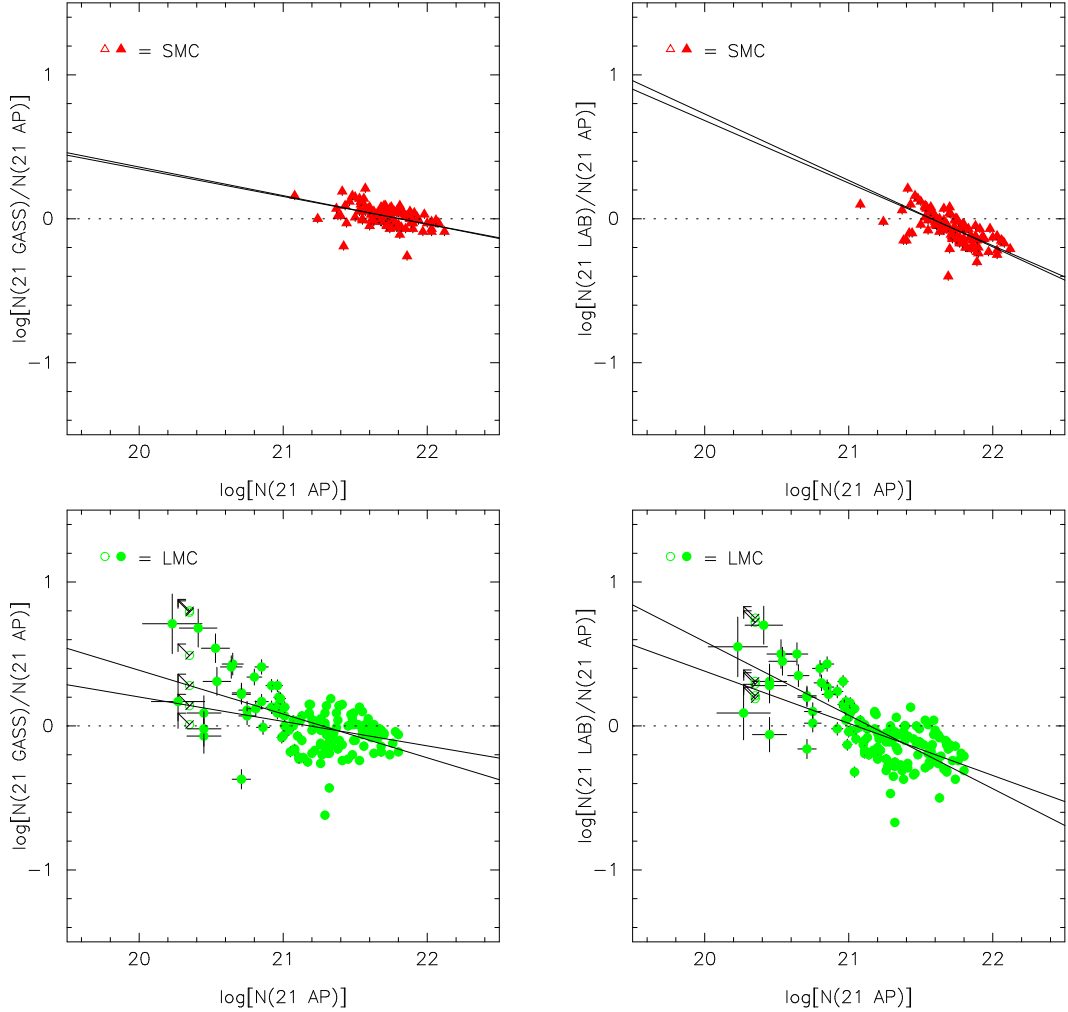}
\caption{Ratios of $N$(\ion{H}{1}) determined from the GASS ({\it left}) and LAB ({\it right}) 21~cm surveys, with respect to the values determined from the higher resolution ATCA+Parkes (AP) spectra.
The solid lines show weighted and unweighted fits to the data.
For the LMC ({\it lower panels}), the fits are restricted to sight lines with $N$(AP) $\ge$ 10$^{21}$ cm$^{-2}$.
The slopes are of order $-$0.2 for the GASS/AP vs. AP comparisons, and of order $-$0.4 for the LAB/AP vs. AP comparisons.}
\label{fig:21comp}
\end{figure} 

\begin{figure}
\epsscale{0.85}
\plotone{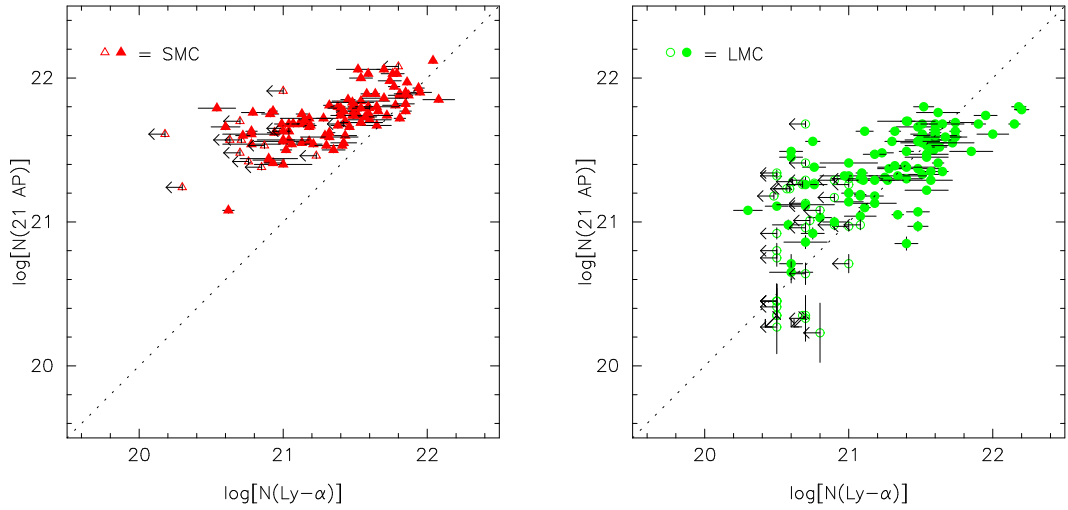}
\caption{$N$(\ion{H}{1}) determined from 21~cm emission vs. $N$(\ion{H}{1}) determined from Lyman-$\alpha$ absorption, for SMC ({\it left}) and LMC ({\it right}) sight lines.
Almost all of the SMC sight lines have $N$(Ly $\alpha$) $\la$ $N$(21~cm); in many cases, the bulk of the emission likely originates beyond the target stars.
A number of the LMC sight lines, however, have $N$(Ly $\alpha$) $\ga$ $N$(21~cm) -- suggestive of either small-scale structure within the 21~cm beam or saturation in the 21~cm emission.}
\label{fig:lavs21}
\end{figure}

\begin{figure} 
\epsscale{0.6}
\plotone{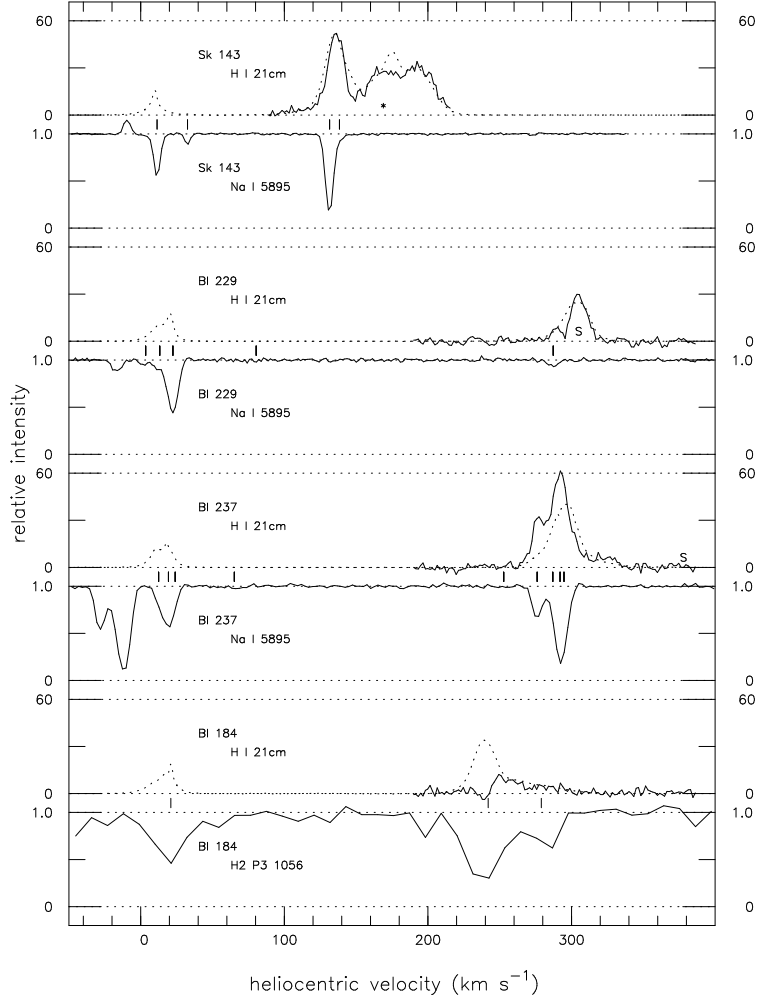}
\caption{For each pair of spectra:  \ion{H}{1} 21~cm emission [{\it upper}; $T_{\rm b}$ (K)] and absorption from \ion{Na}{1} or H$_2$ [{\it lower}] toward Sk~143 (SMC), BI~229 (LMC), BI~237 (LMC), and BI~184 (LMC).
For the 21~cm emission, the solid line gives the combined ATCA+Parkes data (FWHM $\sim$ 1.0-1.5 arcmin; Stanimirovi\'{c} et al. 1999; Kim et al. 2003); the dotted line gives GASS data (FWHM = 14.4 arcmin; Kalberla et al. 2010).
Emission or absorption at $v$ $\la$ 60 km~s$^{-1}$ (SMC) or $v$ $\la$ 100 km~s$^{-1}$ (LMC) is likely due to gas in the Galactic disk or halo; emission or absorption at higher velocities is likely due to gas in the SMC or LMC.
The \ion{Na}{1} and H$_2$ (Welty et al. 2006; Welty \& Crowther, in prep.; this paper; FWHM $\sim$ 4--8 and 20 km~s$^{-1}$, respectively) should be associated with the main atomic components foreground to these targets. 
Relatively strong Magellanic Clouds H$_2$ absorption, with $N$(H$_2$) $\ga$ 4.5 $\times$ 10$^{19}$ cm$^{-2}$, is seen toward Sk~143, BI~237, and BI~184.
Toward Sk~143 and BI~229 [with $N$(21~cm) $>$ $N$(Ly-$\alpha$)], the 21~cm emission at velocities higher than those seen in \ion{Na}{1} absorption is likely from gas beyond those stars.
Toward BI~237 [with $N$(21~cm) $<$ $N$(Ly-$\alpha$)], most of the 21~cm emission is probably in the foreground; either the broader 21~cm beams sample lower average column densities or the emission is somewhat saturated.
Toward BI~184 [with $N$(21~cm) $<$ $N$(Ly-$\alpha$)], the ATCA+Parkes 21~cm profile appears to exhibit self-absorption near 240 km~s$^{-1}$, the velocity of the strongest H$_2$ absorption.}
\label{fig:emabs}
\end{figure} 

\begin{figure}
\epsscale{0.9}
\plottwo{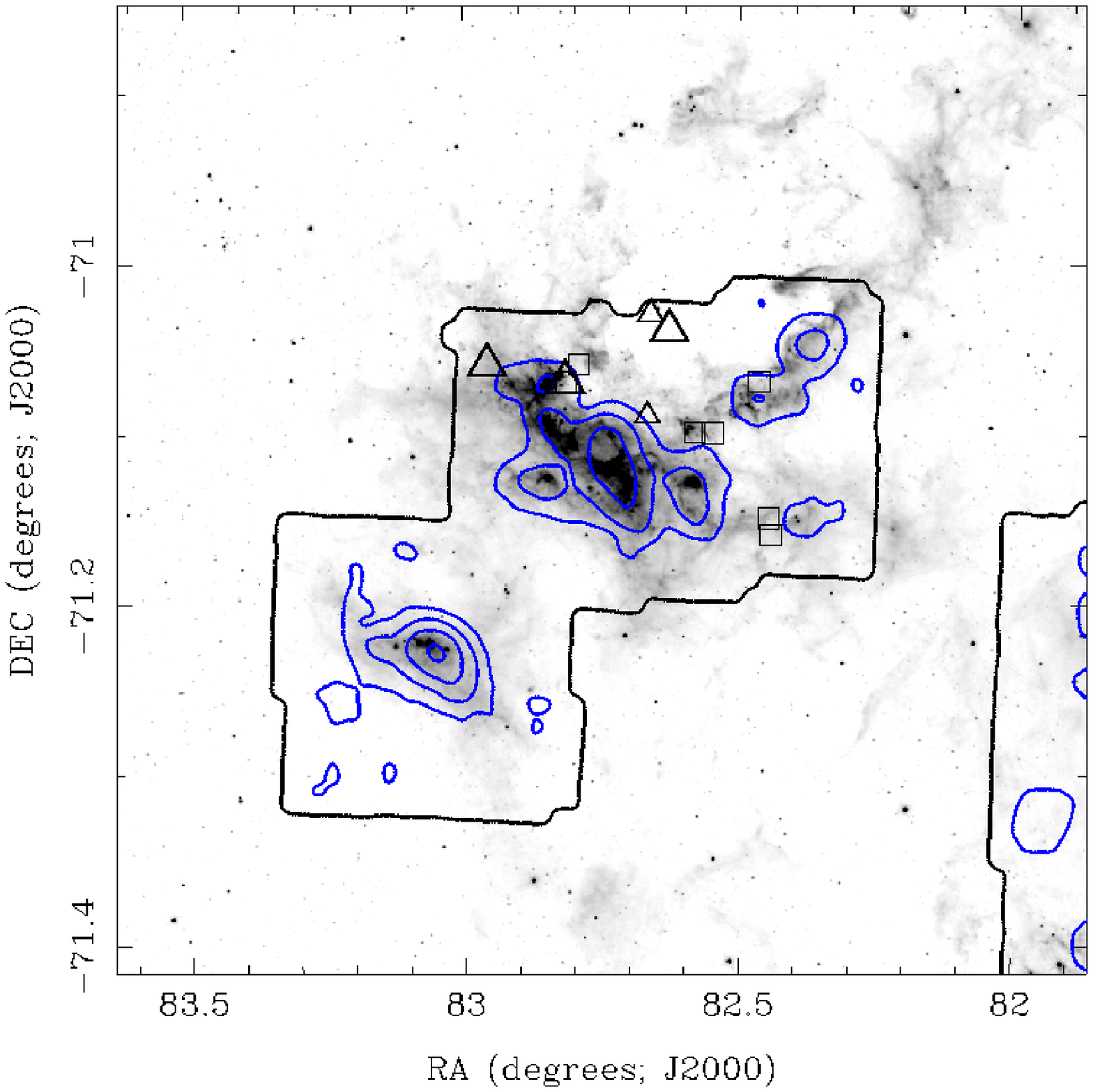}{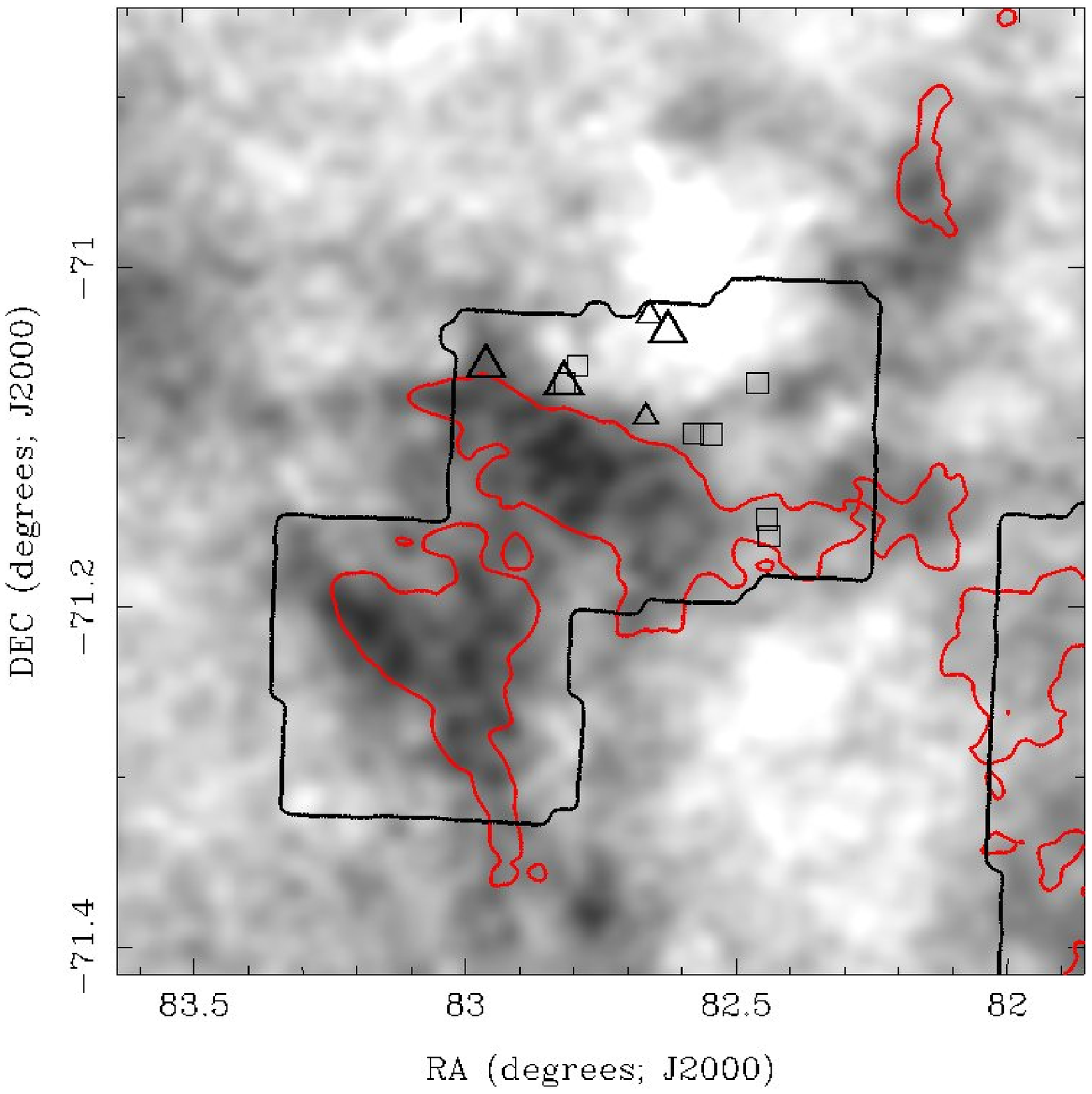}
\caption{CO (1-0), 8$\mu$m, and \ion{H}{1} 21~cm emission in LMC He 206 region.
The left panel compares the CO (1-0) emission (blue contours; at 1, 4, 9, and 16 K km~s$^{-1}$) with the 8$\mu$m emission (grayscale).
The right panel shows the total LMC \ion{H}{1} 21~cm emission from the ATCA+Parkes survey (grayscale; red contour is for $T_{\rm pk}$ = 50 K).
In both panels, the triangles and squares denote sight lines with {\it FUSE} H$_2$ spectra and {\it HST} Ly-$\alpha$ spectra, respectively; the rectangular ``contours'' show the regions surveyed in CO.
The sight lines to Sk$-$71~38 and BI~184 (northern-most adjacent triangles, at the edge of a seeming ``hole'' in the \ion{H}{1}) both may exhibit indications of \ion{H}{1} self-absorption in the ATCA+Parkes 21~cm profiles (Fig.~\ref{fig:emabs}).
The sight lines to LMC1-233 and LMC1-246 (two southwestern-most squares) both have $N$(Ly-$\alpha$) $>$ 1.5 $\times$ $N$(21~cm).}
\label{fig:n206}
\end{figure}

\clearpage

\begin{figure} 
\epsscale{0.3}
\plotone{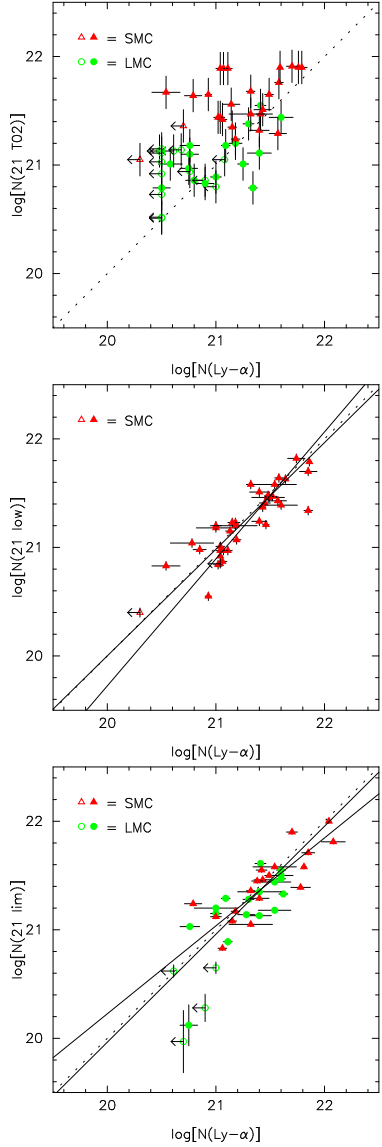}
\caption{$N$(\ion{H}{1}) predicted from 21~cm emission (with or without constraints from \ion{Na}{1} and/or H$_2$ absorption) vs. $N$(\ion{H}{1}) determined from Lyman-$\alpha$ absorption.
Red triangles denote SMC sight lines; green circles denote LMC sight lines; open symbols denote limits.
The top figure shows the values predicted by taking 75\% of the total column density determined from 21~cm emission (for the sample of Tumlinson et al. 2002).
The middle figure shows the values derived from only the lower-velocity SMC 21~cm component(s), for sight lines where the main \ion{Na}{1} and/or H$_2$ absorption falls within that velocity range.
The bottom figure shows the values obtained by restricting the velocity interval in the 21~cm emission to (slightly broader than) that seen in \ion{Na}{1} absorption, for sight lines with data for both Lyman-$\alpha$ and \ion{Na}{1} absorption (see Table~\ref{tab:21vel}).
The solid lines (with slopes $\sim$ 0.8--1.2) in the middle and bottom panels are weighted and unweighted fits to the data.}
\label{fig:h1pred}
\end{figure}
 
\begin{figure}
\epsscale{0.9}
\plotone{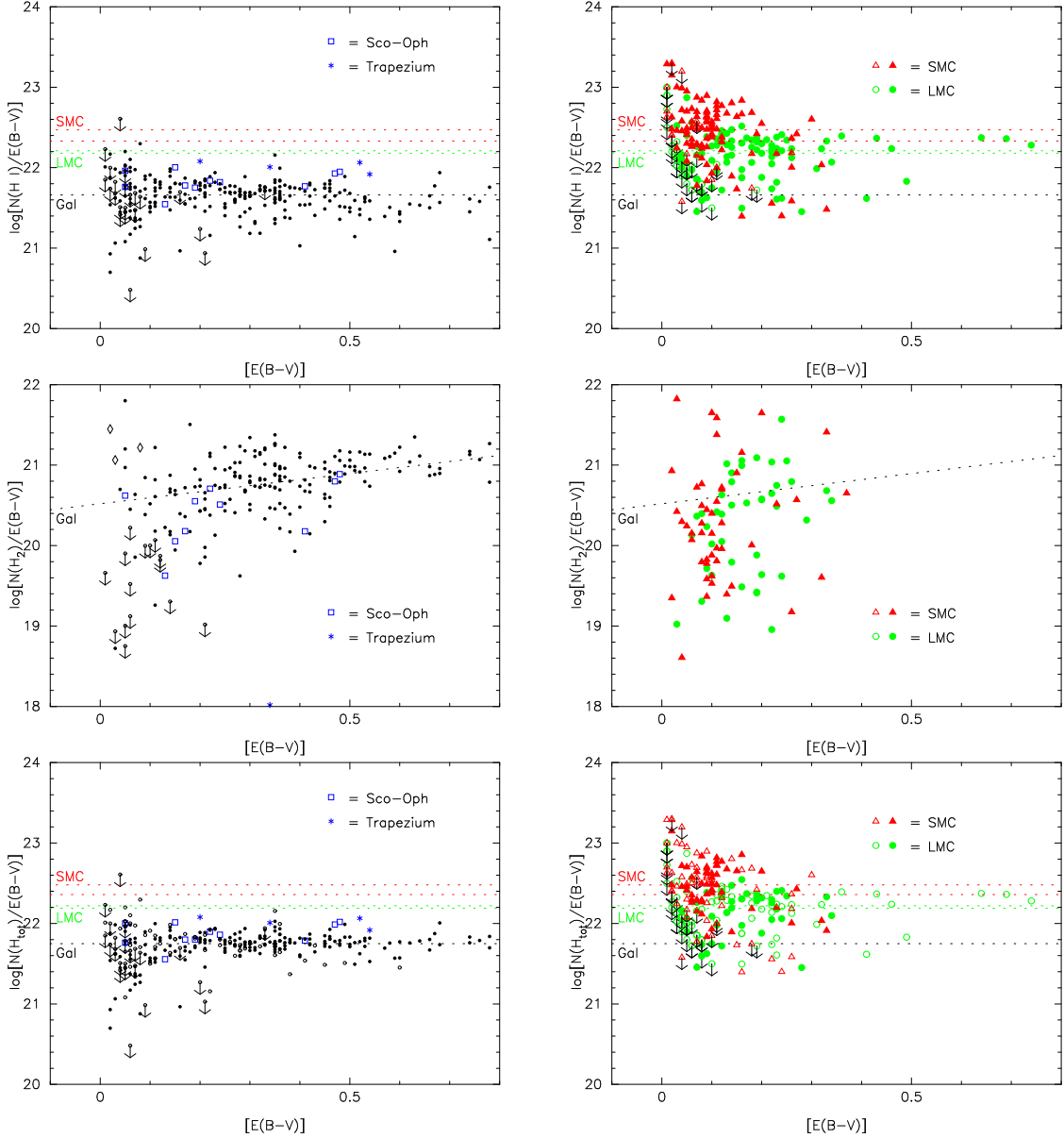}
\caption{Gas-to-dust ratios $N$(\ion{H}{1})/$E(B-V)$, $N$(H$_2$)/$E(B-V)$, and $N$(H$_{\rm tot}$)/$E(B-V)$ vs. $E(B-V)$, for sight lines in the Milky Way (left column) and Magellanic Clouds (right column).
Red triangles denote SMC sight lines; green circles denote LMC sight lines; open symbols denote limits; smaller black circles denote Galactic sight lines.
For H$_{\rm tot}$, the open symbols without arrows denote LMC or SMC sight lines with no H$_2$ data; the H$_2$ estimated from $E(B-V)$ (from the Galactic trend) would increase $N$(H$_{\rm tot}$)/$E(B-V)$ by less than 0.1 dex in nearly all cases.
Uncertainties for the individual points are not given in the figure, in order to more clearly show the relative distributions of the ratios for the three galaxies.
Typical uncertainties for $E(B-V)$ are 0.03-0.04 mag, for $N$(\ion{H}{1}) and $N$(H$_{\rm tot}$) are 0.04-0.08 dex, and for $N$(H$_2$) are 0.1-0.2 dex.
The horizontal dotted lines for \ion{H}{1} and H$_{\rm tot}$ indicate the mean values for each galaxy [log(mean ratio) and mean of log(ratio)], for sight lines with $E(B-V)$ $\ge$ 0.05 mag.
For both \ion{H}{1} and H$_{\rm tot}$, the mean LMC and SMC ratios are higher than the Galactic values, with larger scatter.
For H$_2$, the ratios are more similar for the three galaxies [especially for $E(B-V)$ $\la$ 0.4], again with more scatter for the LMC and SMC; the dotted line shows the mean trend for the Galactic sight lines.}
\label{fig:rvsebv}
\end{figure} 
 
\begin{figure}
\epsscale{0.5}
\plotone{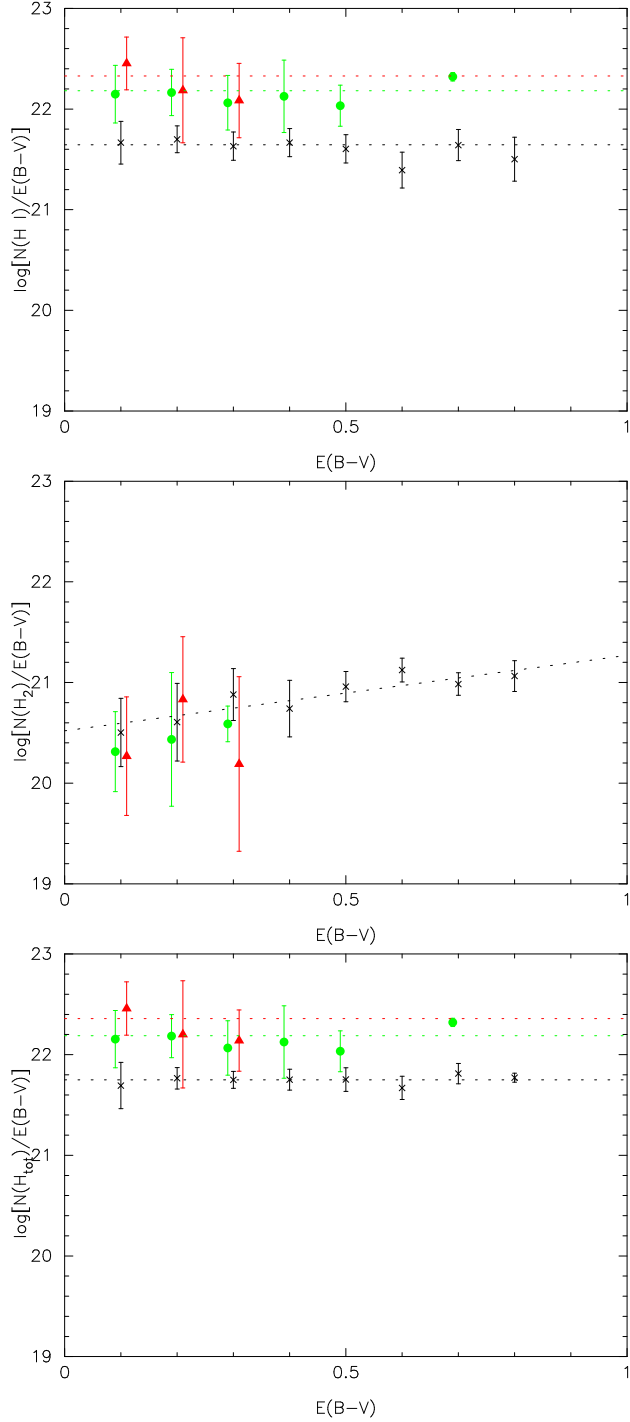}
\caption{Gas-to-dust ratios $N$(\ion{H}{1})/$E(B-V)$, $N$(H$_2$)/$E(B-V)$, and $N$(H$_{\rm tot}$)/$E(B-V)$, averaged over 0.1-mag bins, as functions of $E(B-V)$.
Red triangles denote SMC values, green circles denote LMC values, black crosses denote Galactic values.
Overall average values for \ion{H}{1} and H$_{\rm tot}$ are shown by dotted horizontal lines.
For the Galactic sight lines, the average ratio for H$_{\rm tot}$ is constant; the ratio for \ion{H}{1} is constant for $E(B-V)$ $\la$ 0.5 mag, but then declines slightly at higher reddening; and the ratio for H$_2$ increases with $E(B-V)$.
For H$_{\rm tot}$, adding estimated H$_2$ for the sight lines without H$_2$ data (Galactic, LMC, or SMC) increases the binned average ratios by at most 0.04 dex.}
\label{fig:avgrat}
\end{figure}

\begin{figure}
\epsscale{0.9}
\plotone{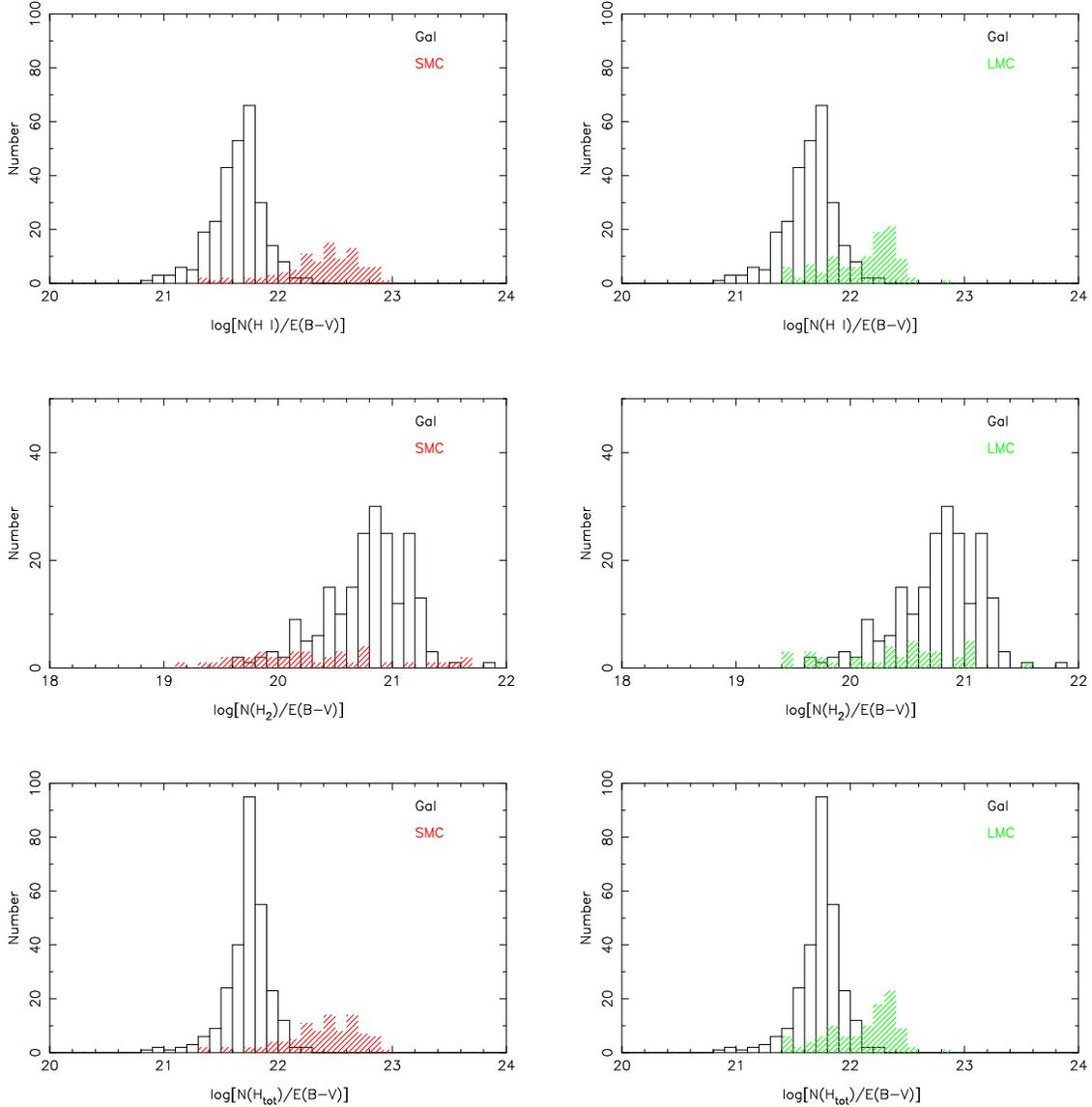}
\caption{Distribution of the gas-to-dust ratios $N$(\ion{H}{1})/$E(B-V)$, $N$(H$_2$)/$E(B-V)$, and $N$(H$_{\rm tot}$)/$E(B-V)$ for sight lines in our Galaxy (black open), the SMC (left; red hatched), and the LMC (right; green hatched).
Sight lines with $E(B-V)$ $<$ 0.05 or log[$N$(H$_2$)] $<$ 18.5 (for the middle panels) have been excluded from the samples.
For $N$(\ion{H}{1}) and $N$(H$_{\rm tot}$), the mean LMC and SMC ratios are higher than the Galactic values, with larger scatter.
For H$_{\rm tot}$, adding estimated H$_2$ for the sight lines without H$_2$ data would shift the distributions by less than 0.1 dex to the right.
For $N$(H$_2$), the ratios are more similar for the three galaxies, again with more scatter for the LMC and SMC.
For the Galactic sight lines, the distribution for $N$(H$_{\rm tot}$) is narrower than that for $N$(\ion{H}{1}).}
\label{fig:hist}
\end{figure}

\clearpage

\begin{figure}
\epsscale{0.85}
\plotone{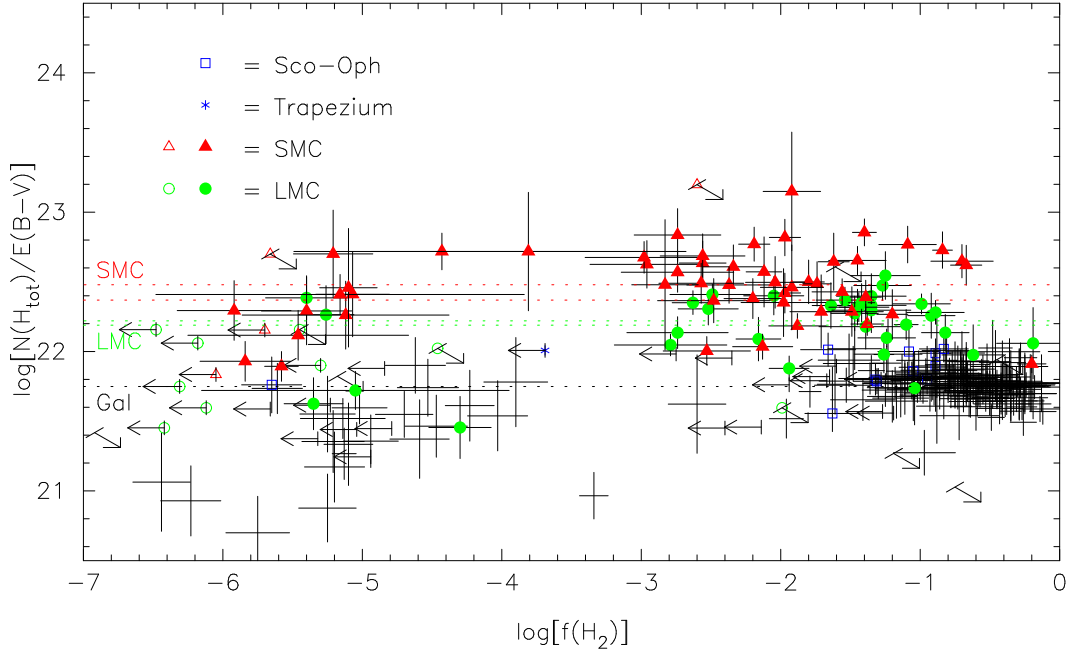}
\caption{Gas-to-dust ratio $N$(H$_{\rm tot}$)/$E(B-V)$ vs. molecular fraction $f$(H$_2$).
Red triangles denote SMC sight lines; green circles denote LMC sight lines; open symbols denote limits; plain crosses denote Galactic sight lines.
Note the possible slight decline of the Galactic $N$(H$_{\rm tot}$)/$E(B-V)$ for log[$f$(H$_2$)] $\ga$ $-$1, perhaps indicative of a lower gas-to-dust ratio in molecular gas.}
\label{fig:rvsf}
\end{figure} 

\begin{figure}
\plotone{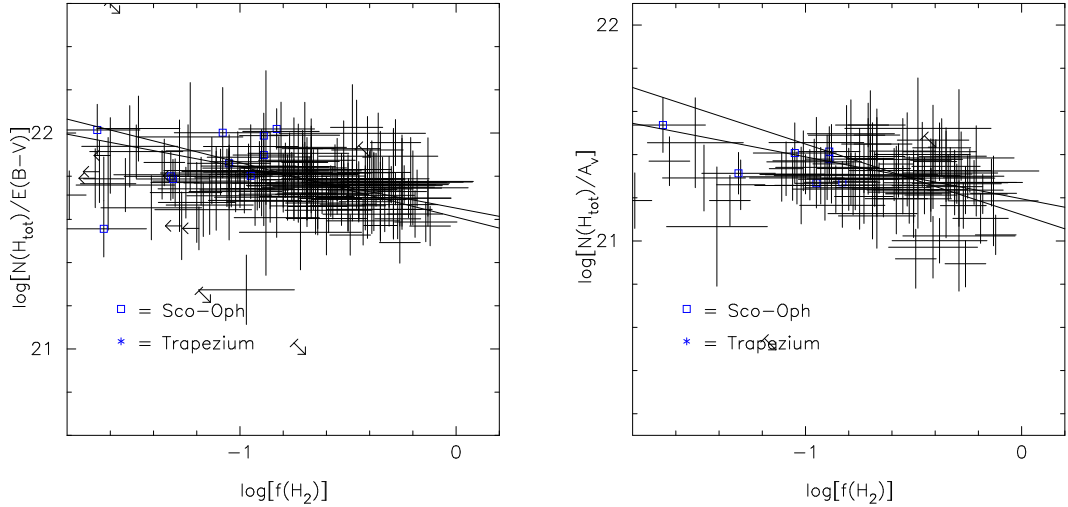}
\caption{Closer view of gas-to-dust ratios $N$(H$_{\rm tot}$)/$E(B-V)$ ({\it left}) and $N$(H$_{\rm tot}$)/$A_{\rm v}$ ({\it right}) vs. molecular fraction $f$(H$_2$) for Galactic sight lines.
The solid lines show weighted and unweighted fits to the data, restricted to $f$(H$_2$) $>$ 0.1.
The slight declines in both ratios over that range in $f$(H$_2$), with slopes $\sim$ $-$0.2, would be consistent with a lower GDR in fully molecular gas (by a factor of 1.5--2.0).}
\label{fig:rvsf2}
\end{figure}

\begin{figure}
\epsscale{0.5}
\plotone{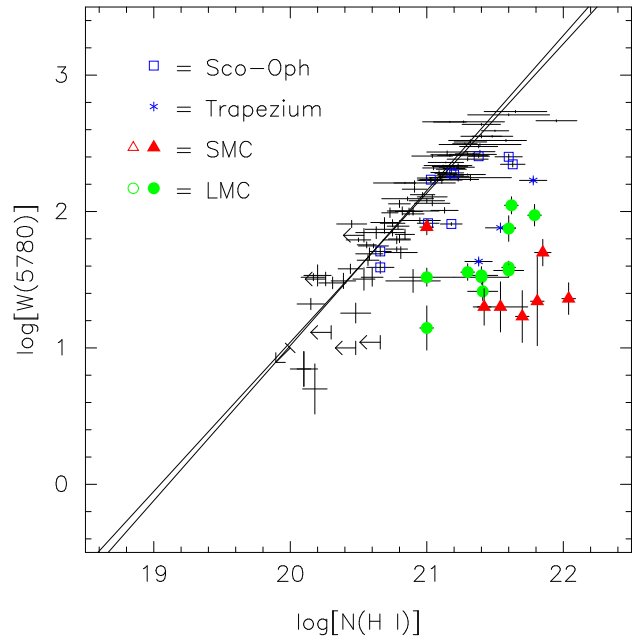}
\caption{Equivalent width of the diffuse interstellar band at 5780.5 \AA\ vs. $N$(\ion{H}{1}).
Red triangles denote SMC sight lines; green circles denote LMC sight lines; open symbols denote limits; plain crosses denote Galactic sight lines.
The solid lines show the approximately linear relationship found for Galactic sight lines, from weighted and unweighted fits to the data; a similar (but offset) relationship may apply for the LMC sight lines.}
\label{fig:5780}
\end{figure} 

\begin{figure}
\epsscale{0.85}
\plotone{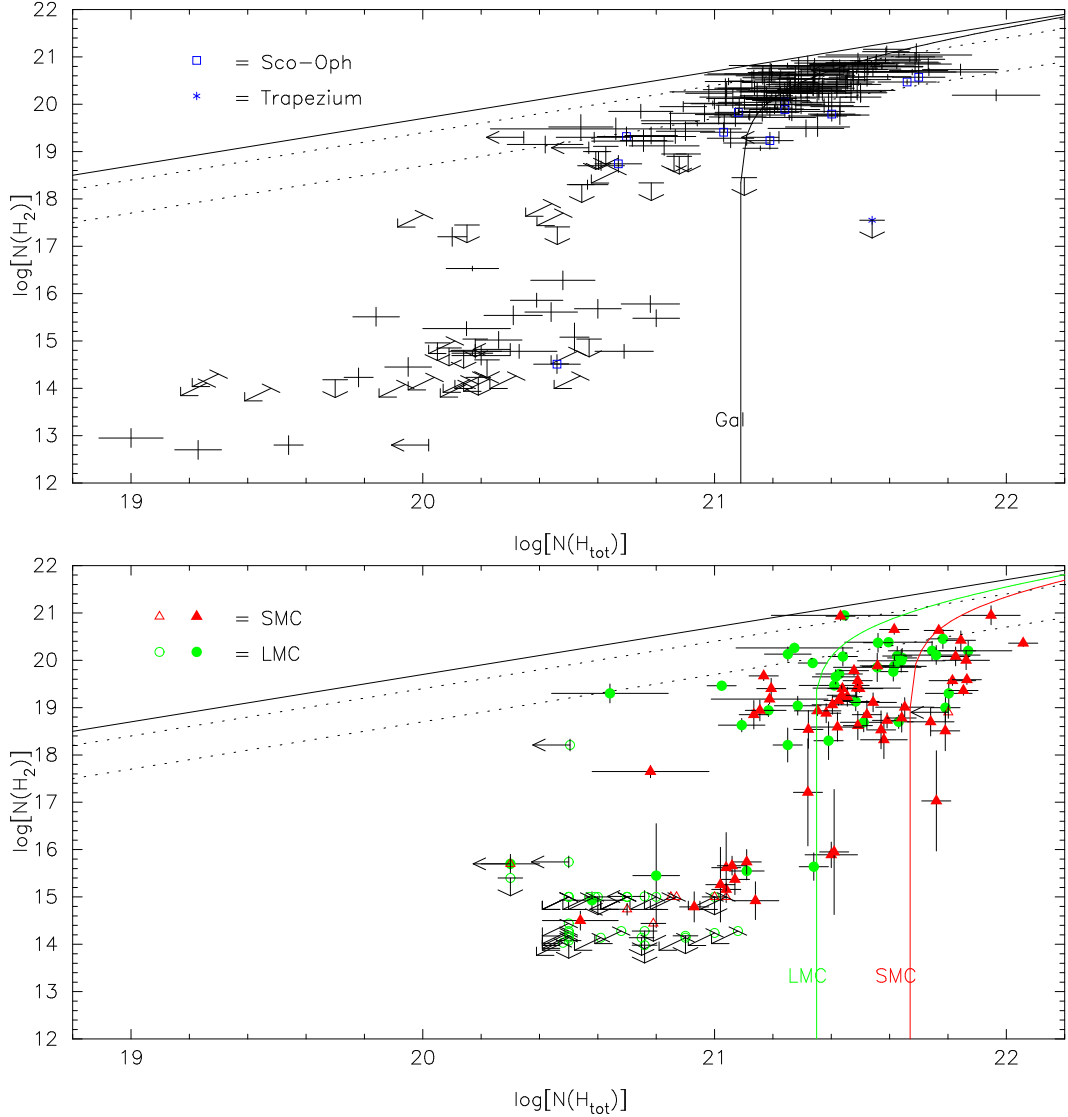}
\caption{Observed $N$(H$_2$) vs. $N$(H$_{\rm tot}$), for Galactic sight lines ({\it top}) and Magellanic Clouds sight lines ({\it bottom}), compared with model predictions from McKee \& Krumholz (2010).
Crosses denote Galactic sight lines; green circles denote LMC sight lines; red triangles denote SMC sight lines.
Black, green, and red curves give predictions for metallicities of 1.0 (Galactic), 0.5 (LMC), and 0.2 (SMC), respectively. 
The solid diagonal line in each panel represents $f$(H$_2$) = 1; the dotted diagonal lines show $f$(H$_2$) = 0.5 and 0.1.
While there is qualitative agreement for the metallicity dependence, many of the observed points (especially for the SMC) lie above and/or to the left of the predicted curves; the atomic-to-molecular transition appears to occur at somewhat lower column densities than predicted by the models.}
\label{fig:mckee}
\end{figure}

\begin{figure}
\epsscale{0.9}
\plotone{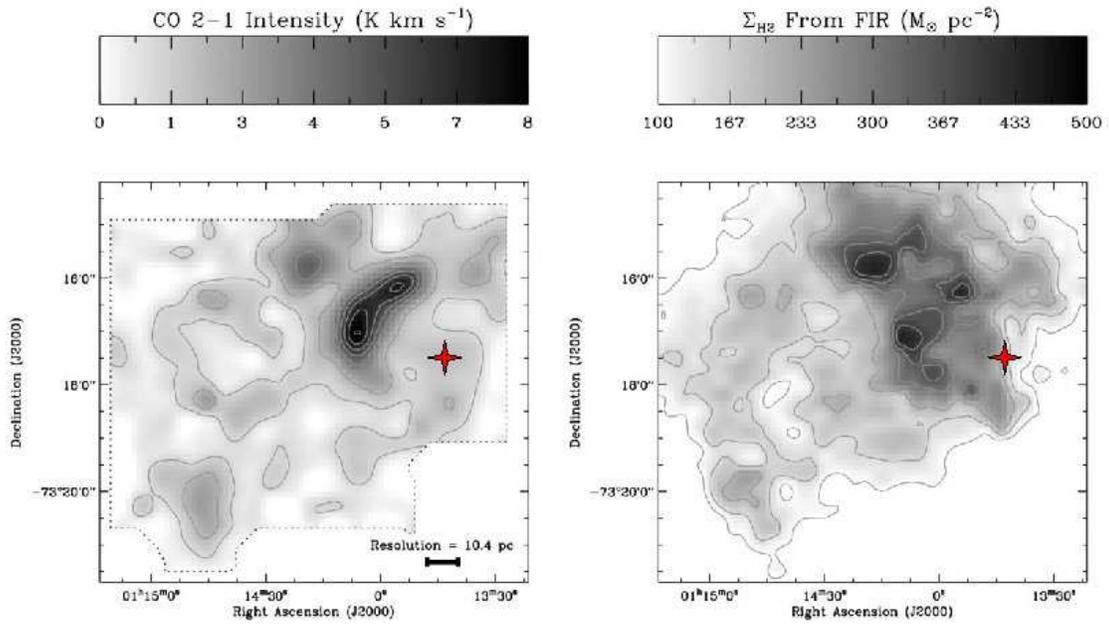}
\caption{CO 2-1 emission ({\it left}) and H$_2$ surface density inferred from far-IR emission ({\it right}) for N83 region in the SMC (adapted from Fig.~5 in Leroy et al. 2009).
Absorption from CH and H$_2$ (at the same velocity as the nearby CO emission) is seen toward the SMC star AV~476 (marked by red star).
The measured $N$(H$_2$) = 9 $\times$ 10$^{20}$ cm$^{-2}$ is nearly an order of magnitude lower than the predicted value at that location.
(Reproduced by permission of the AAS.)}
\label{fig:leroy}
\end{figure}

\clearpage


\clearpage

\voffset 0.5in 

\begin{deluxetable}{lrcllrrclcrcrrcrr}
\rotate
\tabletypesize{\scriptsize}
\tablecolumns{17} 
\tablecaption{SMC Sight lines \label{tab:smclos}}
\tablewidth{0pt} 
 
\tablehead{
\multicolumn{1}{l}{Star}&
\multicolumn{1}{l}{AzV}&
\multicolumn{1}{c}{Reg}&
\multicolumn{2}{c}{RA (J2000) DEC}&
\multicolumn{1}{c}{$V$\tablenotemark{a}}&
\multicolumn{1}{c}{$(B-V)$\tablenotemark{a}}&
\multicolumn{1}{c}{$E(B-V)$\tablenotemark{b}}&
\multicolumn{1}{l}{Type}&
\multicolumn{1}{c}{Ref\tablenotemark{c}}&
\multicolumn{1}{c}{$N$(H I)}&
\multicolumn{1}{c}{Ref\tablenotemark{d}}&
\multicolumn{1}{c}{$N$(H$_2$)}&
\multicolumn{1}{c}{$T_{01}$}&
\multicolumn{1}{c}{Ref\tablenotemark{e}}&
\multicolumn{1}{c}{$N$(21~cm)\tablenotemark{f}}&
\multicolumn{1}{c}{$T_{\rm pk}$\tablenotemark{f}}\\
\multicolumn{1}{c}{ }& 
\multicolumn{1}{c}{ }& 
\multicolumn{1}{c}{ }&
\multicolumn{1}{c}{($^{h~m~s}$)}& 
\multicolumn{1}{c}{(\arcdeg~'~") }& 
\multicolumn{1}{c}{ }& 
\multicolumn{1}{c}{ }&
\multicolumn{1}{c}{tot/SMC}& 
\multicolumn{1}{c}{ }& 
\multicolumn{1}{c}{ }&
\multicolumn{1}{c}{(cm$^{-2}$)}&
\multicolumn{1}{c}{ }&
\multicolumn{1}{c}{(cm$^{-2}$)}&
\multicolumn{1}{c}{(K)}&
\multicolumn{1}{c}{ }&
\multicolumn{1}{c}{(cm$^{-2}$)}&
\multicolumn{1}{c}{(K)}} 
\startdata
Sk  5&  4   & &00 45 03.3&$-$72 41 58& 13.83&   0.10 &      0.26/0.22 &B2 Ib        & 1 &   20.90&1&\nodata&   &   & 21.43&  47\\
     &  6   & &00 45 18.2&$-$73 15 24& 13.31&   0.05 &      0.36/0.32 &O9 III       & 2 &   21.54&1& 19.11 & 77& 1 & 22.00& 134\\
Sk  7& 10   & &00 45 47.0&$-$73 39 55& 12.58&$-$0.02 &      0.13/0.09 &B2.5 Ia      & 2 &   21.54&2&\nodata&   &   & 21.69&  51\\
Sk  9& 14   & &00 46 32.6&$-$73 06 06& 13.55&$-$0.17 &      0.15/0.11 &O5 V         & 3 &   21.76&1& 17.03 & 53& 2 & 22.03& 126\\ 
Sk 10& 15   & &00 46 42.2&$-$73 24 56& 13.12&$-$0.19 &      0.13 0.09 &O6.5 II(f)   & 4 &   21.58&1& 18.32 & 81& 1 & 21.89&  95\\ 
Sk 11& 16   & &00 46 55.0&$-$73 08 34& 13.15&   0.10 &      0.41/0.37 &O9 II        & 1 & \nodata& & 20.22 & 53& 1 & 22.07& 125\\
Sk 13& 18   & &00 47 12.2&$-$73 06 33& 12.44&   0.03 &      0.20/0.16 &B2 Ia        & 2 &   22.04&1& 20.36 & 66& 3 & 22.12& 142\\
Sk 15& 22   & &00 47 38.7&$-$73 07 49& 12.20&$-$0.01 &      0.08/0.04 &B5 Ia        & 2 &$<$21.80&1& 18.9: &   & 1 & 22.08& 136\\
\enddata
\tablecomments{Region = w (in column 3) denotes stars in ``wing'' region.}
\tablenotetext{a}{Photometry is from Massey (2002) or Parker et al. (1992) (preferred) or from Dachs (1970); Azzopardi et al. (1975); Ardeberg \& Maurice (1977); Isserstedt (1978); Ardeberg (1980); Bouchet et al. (1985); Gordon et al. (2003); Lennon et al. (2003); and/or Evans et al. (2006).}
\tablenotetext{b}{Using intrinsic colors of FitzGerald 1970 and assuming a Galactic contribution of 0.03--0.04 mag.}
\tablenotetext{c}{Spectral types are from 1 = Smith Neubig \& Bruhweiler 1997; 2 = Lennon 1997; 3 = Massey et al. 2004, 2005, 2009; 4 =  Walborn et al. 1995, 2000, 2002; 5 = Massey et al. 1995; 6 = Garmany et al. 1987; 7 = Foellmi et al. 2003a; 8 = P. Crowther, priv. comm.; 9 = Ardeberg \& Maurice 1977; 10 = Lennon et al. 2003; 11 = Evans et al. 2004c, 2004d, 2006; 12 = Bouchet et al. 1985; 13 = Massey 2002.}
\tablenotetext{d}{$N$(H I):  1 = this study; 2 = Fitzpatrick 1985a; 3 = Gordon et al. 2003 (adjusted for comparison star values); 4 = Bouchet et al. 1985; 5 = Evans et al. 2004a.}
\tablenotetext{e}{$N$(H$_2$):  1 = this study; 2 = Tumlinson et al. 2002; 3 = Cartledge et al. 2005; 4 = Mallouris et al. 2001.  Values given as mod, wk, or ul are from unmeasured {\it FUSE} data.}
\tablenotetext{f}{$N$(21~cm) and $T_{\rm pk}$ are from the combined ATCA+Parkes spectra (Stanimirovi\'{c} et al. 1999).}
\end{deluxetable}

\clearpage

\begin{deluxetable}{lrcllrrclcrcrrcrr}
\rotate
\tabletypesize{\scriptsize}
\tablecolumns{17}  
\tablecaption{LMC Sight lines \label{tab:lmclos}} 
\tablewidth{0pt}  
  
\tablehead{ 
\multicolumn{1}{l}{Star}& 
\multicolumn{1}{l}{HD}& 
\multicolumn{1}{c}{Reg}&
\multicolumn{2}{c}{RA (J2000) DEC}& 
\multicolumn{1}{c}{$V$\tablenotemark{a}}& 
\multicolumn{1}{c}{$(B-V)$\tablenotemark{a}}&
\multicolumn{1}{c}{$E(B-V)$\tablenotemark{b}}& 
\multicolumn{1}{l}{Type}&
\multicolumn{1}{c}{Ref\tablenotemark{c}}&
\multicolumn{1}{c}{$N$(H I)}&
\multicolumn{1}{c}{Ref\tablenotemark{d}}&
\multicolumn{1}{c}{$N$(H$_2$)}&
\multicolumn{1}{c}{$T_{01}$}&
\multicolumn{1}{c}{Ref\tablenotemark{e}}&
\multicolumn{1}{c}{$N$(21~cm)\tablenotemark{f}}&
\multicolumn{1}{c}{$T_{\rm pk}$\tablenotemark{f}}\\ 
\multicolumn{1}{c}{ }&  
\multicolumn{1}{c}{ }&  
\multicolumn{1}{c}{ }&
\multicolumn{1}{c}{($^{h~m~s}$)}&  
\multicolumn{1}{c}{(\arcdeg~'~") }&  
\multicolumn{1}{c}{ }&  
\multicolumn{1}{c}{ }&
\multicolumn{1}{c}{tot/LMC}&  
\multicolumn{1}{c}{ }&  
\multicolumn{1}{c}{ }&
\multicolumn{1}{c}{(cm$^{-2}$)}&
\multicolumn{1}{c}{ }&
\multicolumn{1}{c}{(cm$^{-2}$)}&
\multicolumn{1}{c}{(K)}&
\multicolumn{1}{c}{ }&
\multicolumn{1}{c}{(cm$^{-2}$)}&
\multicolumn{1}{c}{(K)}}
\startdata
RXJ0439.8-6809 &   & &04 39 49.7&$-$68 09 02& 21.60&        &   \nodata   &\nodata      &   &$<$20.70&1& \nodata &   &   & 20.33&   9\\ 
Sk$-$67 2   &270754& &04 47 04.5&$-$67 06 53& 11.26&   0.07 &   0.26/0.24 &B1 Ia+       & 1 &   21.00&2&   20.95 & 46& 1 & 21.32&  46\\
Sk$-$67 5   &268605& &04 50 18.9&$-$67 39 38& 11.34&$-$0.12 &   0.14/0.11 &O9.7 Ib      & 2 &   21.00&1&   19.46 & 57& 2 & 21.20&  47\\
Sk$-$66 1   &268623& &04 52 19.1&$-$66 43 53& 11.62&$-$0.06 &   0.11/0.09 &B2 Ia        & 1 & \nodata& &   19.19 &102& 3 & 21.11&  44\\
Sk$-$69 19  &      & &04 52 25.0&$-$69 30 21& 12.84&$-$0.07 &   0.15/0.05 &B0.5         & 3 &   21.57&2& \nodata &   &   & 21.29&  68\\
BI 9        &      & &04 52 47.9&$-$68 47 39& 13.70&$-$0.22 &   0.09/0.01 &O7.5 III((f))& 4 &$<$20.73&1& \nodata &   &   & 21.01&  20\\
Sk$-$67 14  &268685& &04 54 31.9&$-$67 15 25& 11.53&$-$0.10 &   0.08/0.04 &B1.5 Ia      & 1 &$<$20.76&1&   15.01 &270& 2 & 21.27&  51\\
Sk$-$67 20  & 32109& &04 55 31.4&$-$67 30 03& 13.79&$-$0.26 &   0.13/0.09 &WN4b         & 5 & \nodata& &   18.67 & 72& 2 & 20.54&  11\\
\enddata
\end{deluxetable}

\voffset 0.5in

\begin{deluxetable}{lrrrrrrrrrrrrrrr}
\rotate
\tabletypesize{\scriptsize}
\tablecolumns{15}
\tablecaption{Ratios (Galactic, LMC, SMC) \label{tab:ratgal}}
\tablewidth{0pt}

\tablehead{
\multicolumn{1}{l}{Ratio}&
\multicolumn{1}{c}{Type}&
\multicolumn{2}{c}{Galactic}&
\multicolumn{2}{c}{LMC (all)}&
\multicolumn{2}{c}{LMC main}&
\multicolumn{2}{c}{LMC2}&
\multicolumn{2}{c}{SMC (all)}&
\multicolumn{2}{c}{SMC bar}&
\multicolumn{2}{c}{SMC wing}\\
\multicolumn{2}{c}{ }&
\multicolumn{1}{c}{Ratio}&
\multicolumn{1}{c}{N}&
\multicolumn{1}{c}{Ratio}&
\multicolumn{1}{c}{N}&
\multicolumn{1}{c}{Ratio}&
\multicolumn{1}{c}{N}&
\multicolumn{1}{c}{Ratio}&
\multicolumn{1}{c}{N}&
\multicolumn{1}{c}{Ratio}&
\multicolumn{1}{c}{N}&
\multicolumn{1}{c}{Ratio}&
\multicolumn{1}{c}{N}&
\multicolumn{1}{c}{Ratio}&
\multicolumn{1}{c}{N}}
\startdata
log[$N$(H I)/$E(B-V)$] & 
wlog &
    21.65$\pm$0.16&237&   22.18$\pm$0.22& 83&   22.10$\pm$0.26& 64&   22.33$\pm$0.07& 21&   
                          22.33$\pm$0.39& 85&   22.39$\pm$0.36& 69&   22.08$\pm$0.47& 17 \\
 &$\Delta$ & & &               0.54&   &             0.46&   &             0.69&   &
                               0.68&   &             0.74&   &             0.43 \\
 & nlin &
    21.67$\pm$0.19&242&   22.21$\pm$0.22& 86&      \nodata    &   &      \nodata    &   &
                          22.47$\pm$0.32& 86&      \nodata    &   &      \nodata    &    \\
 &$\Delta$ & & &               0.54&   &     \nodata     &   &     \nodata     &   &
                               0.80&   &     \nodata     &   &     \nodata      \\
 & \\
log[$N$(H$_2$)/$E(B-V)$] &                 
wlog &
    20.92$\pm$0.27&177&   20.51$\pm$0.57& 38&   20.52$\pm$0.62& 27&   20.62$\pm$0.24& 10&
                          20.64$\pm$0.60& 40&   20.55$\pm$0.53& 28&   20.85$\pm$0.70& 12 \\
 &$\Delta$ & & &            $-$0.41&   &          $-$0.40&   &          $-$0.30&   &
                            $-$0.28&   &          $-$0.37&   &          $-$0.07& \\
 & ebv35&
    20.80$\pm$0.36&101&   20.51$\pm$0.57& 38&   \nodata       &   &   \nodata       &   &
                          20.64$\pm$0.63& 39&   \nodata       &   &   \nodata       &    \\
\enddata
\tablecomments{Values are weighted mean[log(ratio)] (wlog) or unweighted log[mean(ratio)] (nlin), with standard deviations; N is the number of sight lines with values for both quantities (after removing $>$2.5$\sigma$ outliers).
$\Delta$ values are differences with respect to average Galactic values.
Sight lines with log[$N$(H$_2$)] $<$ 18.5 or $E(B-V)$ $<$ 0.05 have been omitted from the samples, for ratios including those quantities.
Because $N$(H$_2$)/$E(B-V)$ increases with $E(B-V)$ in (at least) the Galactic ISM, sight lines with $E(B-V)$ $>$ 0.35 have been omitted from the ebv35 samples [to give similar ranges in $E(B-V)$ for our Galaxy, the LMC, and the SMC].
For the H2est samples, $N$(H$_2$) has been estimated from $E(B-V)$ for sight lines with no H$_2$ data.
Sight lines included in the SMC ``wing'' region and the LMC2 region are noted in Tables 1 and 2.}
\end{deluxetable}

\appendix

\begin{deluxetable}{rllrrl}
\tabletypesize{\scriptsize}
\tablecolumns{6}
\tablecaption{{\it HST} Observing Programs \label{tab:prog}}
\tablewidth{0pt}

\tablehead{
\multicolumn{1}{c}{Program}&
\multicolumn{1}{c}{Instrument}&
\multicolumn{1}{c}{Resolution}&
\multicolumn{1}{c}{SMC}&
\multicolumn{1}{c}{LMC}&
\multicolumn{1}{c}{PI}\\
\multicolumn{2}{c}{ }&
\multicolumn{1}{c}{(\AA)}&
\multicolumn{3}{c}{ }}
\startdata
2233 & FOS G130H  & 0.90  &  0 &  4 & Kudritzki \\
2563 & FOS G130H  & 0.90  &  0 &  1 & Kirshner \\
3030 & GHRS G140L & 0.60  &  0 &  1 & Ebbets \\
4110 & FOS G130H  & 0.90  &  9 &  5 & Kudritzki \\
4260 & FOS G130H  & 0.90  &  0 &  6 & Leitherer \\
5346 & FOS G130H  & 0.90  &  0 &  4 & Lennon \\
5444 & FOS G130H  & 0.90  & 10 & 20 & Robert \\
5460 & FOS G130H  & 0.90  &  0 &  3 & Hillier \\
\enddata
\end{deluxetable}

\begin{deluxetable}{llrrr}
\tabletypesize{\scriptsize}
\tablecolumns{5}
\tablecaption{21~cm Velocity Limits \label{tab:21vel}}
\tablewidth{0pt}

\tablehead{
\multicolumn{1}{c}{Star}&
\multicolumn{1}{c}{v1--v2\tablenotemark{a}}&
\multicolumn{1}{c}{$N$(Ly-$\alpha$)}&
\multicolumn{1}{c}{$N$(21 all)\tablenotemark{b}}&
\multicolumn{1}{c}{$N$(21 lim)\tablenotemark{c}}}
\startdata
Sk 13       & 110-157 &    22.04 & 22.12 & 22.00 \\ 
Sk 18       & 114-152 &    21.70 & 22.06 & 21.90 \\ 
AV 47       & 113-134 &    21.32 & 21.81 & 21.36 \\ 
Sk 40       & 123-154*&    21.54 & 21.84 & 21.58 \\ 
AV 80       & 107-137*&    21.81 & 21.89 & 21.58 \\ 
AV 95       & 100-143 &    21.49 & 21.77 & 21.50 \\ 
AV 207      & 157-190 &    21.43 & 21.60 & 21.46 \\ 
AV 214      & 150-169 &    21.78 & 21.81 & 21.39 \\
\enddata
\tablenotetext{a}{Entries marked with an asterisk include contributions from another weak, isolated component.}
\tablenotetext{b}{Column density integrated over full extent of 21~cm profile.}
\tablenotetext{c}{Column density integrated over limited range in velocities seen in \ion{Na}{1} absorption (extended by 5 km~s$^{-1}$ at both ends).}
\end{deluxetable}


\begin{references}

\reference {ar89} Abgrall, H., \& Roueff, E. 1989, \aaps, 79, 313
\reference {and04} Andr\'{e}, M. K., Le Petit, F., Sonnentrucker, P., et al. 2004, \aap, 422, 483
\reference {afv87} Andreani, P., Ferlet, R., \& Vidal-Madjar, A. 1987, \nat, 326, 770
\reference {ard80} Ardeberg, A. 1980, \aaps, 42, 1
\reference {ard72} Ardeberg, A., Brunet, J.-P., Maurice, E., \& Pr\'{e}vot, L. 1972, \aaps, 6, 249
\reference {am77} Ardeberg, A., \& Maurice, E. 1977, \aaps, 30, 261
\reference {avm75} Azzopardi, M., Vigneau, J., \& Macquet, M. 1975, \aaps, 22, 285
\reference {barb04} Barbaro, G., Geminale, A., Mazzei, P., \& Congiu, E. 2004, \mnras, 353, 760
\reference {bart01} Bartzakos, P., Moffat, A. F. L., \& Niemela, V. S. 2001, \mnras, 324, 18
\reference {bern08} Bernard, J.-P. et al. 2008, \aj, 136, 919
\reference {blair09} Blair, W. P., Oliveira, C., Lamassa, S., et al. 2009, \pasp, 121, 634
\reference {br04} Blitz, L., \& Rosolowsky, E. 2004, \apj, 612, L29
\reference {br06} Blitz, L., \& Rosolowsky, E. 2006, \apj, 650, 933
\reference {bdb01} Bluhm, H., \& de Boer, K. S. 2001, \aap, 379, 82
\reference {bsd78} Bohlin, R. C., Savage, B. D., \& Drake, J. F. 1978, \apj, 224, 132
\reference {bol03} Bolatto, A. D., Leroy, A., Israel, F. P., \& Jackson, J. M. 2003, \apj, 595, 167
\reference {bol11} Bolatto, A. D., Leroy, A. K., Jameson, K., et al. 2011, \apj, 741:12
\reference {bot07} Bot, C., Boulanger, F., Rubio, M., \& Rantakyro, F. 2007, \aap, 471, 103
\reference {bou85} Bouchet, P., Lequeux, J., Maurice, E., Pr\'{e}vot, L., \& Pr\'{e}vot-Burnichon, M. L. 1985, \aap, 149, 330
\reference {bat99} Breysacher, J., Azzopardi, M., \& Testor, G. 1999, \aaps, 137, 117
\reference {bru75} Brunet, J. P., Imbert, M., Martin, N., et al. 1975, \aaps, 21, 109
\reference {bh78} Burstein, D., \& Heiles, C. 1978, \apj, 225, 40
\reference {cart05} Cartledge, S. I. B., Clayton, G. C., Gordon, K. D., et al. 2005, \apj, 630, 355
\reference {cn96} Caulet, A., \& Newell, R. 1996, \apj, 465, 205
\reference {cgm86} Conti, P. S., Garmany, C. D., \& Massey, P. 1986, \aj, 92, 48
\reference {cox06} Cox, N. L. J., Cordiner, M. A., Cami, J., et al. 2006, \aap, 447, 991
\reference {cox07} Cox, N. L. J., Cordiner, M. A., Ehrenfreund, P., et al. 2007, \aap, 470, 941
\reference {cro97} Crowther, P. A., \& Smith, L. J. 1997, \aap, 320, 500
\reference {cro02} Crowther, P. A., Hillier, D. J., Evans, C. J., et al. 2002, \apj, 579, 774
\reference {dach70} Dachs, J. 1970, \aap, 9, 95
\reference {dht01} Dame, T. M., Hartmann, D., \& Thaddeus, P. 2001, \apj, 547, 792
\reference {dan02} Danforth, C., Howk, J. C., Fullerton, A. W., Blair, W. P., \& Sembach, K. R. 2002, \apjs, 139, 81
\reference {dfl84} Danks, A. C., Federman, S. R., \& Lambert, D. L. 1984, \aap, 130, 62
\reference {debxx} de Boer, K. S., Fitzpatrick, E. L., \& Savage, B. D. 1985, \mnras, 217, 115
\reference {dx92} Deharveng, J.-M., \& Caplan, J. 1992, \aap, 259, 480
\reference {dic94} Dickey, J. M., Mebold, U., Marx, M., et al. 1994, \aap, 289, 357
\reference {dic00} Dickey, J. M., Mebold, U., Stanimirovi\'{c}, S., \& Staveley-Smith, L. 2000, \apj, 536, 756
\reference {ds94a} Diplas, A., \& Savage, B. D. 1994a, \apjs, 93, 211
\reference {ds94b} Diplas, A., \& Savage, B. D. 1994b, \apj, 427, 274
\reference {dob08} Dobashi, K., Bernard, J.-P., Hughes, A., et al. 2008, \aap, 485, 205
\reference {dwek98} Dwek, E. 1998, \apj, 501, 643
\reference {ev04a} Evans, C. J., Crowther, P. A., Fullerton, A. W., \& Hillier, D. J. 2004a, \apj, 610, 1021
\reference {ev04d} Evans, C. J., Howarth, I. D., Irwin, M. J., Burnley, A. W., \& Harries, T. J. 2004d, \mnras, 353, 601
\reference {ev06} Evans, C. J., Lennon, D. J., Smartt, S. J., \& Trundle, C. 2006, \aap, 456, 623
\reference {ev04b} Evans, C. J., Lennon, D. J., Trundle, C., Heap, S. R., \& Lindler, D. J. 2004b, \apj, 607, 451
\reference {ev04c} Evans, C. J., Lennon, D. J., Walborn, N. R., Trundle, C., \& Rix, S. A. 2004c, \pasp, 116, 909
\reference {fi85} Feitzinger, J. V., \& Isserstedt, J. 1983, \aaps, 51, 505
\reference {fg70} FitzGerald, M. P. 1970, \aap, 4, 234
\reference {fit85a} Fitzpatrick, E. L. 1985a, \apjs, 59, 77
\reference {fit85b} Fitzpatrick, E. L. 1985b, \apj, 299, 219
\reference {fit86} Fitzpatrick, E. L. 1986, \aj, 92, 1068
\reference {fit88} Fitzpatrick, E. L. 1988, \apj, 335, 703
\reference {fit91} Fitzpatrick, E. L. 1991, \pasp, 103, 1123
\reference {fit90} Fitzpatrick, E. L., \& Garmany, C. D. 1990, \apj, 363, 119
\reference {fit02} Fitzpatrick, E. L., Ribas, I., Guinan, E. F., et al. 2002, \apj, 564, 260
\reference {fs97} Fitzpatrick, E. L., \& Spitzer, L. 1997, \apj, 475, 623
\reference {fw90} Fitzpatrick, E. L., \& Walborn, N. R. 1990, \aj, 99. 1483
\reference {foe03a} Foellmi, C., Moffat, A. F. J., \& Guerrero, M. A. 2003a, \mnras, 338, 360
\reference {foe03b} Foellmi, C., Moffat, A. F. J., \& Guerrero, M. A. 2003b, \mnras, 338, 1025
\reference {fri11} Friedman, S. D., York, D. G., McCall, B. J., et al. 2011, \apj, 727, 33
\reference {fkh10} Fumagalli, M., Krumholz, M. R., \& Hunt, L. K. 2010, \apj, 722, 919
\reference {gar87} Garmany, C. D., Conti, P. S., \& Massey, P. 1987, \aj, 93, 1070
\reference {gs06} Gillmon, K., Shull, J. M., Tumlinson, J., \& Danforth, C. 2006, \apj, 636, 891
\reference {gk11} Gnedin, N. Y., \& Kravtsov, A. V. 2011, \apj, 728:88
\reference {gned09} Gnedin, N. Y., Tassis, K., \& Kravtsov, A. V. 2009, \apj, 697, 55
\reference {gold08} Goldsmith, P. F., Heyer, M., Narayanan, G., et al. 2008, \apj, 680, 428
\reference {gc98} Gordon, K. D., \& Clayton, G. C. 1998, \apj, 500, 816
\reference {gord03} Gordon, K. D., Clayton, G. C., Misselt, K. A., Landolt, A. U., \& Wolff, M. J. 2003, \apj, 594, 279
\reference {gun98} Gunderson, K. S., Clayton, G. C., \& Green. J. C. 1998, \pasp, 110, 60
\reference {her93} Herbig, G. H. 1993, \apj, 407, 142
\reference {isr97} Israel, F. P. 1997, \aap, 328, 471
\reference {iss75} Isserstedt, J. 1975, \aaps, 19, 259
\reference {iss78} Isserstedt, J. 1978, \aaps, 33, 193
\reference {iss79} Isserstedt, J. 1979, \aaps, 38, 329
\reference {iss82} Isserstedt, J. 1982, \aaps, 50, 7
\reference {jax01} Jaxon, E. G., Guerrero, M. A., Howk, J. C., et al. 2001, \pasp, 113, 1130
\reference {jenk09} Jenkins, E. B. 2009, \apj, 700, 1299
\reference {kalb05} Kalberla, P. M. W., Burton, W. B., Hartmann, D., et al. 2005, \aap, 440, 775
\reference {kalb10} Kalberla, P. M. W., McClure-Griffiths, N. M., Pisano, D., et al. 2010, \aap, 521, A17
\reference {kim03} Kim, S., Staveley-Smith, L., Dopita, M., et al. 2003, \apjs, 148, 473
\reference {koor82} Koornneef, J. 1982, \aap, 107, 247
\reference {kmt09} Krumholz, M. R., McKee, C. F., \& Tumlinson, J. 2009, \apj, 693, 216
\reference {len97} Lennon, D. J. 1997, \aap, 317, 871
\reference {len03} Lennon, D. J., Dufton, P. L., \& Crowley, C. 2003, \aap, 398, 455
\reference {ler09} Leroy, A. K., Bolatto, A., Bot, C., et al. 2009, \apj, 702, 352
\reference {ler11} Leroy, A. K., Bolatto, A., Gordon, K., et al. 2011, \apj, 737, 12
\reference {ler07} Leroy, A., Bolatto, A., Stanimirovi\'{c}, S., et al. 2007, \apj, 658, 1027
\reference {lpl10} Liszt, H. S., Pety, J., \& Lucas, R. 2010, \aap, 518, A45
\reference {lr92} Luks, Th., \& Rohlfs, K. 1992, \aap, 263, 41
\reference {mall01} Mallouris, C., Welty, D. E., York, D. G., et al. 2001, \apj, 558, 133
\reference {mb88} Maloney, P., \& Black, J. H. 1988, \apj, 325, 389
\reference {mz00} Marx-Zimmer, M., Herbstmeier, U., Dickey, J. M., et al. 2000, \aap, 354, 787
\reference {massa03} Massa, D., Fullerton, A. W., Sonneborn, G., \& Hutchings, J. B. 2003, \apj, 586, 996
\reference {mas02} Massey, P. 2002, \apjs, 141, 81
\reference {mas04} Massey, P., Bresolin, F., Kudritzki, R. P., Puls, J., \& Pauldrach, A. W. A. 2004, \apj, 608, 1001
\reference {mas95} Massey, P., Lang, C. C., DeGioia-Eastwood, K., \& Garmany, C. D. 1995, \apj, 438, 188
\reference {mas05} Massey, P., Puls, J., Pauldrach, A. W. A., et al. 2005, \apj, 627, 477
\reference {mas00} Massey, P., Waterhouse, E., \& DeGioia-Eastwood, K. 2000, \aj, 119, 2214
\reference {mas09} Massey, P., Zangari, A. M., Morrell, N. I., et al. 2009, \apj, 692, 618a
\reference {mat86} Mattila, K. 1986, \aap, 160, 157
\reference {mau89} Maurice, E., Bouchet, P., \& Martin, N. 1989, \aaps, 78, 445
\reference {mk10} McKee, C. F., \& Krumholz, M. R. 2010, \apj, 709, 308
\reference {meb97} Mebold, U., D$\ddot{\rm u}$sterberg, C., Dickey, J. M., Staveley-Smith, L., \& Kalberla, P. 1997, \apj, 490, L65
\reference {meix06} Meixner, M., Gordon, K. D., Indebetouw, R., et al. 2006, \aj, 132, 2268
\reference {mis99} Misselt, K. A., Clayton, G. C., \& Gordon, K. D. 1999, \apj, 515, 128
\reference {mol93} Molaro, P., Vladilo, G., Monai, S., et al. 1993, \aap, 274, 505
\reference {mor03} Morton, D. C. 2003, \apjs, 149, 205
\reference {nc02} Negueruela, I., \& Coe, M. J. 2002, \aap, 385, 517
\reference {nm86} Niemela, V. S., \& Morrell, N. I. 1986, \apj, 310, 715
\reference {ost10} Ostriker, E. C., McKee, C. F., \& Leroy, A. K. 2010, \apj, 721, 975
\reference {ost01} Ostrov, P. G. 2001, \mnras, 321, L25
\reference {pak98} Pak, S., Jaffe, D. T., van Dishoeck, E. F., Johansson, L. E. B., \& Booth, R. S. 1998, \apj, 498, 735
\reference {park92} Parker, J. Wm., Garmany, C. D., Massey, P., \& Walborn, N. R. 1992, \aj, 103, 1205
\reference {pp09} Pelupessy, F. I., \& Papadopoulos, P. P. 2009, \apj, 707, 954
\reference {planck11a} Planck Collaboration 2011a, \aap, in press (arXiv:1101.2037)
\reference {planck11b} Planck Collaboration 2011b, \aap, in press (arXiv:1101.2046)
\reference {rach09} Rachford, B. L., Snow, T. P., Destree, J. D., et al. 2009, \apjs, 180, 125
\reference {rach02} Rachford, B. L., Snow, T. P., Tumlinson, J., et al. 2002, \apj, 577, 221
\reference {rich00} Richter, P. 2000, \aap, 359, 1111
\reference {rd10} Roman-Duval, J., Israel, F. P., Bolatto, A. et al. 2010, \aap, 518, L74
\reference {rou78} Rousseau, J., Martin, N., Pr\'{e}vot, L., et al. 1978, \aaps, 31, 243
\reference {sbdb77} Savage, B. D., Bohlin, R. C., Drake, J. F., \& Budich, W. 1977, \apj, 216, 291
\reference {sfd98} Schlegel, D. J., Finkbeiner, D. P., \& Davis, M. 1998, \apj, 500, 525
\reference {sk99} Schmidt-Kaler, Th., Gochermann, J., Oestreicher, M. O., et al. 1999, \mnras, 306, 279
\reference {schn08} Schnurr, O., Moffat, A. F. J., St-Louis, N., Morrell, N. I., \& Guerrero, M. A., 2008, \mnras, 389, 806
\reference {svs85} Shull, J. M., \& van Steenberg, M. E. 1985, \apj, 294, 599
\reference {smit99} Smith, V. V. 1999, in New Views of the Magellanic Clouds, IAU Symp., vol. 190, Y.-H. Chu, N. B. Suntzeff, J. E. Hesser, \& D.A. Bohlender, eds., (San Francisco:  PASP), p. 259
\reference {snb97} Smith-Neubig \& Bruhweiler 1997, \aj, 114, 1951
\reference {snb99} Smith-Neubig \& Bruhweiler 1999, \aj, 117, 2856
\reference {sof06} Sofia, U. J., Gordon, K. G., Clayton, G. C., et al. 2006, \apj, 636, 753
\reference {sbhc86} Songaila, A., Blades, J. C., Hu, E. M., \& Cowie, L. L. 1986, \apj, 303, 198
\reference {stan99} Stanimirovi\'{c}, S., Staveley-Smith, L., Dickey, J. M., Sault, R. J., \& Snowden, S. L. 1999, \mnras, 302, 417
\reference {stan04} Stanimirovi\'{c}, S., Staveley-Smith, L., \& Jones, P. A. 2004, \apj, 604, 176
\reference {stas04} Stasi\'{n}ska, G., Graefener, G., Pe{n}a, M., et al. 2004, \aap, 413, 329
\reference {ss03} Staveley-Smith, L., Kim, S., Calabretta, M. R., Haynes, R. F., \& Kesteven, M. J. 2003, \mnras, 339, 87
\reference {ss97} Staveley-Smith, L., Sault, R. J., Hatzidimitriou, D., Kesteven, M. J., \& McConnell, D. 1997, \mnras, 289, 225
\reference {thom82} Thompson, G. I., Nandy, K., Morgan, D. H., Willis, A. J., \& Wilson, R. 1982, \mnras, 200, 551
\reference {tum02} Tumlinson, J., Shull, J. M., Rachford, B. L., et al. 2002, \apj, 566, 857
\reference {vid87} Vidal-Madjar, A., Andreani, P., Cristiani, S., et al. 1987, \aap, 177, L17
\reference {vla93} Vladilo, G., Molaro, P., Monai, S., et al. 1993, \aap, 274, 37
\reference {wak11} Wakker, B. P., Lockman, F. J., \& Brown, J. M. 2011, \apj, 728, 159
\reference {wal77} Walborn, N. R. 1977, \apj, 215, 53
\reference {wal02} Walborn, N. R., Fullerton, A. W., Crowther, P. A., et al. 2002, \apjs, 141, 443
\reference {wal95} Walborn, N. R., Lennon, D. J., Haser, S. M., Kudritzki, R.-P., \& Voels, S. A. 1995, \pasp, 107, 104
\reference {wal00} Walborn, N. R., Lennon, D. J., Heap, S. R., et al. 2000, \pasp, 112, 1243
\reference {wal04} Walborn, N. R., Morrell, N. I., Howarth, I. D., et al. 2004, \apj, 608, 1028
\reference {wal93} Walborn, N. R., Phillips, M. M., Walker, A. R., \& Elias, J. H. 1993, \pasp, 105, 1240
\reference {wal89} Walborn, N. R., Prevot, M. L., Prevot, L., et al. 1989, \aap, 219, 229
\reference {wal10} Walborn, N. R., Howarth, I. D., Evans, C. J., et al. 2010, \aj, 139, 1283
\reference {wat11} Watson, D. 2011, \aap, 533, A16
\reference {way90} Wayte, S. R. 1990, \apj, 355, 473
\reference {wc10} Welty, D. E., \& Crowther, P. A. 2010, \mnras, 404, 1321
\reference {wfgtl06} Welty, D. E., Federman, S. R., Gredel, R., Thorburn, J. A., \& Lambert, D. L. 2006, \apjs, 165, 138
\reference {wfsy99a} Welty, D. E., Frisch, P. C., Sonneborn, G., \& York, D. G. 1999, \apj, 512, 636
\reference {whm03} Welty, D. E., Hobbs, L. M., \& Morton, D. C. 2003, \apjs, 147, 61
\reference {wlbhy97} Welty, D. E., Lauroesch, J. T., Blades, J. C., Hobbs, L. M., \& York, D. G. 1997, \apj, 489, 672
\reference {wlbhy01} Welty, D. E., Lauroesch, J. T., Blades, J. C., Hobbs, L. M., \& York, D. G. 2001, \apj, 554, L79
\reference {wolf10} Wolfire, M. G., Hollenbach, D., \& McKee, C. F. 2010, \apj, 716, 1191
\reference {wb02} Wong, T., \& Blitz, L. 2002, \apj, 569, 157
\reference {wong09} Wong, T., Hughes, A., Fukui, Y., et al. 2009, \apj, 696, 370
\reference {wong11} Wong, T.,Hughes, A., Ott, J., et al. 2011, \apjs, in press (arXiv:1108.5715)
\reference {xc99} Xu, J. \& Crotts, A. P. S. 1999, \apj, 511, 262
\end{references}
\end{document}